\documentclass[acmsmall]{acmart}

\PassOptionsToPackage{table,prologue}{xcolor}

\usepackage{enumitem}
\usepackage{booktabs}
\usepackage{tabularx}
\usepackage{array}
\usepackage{graphicx}
\usepackage{subcaption}
\usepackage{multirow} 
\usepackage{makecell}
\usepackage{longtable}
\usepackage{geometry}
\usepackage[table]{xcolor}
\usepackage{xcolor}
\usepackage{colortbl}

\newcolumntype{L}[1]{>{\raggedright\arraybackslash}p{#1}}

\AtBeginDocument{%
  }

\setcopyright{acmlicensed}
\copyrightyear{2018}
\acmYear{2018}
\acmDOI{XXXXXXX.XXXXXXX}

\acmConference[Conference acronym 'XX]{Make sure to enter the correct
  conference title from your rights confirmation emai}{June 03--05,
  2018}{Woodstock, NY}
\acmISBN{978-1-4503-XXXX-X/18/06}

\begin{document}

\title[A Turing Test for ``Localness'']{A Turing Test for ``Localness'': Conceptualizing, Defining, and Recognizing Localness in People and Machines}

\newcommand{\addcite}[1]{{\color{red}{\noindent{[\textbf{ADDCITE}: #1]}}}}

\author{Zihan Gao}
\email{zihan.gao@wisc.edu}
\orcid{0009-0004-5207-5511}
\affiliation{%
  \institution{Information School, University of Wisconsin-Madison}
  \city{Madison}
  \country{Wisconsin}
  \country{USA}
}

\author{Justin Cranshaw}
\email{justin@cranshaw.me}
\orcid{0009-0002-5856-1735}
\affiliation{%
  \institution{Maestro AI}
  \city{Seattle}
  \country{Washington}
  \country{USA}
}

\author{Jacob Thebault-Spieker}
\email{jacob.thebaultspieker@wisc.edu}
\orcid{0000-0003-1569-4466}
\affiliation{%
  \institution{Information School, University of Wisconsin-Madison}
  \city{Madison}
  \country{Wisconsin}
  \country{USA}
}

\renewcommand{\shortauthors}{Trovato et al.}

\begin{abstract}
As digital platforms increasingly mediate interactions tied to place, ensuring genuine local participation is essential for maintaining trust and credibility in location-based services, community-driven platforms, and civic engagement systems. However, localness is a social and relational identity shaped by knowledge, participation, and community recognition. Drawing on the German philosopher Heidegger's concept of dwelling---which extends beyond physical presence to encompass meaningful connection to place---we investigate how people conceptualize and evaluate localness in both human and artificial agents.
Using a chat-based interaction paradigm inspired by Turing's Imitation Game and Von Ahn's Games With A Purpose, we engaged 230 participants in conversations designed to examine the cues people rely on to assess local presence. Our findings reveal a multi-dimensional framework of localness, highlighting differences in how locals and nonlocals emphasize various aspects of local identity. We show that people are significantly more accurate in recognizing locals than nonlocals, suggesting that localness is an affirmative status requiring active demonstration rather than merely the absence of nonlocal traits. Additionally, we identify conditions under which artificial agents are perceived as local and analyze participants' sensemaking strategies in evaluating localness. Through predictive modeling, we determine key factors that drive accurate localness judgments.
By bridging theoretical perspectives on human--place relationships with practical challenges in digital environments, our work informs the design of location-based services that foster meaningful local engagement. Our findings contribute to a broader understanding of localness as a dynamic and relational construct, reinforcing the importance of dwelling as a process of belonging, recognition, and engagement with place.

\end{abstract}




\maketitle

\section{Introduction}

\begin{quote}
    ``Well deserving, but poetically, man dwells on this earth.''  
    — \textit{In Lovely Blue}, Friedrich Hölderlin
\end{quote}

This poetic reflection, central to Martin Heidegger's philosophy on dwelling, points to a fundamental tension in how humans connect with place: between mere physical presence and authentic belonging \cite{heidegger1971building,heidegger1975poetry}. For Heidegger, true dwelling transcends simple habitation to encompass a way of being that is simultaneously receptive to place and actively engaged in creating meaningful connections. As human movement patterns show increasing rootlessness, with more and more people moving and uprooting their lives \cite{gao2024journeying}, individuals face complex challenges in developing ``dwelling'' --- a state that combines temporal presence, active participation, and contemplative engagement with place.

This philosophical framing of dwelling, and the human need to establish dwelling, points to an emerging challenge for digital platforms: effectively recognizing and supporting authentic local presence. 
The complexity of authentic local presence authenticity manifests across multiple dimensions of human-place relationships, which creates barriers for effective computational tools to achieve this goal. 
For instance, research on ``sense of place'' has revealed how individuals develop emotional and psychological connections with place \cite{tuan1977space,jorgensen2001sense,raymond2010measurement,harrison1996re,lentini2010space}, while parallel studies have explored how people understand and navigate physical spaces through both spatial features and meaningful social interactions \cite{harrison1996re,quercia2014shortest,hsu2019smell,graham2022geographies,paananen2021investigating}. These investigations highlight the multifaceted nature of place experience, encompassing geometric, sensory, cultural, personal, and relational aspects \cite{lentini2010space}. Moreover, these human-place bonds are not static; when people change locales, they face diverse challenges that require rebuilding authentic connections --- true ``dwelling'' --- in their new environments \cite{gao2024journeying,taylor2015data}. This rich theoretical foundation suggests that any system for assessing local authenticity must embrace and account for these inherent complexities.

For nearly three decades, the fields of CSCW and HCI have explored various aspects of how people develop meaningful relationships with places  \cite{harrison1996re}. This research spans multiple directions: exploring the value of ``place'' \cite{harrison1996re},  designing tools to support specific local communities \cite{hu2013whoo,malmborg2015designing,alvarado2020fostering,graham2022geographies}, establishing the importance of localness in user-generated content \cite{johnson2016not,thebault2018distance}, exploring the potential for new kinds of technologies that represent the qualitative experiences of being a local more effectively \cite{cranshaw2016journeys,ludford2007capturing, quercia2015smelly, hsu2019smell, panciera2013soil}. These studies have significantly advanced our understanding of human-place relationships, operating on the premise that user-generated content can provide authentic local information. This assumption extends beyond academic research to commercial applications, where location-based services and mapping platforms increasingly rely on user-generated local content as a cornerstone of their functionality \cite{colley2017geography, osm_apple, osm_lyft, kariryaa2018defining}.

Local user-generated content offers unique, community-driven insights that are often overlooked by broader media \cite{goodchild2007citizens}. This locally sourced knowledge plays an important role in enhancing location-based services, including local recommendations and local information retrieval \cite{ludford2007capturing,kariryaa2018defining,white2012characterizing}, as well as supporting civic engagement through digital platforms \cite{balestrini2017city,garbett2016app,vlachokyriakos2014postervote}. Additionally, user-generated content contributes to digital placemaking efforts, fostering local community building and a stronger sense of place \cite{sun2015importance,cranshaw2016journeys,sun2017movemeant,sun2018multi}.
Research has consistently shown that the quality of this locally sourced content strongly correlates with contributors' authentic local familiarity \cite{thebault2018distance, johnson2016not, eckle_quality_2015}. This relationship highlights the critical importance of understanding and recognizing genuine local connections in digital environments. 
However, current computational approaches to assessing localness authenticity fall short of capturing the full complexity of these connections. Most existing systems rely on simplified proxy measures: length of residence \cite{thebault2018geographic}, geometric distance \cite{hecht2010localness, jurgens2015geolocation}, check-in frequencies \cite{johnson2016geography, popescu2010mining}, or profile-based location data \cite{hecht2011tweets}. These metrics fail to encompass the deeper contemplative and participatory aspects of dwelling \cite{kariryaa2018defining, graham2022geographies, johnson2016not}. As digital platforms increasingly become the primary venues for communities to share and validate local knowledge, the challenge of accurately recognizing authentic localness has become particularly critical.

This technological challenge also compounds existing concerns about coordinated campaigns that exploit local credibility for spreading misinformation \cite{starbird2019disinformation, kumar2018false}. The emergence of Large Language Models (LLMs) further complicates this landscape in two significant ways. First, these systems' ability to generate sophisticated local-seeming content makes it increasingly difficult to distinguish genuine local participation from impersonation \cite{salewski2023context,puccetti2024ai}. Second, while LLMs show promise for processing local context computationally, questions remain about their true understanding of this context and their capacity to support authentic local engagement \cite{yin2021broaden, yin2022geomlama,sap2019atomic,acharya2020towards}. As platforms face growing pressure both to computationally assess local participation authenticity and to understand LLMs' capabilities and limitations in supporting local engagement, the need for better approaches becomes increasingly apparent. 
This calls for both advanced methods of evaluating authentic localness --- dwelling --- in digital environments and a critical examination of how computational systems can appropriately support these processes.

The intersection of localness concepts, computational localness authenticity assessment challenges, and emerging LLM capabilities points to several critical research needs. 
First and foremost, a fundamental necessary step is developing a deeper understanding of how people conceptualize localness itself. This foundational knowledge is essential for designing effective computational authenticity assessments and evaluating how digital systems support or potentially misrepresent authentic local participation \cite{johnson2016not, graham2022geographies}. 
While platforms struggle to distinguish authentic local participation, scholarly understanding of how humans themselves navigate this challenge is lacking. Understanding whether and how people can reliably differentiate between local and nonlocal participants --- and between human and AI-generated content --- may serve as inspiration. 
Beyond specific differentiation judgements, understanding the sensemaking process --- how individuals gather, filter, and synthesize information --- could inform practical pathways for computational localness authenticity assessment \cite{kariryaa2018defining}, just as prior work has done for other complex human judgement goals \cite{venkatagiri2019groundtruth, mohanty2019photo}. 
Moreover, as LLMs increase in sophistication at generating local-seeming content, this further complicates the issue and heightens the need to explore how LLM's ability to represent localness affects local authenticity recognition, with important implications for both detecting impersonation and leveraging AI to support genuine local connections \cite{bender2021dangers, yin2022geomlama}.

Based on these needs, we investigate the following research questions:
\begin{itemize}
    \item RQ1: What defines localness according to participants?
    \item RQ2: How accurately can participants distinguish local vs. nonlocal and human vs. LLM partners?
    \item RQ3: When do people perceive LLMs as local or as human?
    \item RQ4: How do participants engage in localness sensemaking during judgments?
    \item RQ5: What factors predict accurate localness judgments?
\end{itemize}

To conduct this work, we developed the ``Localness Imitation Game,'' a novel experimental framework combining elements of Alan Turing's germinal ``Imitation Game'' \cite{turing1950} with Von Ahn's Games With A Purpose techniques\cite{von2004labeling, von2006games}. This approach enables systematic investigation of how people assess localness authenticity through natural conversation while supporting comparison across human and LLM interactions. Inspired by prior techniques of developing computational legibility of complex issues \cite{von2008designing}, our methodology draws on sensemaking theory, recognizing localness authenticity inference as a complex cognitive process involving gathering, filtering, and synthesizing multiple cues \cite{pirolli2005sensemaking}. The multi-turn chat format supports observation of how people progressively build understanding and make judgments about localness through natural dialogue, while enabling systematic comparison between human-human and human-LLM interactions.


Through this methodological approach, our study makes five key contributions to social computing research:

\begin{itemize}
    \item \textbf{Localness Is Multidimensional and Socially Constructed:} 
    We develop a conceptual framework identifying three core dimensions of localness: \textit{Cognitive} (e.g., knowledge of landmarks, historical facts, or recommendations), \textit{Physical} (e.g., birthplace, length of residence), and \textit{Relational} (e.g., emotional ties, community participation). While locals emphasized relational and participatory aspects of localness, nonlocals were more likely to focus on physical and temporal markers. These contrasting perspectives reveal that localness is not a static status, but a dynamic identity shaped by one's self-perception and lived experience.

    \item \textbf{Human Accuracy in Localness Judgments Is Uneven:} 
    Analysis of 932 chat rounds shows that participants were generally successful at identifying local human partners, but often struggled to classify nonlocals correctly. Judgments were significantly more accurate in human-human interactions than human-LLM, particularly when conversations included hyperlocal references (e.g., street nicknames, colloquial expressions) or relational details.

    \item \textbf{LLMs Are Frequently Perceived as Nonlocal:} 
    Despite producing fluent local-seeming content, LLMs were usually identified as nonlocal. Participants flagged shallow emotional references, vague spatial descriptions, and lack of personal memory as signs of inauthenticity. These results highlight a persistent gap between language generation and lived experience, and motivate further development of LLMs' localness representation.

    \item \textbf{Sensemaking Strategies Shape Judgment Outcomes:} 
    Participants employed distinct sensemaking tactics during interactions—such as probing for neighborhood-specific landmarks, asking about routine behaviors, and cross-checking answers across chat turns. Successful judgments often drew from multiple layers of evidence, combining factual accuracy with signs of relational and affective presence. In contrast, incorrect judgments typically relied on single cues (e.g., mention of a known place without depth).

    \item \textbf{Knowledge and Emotion Are Key Predictors of Accuracy:} 
    Predictive modeling identified that indicators from the \textit{Knowledge} (e.g., quality of recommendations, contextual fluency) and \textit{Emotional} (e.g., expressions of belonging or comfort) dimensions were the strongest predictors of accurate localness assessments. These findings suggest design strategies for future systems that aim to model, infer, or support authentic local participation.
\end{itemize}

Together, these contributions advance our understanding of how authentic local presence --- true ``dwelling'' --- manifests in digital environments while providing concrete guidance for designing systems that better support genuine local participation in an era of increasing digital mediation and sophisticated impersonation capabilities.

\section{Related Work}

Our work here builds on and synthesizes four interconnected bodies of literature. 
To characterize and draw connections between these bodies of literature, we begin with fundamental theories of space, place, and localness, which provide the conceptual foundation for understanding how people develop connections with their environments. 
We then examine how these concepts have been operationalized in location-based systems, revealing both the potential for and limitations of current approaches to capturing local knowledge. 
The emergence of LLMs introduces new possibilities and challenges in representing local knowledge, leading us to review current understanding of LLMs' capabilities and limitations in handling local context. 
Finally, we draw on sensemaking literature to understand how people process and evaluate complex information, particularly in collaborative and AI-augmented contexts. 
This combination of perspectives enables us to critically examine how humans assess localness authenticity and how computational systems might support this process.

\subsection{Space, Place and Localness}

The concepts of space and place are foundational in understanding localness. Space refers to the physical dimensions of a location, while place encompasses social relationships and emotional attachments, creating a sense of belonging and identity \cite{tuan1977space,relph1976place}. This transformation from space to place is socially constructed, and reflects and human values \cite{relph1976place}. \citet{jorgensen2001sense,stedman2008we} emphasize that a ``sense of place'' includes the psychological and emotional connections that people have with their surroundings, influenced by personal experiences and memories. \citet{manzo2005better} adds that these experiences of evolving identity, personal growth, and feelings of safety or threat are what make places important to people. Within HCI, \citet{harrison1996re} argue that a sense of place in collaborative systems is derived not just from spatial properties but from shared cultural understandings and behaviors. They emphasize the importance of designing systems that foster a sense of place through social meaning rather than merely physical structures. This aligns with the work of \citet{lentini2010space}, who describe place as involving geometrical, sensorial, cultural, personal, and relational experiences. These various dimensions from \citet{lentini2010space} highlight how people engage with their surroundings through spatial qualities, sensory input, cultural norms, personal growth, and social interactions.

Similarly, localness --- or someone becoming local --- emerges from the interactions between humans and places, with the accumulation of platial and spatial connections and experiences. However, existing literature on  technical definitions of locals and non-locals often relies on simplistic measures such as zip codes \cite{bandy2021errors}, check-in frequency in cities \cite{sanchez2022travelers}, locations of information searching logs \cite{white2012characterizing}, and self-reported profiles \cite{wu2011mining}. While these methods enable efficient analysis, they are limited in scope and do not fully capture the nuanced, rich complexity  of what it means to be ``local''. \citet{kariryaa2018defining} collected existing computational definitions of localness in the computing literature and evaluated their effectiveness, indicating the need for more sophisticated systematic definitions.

\subsection{Localness in Location-based Systems}

Location-based systems have significantly advanced our ability to capture diverse forms of local data, contributing to a deeper understanding of urban environments. Such systems are useful for urban planning \cite{silva2019urban}, placemaking \cite{garbett2016app}, and civic empowerment \cite{taylor2015data}. \citet{taylor2015data} introduced the concept of "data-in-place," highlighting the importance of embedding data within the local timeframes, regions, and social settings of communities. This approach underscores the significance of considering both material and social dimensions of data in community contexts. For example, \citet{quercia2014shortest} collected personal, sentimental, and aesthetic experiences of different locations to recommend routes based on subjective aspects of urban spaces. Their work set the stage for further explorations into the experiential qualities of urban environments. Building on this, \citet{quercia2015smelly} investigated urban smellscapes, demonstrating the value of integrating sensory data into urban planning to enhance the livability of cities. Similarly, \citet{hsu2019smell} developed Smell Pittsburgh, a community-empowered mobile smell reporting system, to provide a richer understanding of urban air quality from a personal perspective. \citet{eissfeldt2019supporting} proposed a citizen participatory noise sensing concept to gather sensor data on urban noise mobility, highlighting the role of local community engagement in data monitoring.

Crowdsourcing and social media have, broadly, proven effective for collecting local data, offering insights into the collective sentiments and experiences of urban spaces \cite{cranshaw2012livehoods}. Location-based social networks enable new user-generated content mechanisms for understanding urban spaces. For example, \citet{yuan2019assessing} assessed placeness using geotagged Yelp data, showing how user reviews and ratings reflect the unique characteristics of different places. \citet{aiello2016chatty} constructed sound maps of urban areas from social media data, integrating auditory experiences into urban planning. \citet{jenkins2016crowdsourcing} also used geo-tagged social media and Wikipedia data to develop a collective sense of place, illustrating the potential of crowdsourced content to enrich our understanding of urban environments. These studies illustrate how user-generated content can inform our understanding of urban environments by reflecting collective experiences and local expertise.

Recognizing the broader value of local knowledge, researchers have explored extracting local-related information from social media to build hyperlocal online communities \cite{hu2013whoo}, collecting information from locals and non-locals on social media \cite{ference2013location}, information retrieval systems \cite{white2012characterizing}, map applications \cite{wu2011mining}, and providing targeted information recommendations for locals and non-locals about points of interest \cite{sanchez2022travelers} and search engine advertisements \cite{bandy2021errors}. Importantly, \citet{kariryaa2018defining} summarized and evaluated existing computational concepts of localness, demonstrating that existing definitions have weaknesses. Importantly, however, most technical tools that leverage computational notions of localness do so in urban settings, which is likely due to well established patterns in social media and user-generated content platforms that find that there is much less content in lower-income and non-urban settings \cite{johnson2016not,thebault2018geographic,li2013spatial,hecht2010tower}. These disparities limit the ability of computational systems to generalize across different geographic and socioeconomic contexts, raising critical questions about how digital platforms define and support localness beyond data-rich urban environments.

Beyond analyzing existing data, researchers have also explored ways to actively cultivate digital placemaking --- using technology to shape how people engage with place and community. Some approaches focus on fostering community presence in otherwise impersonal or transient spaces. For example, \citet{cranshaw2016journeys} developed Journeys \& Notes, a mobile platform that facilitates asynchronous social interactions among travelers by allowing them to log personal experiences along travel routes. By layering social narratives onto these spaces, the system aimed to transform them from anonymous pathways into meaningful, shared environments. Other research has emphasized using digital systems to enhance community awareness and local engagement. A series of studies by Sun and colleagues explored how technology can facilitate social connections within urban environments. Their early work \cite{sun2015importance} emphasized the importance of integrating open-ended, playful interactions into location-based systems to deepen users’ sense of place. Building on this foundation, they developed MoveMeant \cite{sun2017movemeant}, a system that enabled users to share location histories anonymously, allowing individuals to develop a heightened awareness of community presence and movement patterns. A subsequent large-scale deployment \cite{sun2018multi} demonstrated that such interactive systems could go beyond passive information sharing to actively foster community ties, making shared local experiences more visible and reinforcing social connections within a place.

By integrating insights from both passive data collection and active digital placemaking, location-based systems have the potential to foster meaningful connections between people and places. However, the effectiveness of these systems hinges on how localness is conceptualized and assessed, particularly in contexts where local data signals may be sparse or unreliable. There is a call of a robust mechanism for evaluating localness authenticity --- the extent to which an individual genuinely embodies local presence beyond surface-level signals. Our study addresses this gap by investigating computational localness authenticity assessment, offering how people verify and affirm local presence in interactive settings. By grounding our work in the concept of dwelling, which emphasizes authentic engagement with place rather than mere physical presence, we offer new insights into the relational and interpretative nature of localness. 

\subsection{Localness and LLMs}

LLMs have rapidly gained prominence due to their ability to incorporate linguistic context into predictions, producing responses that often appear surprisingly coherent and intelligent. 
Previous studies have shown that LLMs can effectively \textit{answer questions} based on provided context \cite{choi2018quac, zhou2018dataset}, indicating that some degree of correct information is encoded through this contextual prediction. However, research has also identified fundamental limitations in LLMs' contextual understanding—particularly regarding commonsense and localized knowledge. \citet{petroni2019language} examined LLMs as knowledge bases and found that while these models can store vast amounts of factual information, they struggle with commonsense reasoning that is contingent on specific contexts.
This limitation has spurred research on integrating commonsense reasoning into LLMs, such as ATOMIC \cite{sap2019atomic} and COMET \cite{bosselut2019comet}, which enhance inferential capabilities through structured knowledge graphs. 
ATOMIC provides a large-scale knowledge graph for if-then reasoning, enhancing models' ability to understand social interactions and causal relationships. COMET extends this by automatically constructing knowledge graphs, further improving inferential capabilities. However, while these models improve on context-aware question answering and general commonsense reasoning, they remain inadequate for encoding culturally and geographically specific nuances. 

Local and cultural specificity is particularly challenging for LLMs because these models are trained on large-scale, predominantly global datasets that often obscure regional variations in language, culture, and community knowledge. Studies have shown that even when LLMs succeed in commonsense reasoning, they struggle with domain-specific or geographically situated information. For example, research by \citet{acharya2020towards} and \citet{yin2022geomlama} suggest that highlights the difficulty LLMs face in capturing geo-diverse knowledge, demonstrating that models trained predominantly on Western-centric data perform significantly worse when applied to non-Western contexts. Similarly, studies such as \citet{liu2021visually} and \citet{shwartz2022good} indicate that localness is not a uniform concept but is shaped by different interpretations among local and non-local individuals. This suggests that LLMs, which primarily learn from broad linguistic patterns, may not effectively encode the nuanced, situated knowledge necessary for assessing local authenticity and supporting local connections.

Existing research highlights gaps in LLMs’ ability to encode geographically grounded commonsense knowledge, but relatively little work has explored how these limitations affect human perceptions of localness in interactive settings. Our study addresses this gap by investigating how humans assess localness in both human and AI-generated interactions, shedding light on the specific ways LLMs succeed or fail in demonstrating local presence. Our work examines localness as a socially negotiated concept, assessing how localness is signaled and recognized in human-AI interactions. By bridging research on LLM limitations with theories of human-place relationships, our study contributes to both computational approaches to localness and the design of AI systems that aim to detect and foster authentic community engagement.

\subsection{Sensemaking for Localness Authenticity Assessmemt}

Sensemaking is the process through which individuals or organizations interpret and construct meaning from information, particularly in complex or uncertain situations \cite{dervin1983overview, dervin1992mind}. It plays a crucial role in human decision-making, particularly in situations requiring the synthesis of multiple, sometimes ambiguous, pieces of information. The sensemaking process is dynamic and iterative, characterized by cycles of information gathering, schema creation, hypothesis generation, and refinement \cite{russell1993cost, pirolli2005sensemaking}. While experts often navigate sensemaking tasks flexibly by detecting patterns and handling ambiguity, novices tend to rely on simpler heuristics and require structured support \cite{foong2017novice, venkatagiri2019groundtruth}. Understanding these differences has been instrumental in designing systems that facilitate structured sensemaking, particularly in information-rich and high-stakes environments.

Advances in visual analytics and interactive data visualization have significantly enhanced the sensemaking process by allowing users to manipulate and explore data visually, thereby supporting both individual and collaborative sensemaking efforts \cite{bradel2014multi}. In certain scenarios, such as collaborative sensemaking \cite{zhao2017supporting}, real-time sensemaking of conversations poses unique challenges due to the highly synchronous, linear, and rapid exchange of verbal information \cite{deshpande2005building,goyal2013leveraging}. Without structured tools, individuals may struggle to track and interpret key information, leading to miscommunication, failures in grounding, and inefficiencies in decision-making. To mitigate these challenges, research has explored external representations that transform sequential verbal exchanges into structured, manipulable formats, facilitating better organization and interpretation of conversational data \cite{goyal2013effects}.

Recent advances in generative AI have further expanded the possibilities for computational sensemaking by enabling more structured, interactive, and personalized experiences \cite{ma2024beyond}. Techniques such as multimodal visualization improve the understanding of complex information by converting text-based outputs into structured representations like node-link diagrams \cite{jiang2023graphologue} or nonlinear interfaces that align with human cognitive processes \cite{suh2023sensecape}. Similarly, structured problem-solving approaches, such as the Tree of Thoughts framework, organize multi-step reasoning tasks in a hierarchical manner, mimicking human cognitive strategies and improving structured deliberation \cite{yao2024tree, long2023large}. Personalization in AI-driven sensemaking has also gained attention, with systems like Citesee integrating historical context into citation analysis to help researchers track idea evolution \cite{chang2023citesee}, and \citet{kang2023synergi} developing mixed-initiative tools that blend user-driven exploration with AI-supported synthesis for academic research.

Our study frames localness assessment as a sensemaking task, examining how individuals process and evaluate localness authenticity in real-time conversations. Our experimental system is designed to scaffold sensemaking by structuring participants’ engagement with conversational data --- they must follow the system flow to extract, filter, and integrate information to determine the authenticity of local presence. Building on sensemaking theory, this study aims to explore theoretical understandings of human-place relationships and practical approaches for designing tools that support structured reasoning in localness verification.

\section{Method}

\subsection{Theoretical Framework}

Our methodology draws on two complementary theoretical frameworks to address the complex challenge of studying localness authenticity assessments: the structured evaluation approach of the Imitation Game and the cognitive lens of sensemaking theory. The combination of these frameworks enables us to systematically examine how people evaluate and verify claims of local authenticity in conversation.

In 1950, renowned computer scientist Alan Turing posed \textit{The Imitation Game} \cite{turing1950}. The common understanding of Turing's ``imitation game'' sets up a situation where an ``interrogator'' is communicating with one human interlocutor and one computational interlocutor through text. The interrogator in Turing's ``imitation game'' has incomplete information, and the game is to guess correctly about which interlocutor is human. 56 years later, in 2006, \cite{von2006games} argued ``computers still don't possess the basic conceptual intelligence or perceptual capabilities that most humans take for granted'' and posed that Games With A Purpose, games where two human players have incomplete information and are trying to reach agreement on labeling images, may be a way to ``address problems that computers can't yet tackle on their own and eventually teach computers many of these human talents.'' Games With A Purpose \cite{von2006games} and Turing's ``imitation game'' before it, have a shared core insight: structuring a communication dyad with incomplete information can be an effective way to evaluate and characterize complex, multi-faceted concepts like ``label everything in this image'' or ``choose which interlocutor is human.'' 

Here, we take inspiration from the core insight of Turing's ``imitation game'' which echoes through \citet{von2006games}'s Games With A Purpose, and develop a similar structure to evaluate and characterize ``localness.'' After all, as noted above, the complexity, multifacetedness, and social construction of localness makes it difficult to fully conceptualize or operationalize computationally. In other words, we pose that while localness is complex and difficult to articulate fully, it is likely that people have an intuitive understanding of localness, just as they do for objects in an image, or for recognizing another human. The game-like structure provides a systematic way to surface these implicit localness authenticity assessments strategies.

We integrate this structural approach with sensemaking theory, which provides a cognitive framework for understanding how people gather, filter, and synthesize information to make complex judgments \cite{pirolli2005sensemaking,russell1993cost,zhang2014towards}. Our methodology operationalizes sensemaking theory through three interconnected phases that map directly to established sensemaking processes:
\begin{itemize}
    \item \textbf{Information Gathering}: This phase mirrors \citet{pirolli2005sensemaking}'s ``foraging loop,'' where participants actively probe for information through chat-based interactions. By structuring information foraging within the Imitation Game format, we create natural foraging behaviors as participants seek out diagnostic signals of localness.
    \item \textbf{Information Structuring}: This phase aligns with the ``schema development'' process in sensemaking theory, where participants filter and organize gathered evidence. The chat format allows participants to iteratively build and test mental models of what constitutes authentic local knowledge and behavior, guiding their subsequent queries.
    \item \textbf{Knowledge Consolidation}: This final phase corresponds to the ``hypothesis refinement'' stage of sensemaking, where participants synthesize their accumulated evidence to make localness judgments. The requirement to provide explicit rationales for these judgments allows us to examine how participants weigh and integrate different types of evidence.
\end{itemize}

The Imitation Game structure creates natural foraging and synthesizing loops, allowing us to observe how participants navigate between gathering new information and refining their existing understanding. It provides a structured framework for analyzing both the process and outcome of localness authenticity assessment attempts, informing our analysis of how local authenticity is progressively constructed through conversation. Additionally, it offers structure for examining how large language models might succeed or fail in demonstrating local authenticity.

\subsection{System}

Drawing on our theoretical framework, we developed an interactive platform to support systematic investigation of localness assessment through natural conversation. The system design evolved through multiple iterations of pilot testing to balance validity with experimental control. Here we describe the final system architecture and key design decisions.

\begin{figure}
    \centering
    \includegraphics[width=1\linewidth]{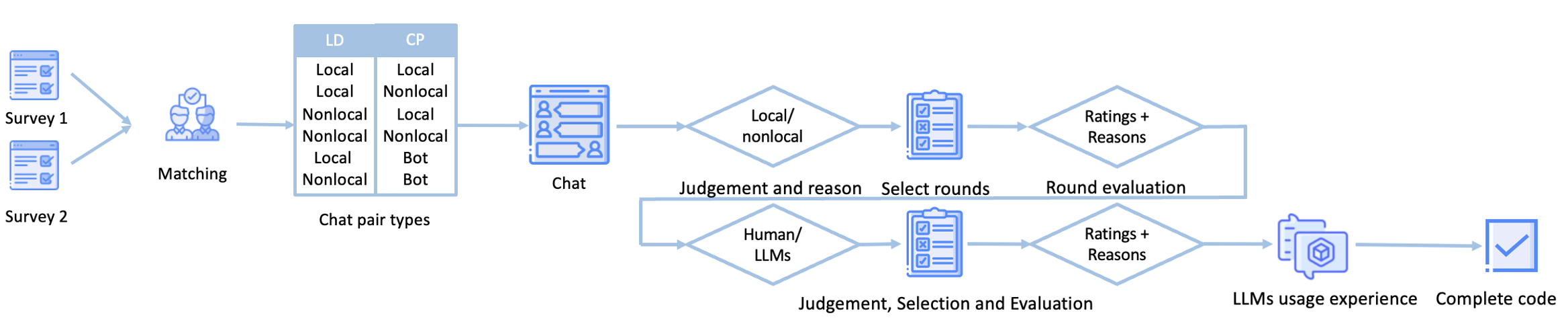}
    \caption{System overview}
    \label{fig:system}
\end{figure}

\subsubsection{Core System Architecture}

The system comprises five integrated components (Figure~\ref{fig:system}), each designed to support specific aspects of our research questions while maintaining natural interaction:

\textbf{Initial Assessment: }
Participants initially filled out two primary survey instruments. The first focused on demographic and contextual information about the participant, including name, email, age, gender, location, native language, English fluency, duration of residence, whether they considered themselves local to where they lived (yes or no), how local they felt (1-5 point scale), and how they interpreted localness in an open-text response. For non-local participants, we also asked them to share where they consider themselves local to. Once participants completed the first survey, we also asked participants to complete a standardized sense-of-place assessment \cite{raymond2010measurement}, as a way to triangulate their self-reported localness status.

\textbf{Matching Mechanism: }
Upon completing the surveys, participants are placed in a waiting pool to be paired for chat. Each chat pair consists of a Localness Decider (LD) and a Chat Partner (CP). The LD asks questions of their chat partner, in order to determine if their chat partner is local or non-local. In this way, the task is uni-directional, the LD is the person in the dyad who is supposed to be deciding the localness of their chat partner. The CP responds to the LD's questions, primarily. We do ask the CP if they believe the LD is local or not, but this is not a primary dimension of analysis. In some cases, the CP will be a randomly selected LLM, rather than a human chat partner. Regardless of whether the LD is chatting with a human or an LLM, they are not notified, and we asked them to judge if their CP is a human or an LLM at the end. Because of the structure of this dyad, there are six possible LD-CP pairings: local LD-local CP(LL), local LD-nonlocal CP(LN), local LD-bot (LB), nonlocal LD-local CP(NL), nonlocal LD-nonlocal CP (NN), and nonlocal LD-bot (NB).

\textbf{Chat Interface: }
Once paired, participants are directed to the system's chat component. The LD is instructed to determine if their CP is a local resident of the LD's city and state. LDs must ask at least three questions and can end the conversation once they have gathered sufficient information.
The CP, on their chat interface, is instructed to convincingly portray a local resident of the LD's city and state, responding to the LD's questions and waiting for the LD to initiate and end the conversation. When the CP is an LLM, it is prompted to act like a local human of the LD's city and state, concealing its AI identity and providing local responses. The system calculates response delays based on human typing speeds to simulate realistic conversation timing. Only LDs can end the conversation, which can be done after a minimum of three minutes or three rounds of interaction, whichever comes first.

\textbf{Localness Judgment Collection: }
Following the chat, we ask both the LD and CP to judge whether their chat partner is local or non-local, and provide their reasoning in an open-ended format. In cases where the CP reports being local to a different place than the LD's area, we also asked the CP if they are local to the LD, in order to account for the possibility that the CP is local to more than one place (e.g. if they moved). 
Participants then review a log of their conversation, selecting the statements that that support their local/non-local judgment. Each round of the conversation is shown with both the LD's question and the CP's answer. Participants rate each selected statement on a 1-5 scale, where 5 indicates that the statement strongly reflects how local their partner is. For each statement, we provide an optional open-ended text box to provide additional reasoning for each of their ratings.

\textbf{LLM Judgment Collection: }
After providing evidence for their localness decision, we then ask participants to complete a similar process, focused on judging whether their chat partner was a human or an LLM. We first ask them to provide their overall reasons, and then review and select conversation statements that support their human/LLM judgment. We again ask them to rate these statements for human-likeness on a 1-5 scale (5 being most human-like), and optionally explain their reasoning. We also ask participants to report how frequently they use chatbots on a 1-5 scale.

\subsubsection{LLM Integration}

For LLMs as CP conditions, we randomly select from three models: \texttt{LLaMA2-70b}, \texttt{gpt-4}, or \texttt{claude3-opus} as CP.
We developed a calibrated prompting and response strategy to maintain experimental validity. LLMs receive location-specific context and instructions to act as local residents while concealing their AI nature. Through multiple rounds of pilot testing (n=47), we iteratively refined prompts to avoid common behavioral tells while maintaining consistent performance across models. The final prompt templates are detailed in Appendix~\ref{appen:prompt}, including location-specific knowledge bases and behavioral parameters.

To enhance behavioral realism, we implemented calibrated delay of LLMs response derived from pilot study human typing patterns ($mean=12s/message$, $\sigma=3.25$). This timing model accounts for message length and complexity, with additional variance introduced to mirror natural human response patterns. Our pilot testing revealed that this tuned combination of semantic and temporal prompting was beneficial for creating authentic-seeming conversational patterns.

\subsection{Experiment Design}

\subsubsection{Study Design}
Building on our theoretical framework and system capabilities, we implemented a between-subjects experimental design focusing on localness assessments across human and LLM interactions. The design creates six experimental conditions through the combination of two Localness Decider (LD) types (local/nonlocal) and three Chat Partner (CP) types (local human/nonlocal human/LLM): Local LD-Local CP (LL), Local LD-Nonlocal CP (LN), Nonlocal LD -Local CP (NL), Nonlocal LD-Nonlocal CP (NN), Local LD-Bot (LB), and Nonlocal LD-Bot (NB).

This design enables systematic comparison of localness assessments strategies. The between-subjects approach avoids learning effects that might occur if participants engaged in multiple assessments attempts.

\subsubsection{Participant Recruitment and Demographics}

We recruited participants through our university's research participant mailing list ($40,000+$ faculty/staff, $20,000+$ students) in an upper-Midwestern U.S. state. To ensure sufficient local-local pairings, we restricted recruitment to the university's county. This geographic focus allowed us to study localness within a well-defined community context while maintaining a diverse participant pool. Participants were entered into raffle for a \$30 gift card, with a 1 in 15 chance of winning. 

Our power analysis, based on pilot data and assuming medium effect sizes ($d = 0.5$), indicated a target of 25-30 completed interactions per condition. This sample size provides 80\% power to detect significant differences in assessments accuracy between conditions while accounting for potential data loss. 

The final participant pool included 230 participants (155 female, 70 male, 4 non-binary, 1 undisclosed) aged 18--76 years ($\text{mean} = 34.21$, $\text{SD} = 15.93$). Participants formed 145 dyads: 85 human-human (170 participants) and 60 human-LLM (60 participants). 178 participants are native English speakers and 196 reporting high English fluency (self-rated $\geq4/5$). Most (112 participants) interacted with LLMs less than monthly, while 34\% (79) used them weekly/daily.

Our participants included 119 local residents and 111 non-local residents. We validated distinct local/nonlocal profiles using two measures: sense of place scale scores and residence duration. Independent samples t-tests revealed significant differences between local and non-local participants. Local participants demonstrated higher sense of place scores ($mean = 40.52$, $std = 4.87$) compared to non-local participants ($mean = 36.21$, $std = 5.85$), confirming the distinctiveness of these categories ($t_{228} = 5.31$, $p < 0.01$, $Cohen's~d = 0.82$).
The analysis of residence duration further supported this distinction. Local participants reported significantly longer residence periods ($mean = 11.13$ years, $std = 10.39$) compared to non-local participants ($mean = 4.36$ years, $std = 5.00$), $t_{228} = 6.29$, $p < 0.01$, $Cohen's~d = 0.76$.

\subsubsection{Condition Assignment and Study Flow}
Participants were dynamically assigned to one of the six conditions (LL, LN, LB, NL, NN, and NB) based on their self-reported local status and current condition needs, maintaining balanced cell sizes while maximizing successful pairings. The study flow followed three main phases: (1) assessment: participants completed demographic surveys and the sense-of-place assessment, establishing their local status and background, (2) interaction: participants then engaged in chat conversations according to their assigned roles, with minimum requirements of three questions or three minutes of interaction, (3) evaluation: participants provided judgments and supporting evidence through structured ratings and open-ended responses.

\subsection{Analysis Framework}

Our investigation of how people recognize and assess authentic local presence required a systematic analytical approach that could capture both conceptual understanding and practical evaluation processes. Drawing on multiple complementary methods, our methodology addresses specific aspects of our research questions while building a cohesive understanding of how people evaluate localness.

The breadth of our research questions necessitated diverse yet integrated analytical approaches. Understanding how participants define and conceptualize localness (RQ1) required qualitative analysis of their articulated understanding. Examining participants' ability to recognize genuine local presence (RQ2) and assess LLMs' capabilities (RQ3) demanded statistical evaluation of judgment patterns across different interaction types. To understand the underlying decision processes (RQ4), we traced how participants gather and evaluate evidence through complementary analyses: temporal analysis revealed how strategies evolved across conversations, while sensemaking analysis mapped how participants synthesized evidence into judgments about local presence. Finally, identifying factors that predict successful recognition of local presence (RQ5) required computational modeling to reveal complex patterns in assessment strategies and determine which elements most strongly influence accurate localness evaluation.

\subsubsection{Qualitative Analysis of Local Presence} 

To understand how people conceptualize localness (RQ1), we first conducted systematic analysis of participants' responses to the Localness Interpretation question from initial assessment. This foundational analysis was not only for understanding how people define localness, but also for developing the analytical framework we used to examine evaluation processes in later research questions.

Through iterative open coding, we developed a preliminary coding framework capturing key themes in how participants conceptualize localness. Two researchers independently applied this framework to the full dataset, achieving strong inter-rater reliability ($Cohen's~kappa = 0.825$). Regular team discussions refined the coding scheme, resulting in a hierarchical framework that captures the multiple dimensions of localness considered during assessments. This framework provided a structured vocabulary for examining how participants' conceptual understanding of localness manifested in their strategies.

This framework then served as the foundation for analyzing the rich textual data generated during locaness authenticity assessments interactions, including conversation transcripts and judgment rationales. Two researchers independently coded a 20\% sample of each data type to establish reliability ($Cohen's~kappa = 0.791$ for conversations, $0.803$ for judgment rationales), followed by full dataset coding. Throughout this process, we maintained detailed coding memos documenting decision rules and edge cases. 

The coding process systematically documented the occurrence and frequency of localness features across multiple levels (domains, dimensions, components, and sub-components) in each localness assessments interaction. This annotation process transformed qualitative data into quantifiable patterns, enabling systematic manipulations. These annotated datasets form the foundation for our subsequent sensemaking analysis (Section~\ref{sec:sensemaking}) and prediction analysis (Section~\ref{sec:predict}), allowing us to trace how participants gather, filter, and synthesize localness information throughout the localness assessments process.

\subsubsection{Statistical Analysis of Recognition Patterns}

Building on our understanding of how participants conceptualize localness, we next examined their ability to recognize it in practice (RQ2) and assess LLM capabilities (RQ3). This analysis required statistical approaches that could reveal patterns in how different groups evaluated local presence while accounting for the complex nature of our experimental conditions.

For comparing assessments patterns between groups, we employed \texttt{Mann-Whitney U test} for two-group comparisons, chosen for its robustness with our ordinal data and non-normal distributions.
For comparisons involving more than two groups, such as examining differences across human-human and human-LLM interactions, we used the \texttt{Kruskal-Wallis test} followed by \texttt{Dunn's post-hoc analysis} for targeted comparisons.

To ensure comprehensive understanding of group differences, we supplemented these tests with standardized effect sizes (\texttt{Cohen's d}), allowing us to quantify not only the statistical significance but also the practical importance of observed differences in localness inference patterns.
All analyses used $\alpha = 0.05$ as the significance threshold, with reported p-values adjusted for multiple comparisons where appropriate.

\subsubsection{Analyzing Temporal Evolution in Conversation Strategies}

To understand how participants’ strategies evolve over the course of their conversations, particularly the differences between correct and incorrect localness judgments (RQ4), we needed an analytical approach that accounts for the structured nature of our data. Conversations unfold over multiple rounds, where each round represents a unique exchange between participants. This introduces key challenges: conversational behaviors are repeated within dyads, meaning they are not independent; some conversational strategies occur very frequently, while others appear only sporadically or not at all in certain rounds; and the overall level of conversational engagement varies between individuals. These factors make traditional statistical approaches, such as linear regression, inadequate.

A central challenge is the presence of zero-inflation—many conversational features are simply absent in certain rounds. A standard regression or count-based model, such as Poisson regression, would assume that all non-occurrences are the result of a continuous distribution, rather than recognizing that many of these zeros are structural (i.e., certain conversational strategies are never used by some participants). To address this, we apply a Bayesian mixed-effects zero-inflated negative binomial (ZINB) model, which allows us to distinguish between cases where a conversational strategy is simply not used and cases where it is actively employed but varies in frequency..

This model separates conversational patterns into two components. The zero-inflation component accounts for the probability that a given conversational strategy is completely absent in a round. If a feature is never used, this suggests that a participant either does not recognize it as relevant or does not feel confident enough to employ it. This absence is influenced by factors such as the participant’s local status, their individual conversational tendencies, and the round number (since conversational strategies may shift over time).

The second component, the count process, models how frequently a strategy is used when it does appear. This captures changes in questioning intensity across rounds, allowing us to assess whether participants who correctly identify localness ask different types of questions, at different rates, compared to those who misjudge their partners. It also lets us examine whether strategies become more refined or disappear as conversations progress.

By employing a Bayesian approach, we quantify the uncertainty in our estimates, providing a clearer picture of the likelihood that certain conversational strategies contribute to accurate localness judgments. The hierarchical nature of the model accounts for differences between individual participants and dyads, ensuring that our findings reflect meaningful patterns rather than individual outliers.

Through this analysis, we aim to determine whether correct and incorrect judgments systematically differ in how often participants fail to use key conversational features, as well as how the use of certain strategies evolves over time. By separating these two aspects—whether a strategy is used at all, and how often it appears when it is used—we gain a comprehensive understanding of how localness assessments unfold dynamically in conversation.

\subsubsection{Analysis of Sensemaking Processes}
\label{sec:sensemaking}

While our temporal analysis revealed when different aspects of localness emerged in conversations, understanding why participants focused on certain elements required deeper investigation into their decision-making processes (RQ4). To do this, we applied sensemaking theory \cite{pirolli2005sensemaking, russell1993cost}, which describes how people iteratively gather, filter, and synthesize information to form judgments. This theoretical lens guided both our data collection design and analytical approach, allowing us to systematically trace how participants moved from initial exploration to structured assessments of local presence.

The first phase, information gathering, focuses on how participants actively seek relevant details about their chat partner’s local ties. To analyze this, we employed a three-pronged approach: (1) framework coding of questions and responses using our validated localness hierarchy ($Cohen's~kappa = 0.791$), (2) frequency analysis to identify which localness features participants prioritized, and (3) Bayesian mixed-effects ZINB modeling to track how questioning strategies evolved over the course of conversations. This combination allowed us to see not only what participants asked but how their questioning adapted as they engaged with their chat partners.

The second phase, information filtering, examines how participants evaluated and structured the responses they received. Not all information was weighted equally—some details were immediately recognized as diagnostic of localness, while others were dismissed or required further verification. To capture this process, we analyzed participants' self-selected key conversation rounds using the same three-pronged approach: framework coding, frequency analysis, and Bayesian modeling. This revealed the patterns behind participants’ choices of what evidence mattered most and how they systematically winnowed down broad exploration into specific, meaningful indicators.

The final phase, knowledge consolidation, focuses on how participants synthesized the evidence they gathered into explicit judgments about their chat partner’s localness. Here, we used two complementary analytical techniques. First, framework coding of judgment rationales allowed us to identify which localness features were most frequently cited as justification for decisions. Second, we conducted grounded theory analysis of these rationales to uncover deeper patterns in decision strategies. Through iterative open coding followed by axial coding to identify relationships between approaches, we identified distinct patterns in how correct and incorrect authenticators evaluated evidence. This analysis revealed key strategies like cross-referencing multiple evidence types, detecting inconsistencies, and evaluating contextual appropriateness—insights that would not have emerged from framework coding alone.

\subsubsection{Predicting Localness Detection Accuracy}

To determine which factors most strongly predict participants’ ability to recognize local presence (RQ5), we developed a machine learning approach that balances predictive power with interpretability. Using XGBoost, a gradient-boosted decision tree model, we analyzed how conversational elements influenced localness detection accuracy. We then applied SHAP (SHapley Additive exPlanations) analysis to unpack the specific features driving these predictions.

We selected XGBoost for three key reasons. First, it captures non-linear relationships in conversational data, allowing it to model the complex interplay of questioning strategies, response patterns, and participant backgrounds. Second, it handles feature interactions effectively, enabling us to understand how different conversational cues combine to signal localness. Third, it addresses class imbalances, ensuring that detection accuracy is not skewed by differences in how often participants correctly or incorrectly classified their chat partners.

To weight participants' judgments appropriately, we applied their ratings of selected conversation rounds where strong localness indicators (ratings of 1 or 5) received greater weight than ambiguous ratings (2 or 4), while neutral ratings (3) were excluded from model training. We optimized XGBoost using a grid search with three-fold cross-validation, ensuring robust performance, and evaluated model accuracy using standard classification metrics, including precision, recall, and ROC curves.

To make the model’s predictions interpretable, we applied SHAP analysis, a game-theoretic approach that assigns importance scores to different features at multiple levels of our localness framework. Rather than treating the model as a ``black box,'' SHAP allows us to see which specific conversational elements --- ranging from broad domains to fine-grained sub-components --- influence detection accuracy. It revealed which conversational elements had the strongest influence on successful localness detection. By analyzing feature importance across hierarchical levels, we identified key factors that differentiate accurate from inaccurate localness judgments. This insight enhances our understanding of how local presence is assessed in conversational settings, offering valuable implications for both human and AI-driven verification systems.

\subsection{Ethical Considerations}
The dataset involves personal identifiers. To ensure anonymity, we replaced specific geographic identifiers (such as city, town, and street names) with general terms like ``[city name],'' ``[town name],'' and ``[street name].'' Additionally, all participant data, including personal identifiers and location details, were anonymized during data analysis to protect privacy. Participants were informed of the study's purpose and provided consent, with the option to withdraw at any time without consequence. Overall, our study adhered to institutional ethical guidelines, including IRB approval, ensuring that both participant privacy and data security were maintained throughout the research process.
\section{Result}

\subsection{Participants Overview}

\begin{table}[t!]
    \caption{Participant demographics and interaction outcomes across experimental groups}
    \small
    \centering
    \begin{tabularx}{0.65\textwidth}{lllllll}
        \toprule
        Metric & LL & LN & NL & NN & LB & NB \\
        \midrule
        Group Size & 27 & 18 & 17 & 23 & 30 & 30 \\
        Avg. Conversation Rounds & 6.52 & 6.83 & 7.35 & 8.33 & 7 & 5.9 \\
        Local Judgment Accuracy & 38/54 & 17/36 & 27/34 & 21/46 & 14/30* & 11/30*  \\
        LLM Judgment Accuracy & 25/54 & 20/36 & 17/34 & 26/46 & 24/30 & 22/30 \\
        \bottomrule
    \end{tabularx}
    \label{tab:participant}
    \begin{minipage}{0.65\textwidth}
        \small
        \textit{Note: {*}Local Judgment Accuracy reflects correct identifications (local/nonlocal) in human-human dyads. For LB/NB groups, Local Judgment indicates ``local'' classifications of LLM partners.} 
    \end{minipage}
\end{table}

Our participants completed the entire procedure in an average of $17.61$ minutes ($std = 16.37$), with conversation segments averaging $12.59$ minutes ($std = 11.92$).
The logs from these 230 participants resulted in a comprehensive dataset of 932 conversation rounds. 
Participants were grouped into six different categories. Table~\ref{tab:participant} provides a summary of key metrics, including the number of participant pairs, the average number of conversation rounds, and the judgment results from both participants and LLMs. In the LB and NB settings, since the participants were interacting with LLMs, there are no correct or incorrect judgments. Therefore, the Local Judgment Result for these settings indicates the number of judgments that were marked as ``local'' out of the total.
\subsection{RQ1: What Defines Localness According to Participants?}
\label{sec:rq1}

With a diverse participant pool representing different levels of local attachment, we analyze how locals and nonlocals conceptualize ``being local''. Our mixed-methods approach reveals a multi-dimensional framework of localness and systematic differences between groups. As noted above, to address this first research question, we rely on thematic analysis of our participants' responses to our open-ended question on how they interpret localness, as a concept.

\begin{figure}
    \centering
    \includegraphics[width=0.5\linewidth]{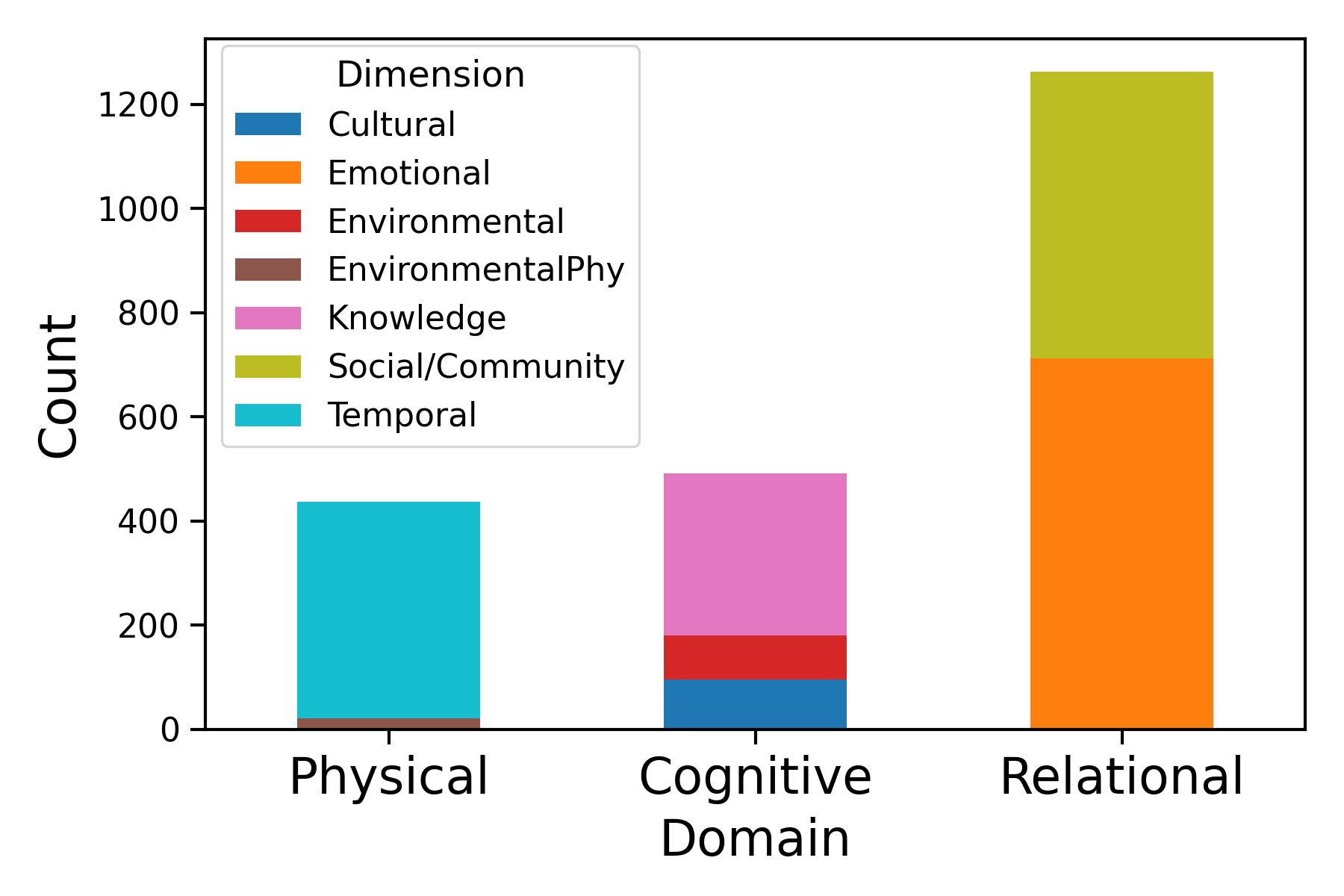}
    \caption{Frequency of Localness Domains and Dimensions}
    \label{fig:fre-dom-dim}
\end{figure}

\begin{longtable}[ht]{%
  >{\raggedright\arraybackslash}p{0.15\linewidth}%
  >{\raggedright\arraybackslash}p{0.15\linewidth}%
  >{\raggedright\arraybackslash}p{0.4\linewidth}%
  >{\raggedright\arraybackslash}p{0.1\linewidth}%
}
\caption{Top 3 levels of Localness Framework (Counts Summed by Component)} \label{tab:framework} \\
\toprule
\textbf{Domain} & \textbf{Dimension} & \textbf{Component} & \textbf{Count} \\
\midrule
\endfirsthead

\multicolumn{4}{c}%
  {{\tablename\ \thetable{} -- continued from previous page}} \\
\toprule
\textbf{Domain} & \textbf{Dimension} & \textbf{Component} & \textbf{Count} \\
\midrule
\endhead

\midrule
\multicolumn{4}{r}{{Continued on next page}} \\
\endfoot

\bottomrule
\endlastfoot

\multirow[t]{7}{*}{Relational}  & \multirow[t]{3}{*}{Emotional} 
  & Feeling of Home        & 219 \\ 
                               &                        & Identity Connection    & 210 \\ 
                               &                        & Sense of Belonging     & 283 \\ 
\cmidrule{2-4}
                               & \multirow[t]{4}{*}{Social/Community} 
  & Active Participation   & 157 \\ 
                               &                        & Civic Engagement       & 32 \\ 
                               &                        & Community Investment   & 123 \\ 
                               &                        & Personal Relationships & 239 \\
\midrule
\multirow[t]{6}{*}{Physical}    & \multirow[t]{3}{*}{Environmental} 
  & Ecological Understanding & 9 \\ 
                               &                        & Geographic Familiarity   & 8 \\ 
                               &                        & Natural Environment      & 4 \\ 
\cmidrule{2-4}
                               & \multirow[t]{3}{*}{Temporal} 
  & Being Born/Native        & 107 \\ 
                               &                        & Formative Years        & 136 \\ 
                               &                        & Long-term Residence      & 173 \\
\midrule
\multirow[t]{11}{*}{Cognitive} & \multirow[t]{3}{*}{Cultural} 
  & Food Culture         & 27 \\ 
                               &                        & Language/Dialect     & 35 \\ 
                               &                        & Local Customs/Norms   & 34 \\ 
\cmidrule{2-4}
                               & \multirow[t]{3}{*}{Environmental} 
  & Ecological Understanding & 21 \\ 
                               &                        & Geographic Familiarity   & 39 \\ 
                               &                        & Natural Environment      & 25 \\ 
\cmidrule{2-4}
                               & \multirow[t]{5}{*}{Knowledge} 
  & Change Awareness       & 47 \\ 
                               &                        & Hidden Gems            & 29 \\ 
                               &                        & Historical Knowledge   & 57 \\ 
                               &                        & Local Recommendations   & 129 \\ 
                               &                        & Navigation/Wayfinding  & 49 \\
\end{longtable}

Our thematic analysis of participants' localness interpretation reveals localness as a multi-layered construct comprising three primary domains (\texttt{Physical}, \texttt{Cognitive}, and \texttt{Relational}, see Figure~\ref{fig:fre-dom-dim}), which encompass seven dimensions, 24 components, and 88 sub-components. We include the first three levels of features and their frequency in Table~\ref{tab:framework} (the full framework with 88 sub-components is in Table~\ref{tab:localness} in Appendix~\ref{appen_frame}). We characterize each of these domains by unpacking the dimensions and components represented in our analysis, below.

\subsubsection{The Relational Domain of Localness}
The first domain in our analysis is \textit{relational} aspect of localness, which represents the salience of ``connecting'' through ``social/community'' and ``emotional'' dimensions. These dimensions emphasize interaction, participation, and belonging.
In our findings, the ``social/community'' dimension captures the depth of interpersonal connections and engagement within a community, including active participation in local events and initiatives, and fostering integration and contribution to communal life. The ``civic engagement'' dimension adds a layer of responsibility through advocacy and participation in governance, suggesting a theme of collective action as well. The ``community investment'' component highlights a long-term commitment to improving local spaces and supporting shared goals, while personal relationships deepen ties through trust, familiarity, and meaningful interactions with neighbors and friends. As one participant explains: \begin{quote}
    ``Being local to me means that one is a part of the community... Since I have started volunteering in village organizations and making friends with other citizens, I feel that I am now a local.''
\end{quote}  
The ``emotional'' dimension in our results captures the psychological and affective bonds individuals develop with a place, rooted in familiarity, identity, and belonging. It encompasses feelings of home, identity connection, and a sense of belonging, each contributing to an individual’s emotional attachment. One participant expressed: \begin{quote}
    ``Being local means feeling like you belong in a community. You know what's going on... and feel at home there.''
\end{quote} 

\subsubsection{The Physical Domain of Localness}
The second domain in our analysis is the \textit{physical} aspect of localness, which represents the salience of ``being'', in how we understand local identity. This domain is based on the tangible, observable ways people occupy and interact with a place, and dimensions within this domain ``temporal presence`` and ``environmental interaction``.
In our results, the ``temporal`` dimension highlights the role of time in fostering a deep connection to place, encompassing experiences from birth to long-term residence. Being born in a place establishes an intrinsic connection, with lifelong familiarity and a sense of historical continuity grounding an individual in their local identity. This innate bond is further shaped during formative years, when core memories, cultural absorption, and early life experiences embed the rhythms and traditions of a place into a person’s identity. Over time, these foundations are reinforced through long-term residence, where accumulated experiences, emotional investment, and witnessing changes in the community deepen the sense of belonging.

Further, in our data, the ``environmental interaction`` dimension focuses on direct physical engagement with spaces, like ecological understanding, geographic familiarity, and interaction with the natural environment. One participant stated: \begin{quote}
    ``I garden at a community garden... Gardening there now makes me feel more connected and I have a sense of place and feel responsible for the area.'' 
\end{quote}

\subsubsection{The Cognitive Domain of Localness}
The third domain in our analysis is the \textit{cognitive} domain, which embodies the ``knowing'' aspect through ``knowledge'', ``cultural'', and ``environmental'' dimensions. 
In our data, the ''knowledge'' dimension represents the intellectual engagement individuals develop with a place, encompassing an understanding of its history, navigation, and local intricacies. For example, ``change awareness`` knowledge captures the ability to recognize past transformations, anticipate future developments, and understand the ongoing evolution of a place, grounding individuals in the dynamic shifts in how places are understood and chane over time. Knowledge of ``hidden gems`` emphasize insider knowledge of unofficial landmarks and local secrets, reflecting a deeper integration into the cultural fabric of the place. ``Historical`` knowledge connects individuals to traditions and past events, and ``local recommendations`` and ``navigation proficiency`` components of this dimension focus on practical expertise, such as knowing the best places for specific needs or confidently navigating without assistance, underscoring an intimate familiarity with the place. One participant stated: \begin{quote}
    ``Locals know where to get the best bread, where you have to go early to get donuts before they sell out, not to walk the wrong direction when going around the square for the farmers market.''
\end{quote} 
The ``cultural understanding'' dimension of our \textit{cognitive} domain represents people's awareness of local norms, traditions, and languages. For example, in our data, the ``food culture`` component includes sub-components like awareness of local food sources (38 instances) and knowledge of local cuisine (51 instances). Language and dialect components include sub-components like understanding local humor and expressions (31 instances). One participant described this ``cultural understanding'' this way: \begin{quote}
    ``Understanding the 'umbrella' culture of a place... the words and phrases that have special meaning, and knowing the history of that place.''
\end{quote}
Our ``environmental comprehension`` dimension is characterized by understandings of the natural and spatial aspects of a place, encompassing ``ecological understanding``, ``geographic organization``, and the ``natural environment``. The ``ecological understanding`` component in our data focuses on awareness of seasonal patterns, environmental cycles, and local watersheds, reflecting the people's understanding around the interconnectedness of ecosystems and environmental issues. Our ``geographic familiarity'' component centers on spatial organization, including recognition of landmarks, physical boundaries, and land features, which enables individuals to navigate and relate to their surroundings. The ``natural environment`` component highlights knowledge of flora and fauna, ecological systems, and natural patterns, fostering a deeper connection to the physical and biological characteristics of a place. One participant noted: 
\begin{quote}
    ``I think about 'local' in ecological terms... I can name some of its trees, flora and fauna species.''
\end{quote}

Importantly, these three domains in our framework are deeply interconnected. The \textit{cognitive understanding} domain enhances \textit{relational} engagement by providing the knowledge needed to participate in social and cultural activities. The \textit{physical} domain, and individual physical presence facilitates learning and emotional bonding, as sustained interaction with a place fosters familiarity and attachment. \textit{Relational} ties, in turn, motivate individuals to invest in both knowledge and physical presence, creating a feedback loop that strengthens local identity over time. One participant characterized the interreliance between these different domains this waycaptures these interactions: \begin{quote}
    ``Local means having opinions on local politics [\texttt{cognitive} domain], being civically engaged (Relational), knowing where to find things [\texttt{cognitive} domain], and remembering things about the place that are no longer there [\texttt{physical} and \texttt{cognitive} domains].''
\end{quote}

\subsubsection{Differences between local and nonlocal people}

With our thematic framework of how people understand localness in hand, we then wanted to interrogate if locals and nonlocals share the same perspective or if their interpretations differ significantly. Overall, our analysis found systematic differences in how local and non-local participants conceptualize localness, particularly in their emphasis on active versus historical connections.

We used Mann Whitney U test to examine how local and non-local participants interpret localness differently across multiple hierarchical levels: domains, dimensions, components, and sub-components. To control for false discovery rate due to multiple comparisons, we applied the Benjamini-Hochberg\ correction to the p-values. More details of statistical results appear in Appendix~\ref{appen_inte}.

Our analysis revealed local participants view localness as an active, ongoing process that requires current community involvement. This manifests most strongly in the Relational domain, where local participants mentioned significantly more frequent ($M = 6.963$) than non-local participants ($M = 4.493$), with a mean difference of 2.47 ($MW-p = 0.038$, $Cohen's d = 0.403$). Within this domain, locals also more emphasized components like Active Participation ($MW-p = 0.005$, $Cohen's d = 0.561$), and sub-componnets like Attending local events ($MW-p = 0.018$, $Cohen's d = 0.465$) and Engaging in community initiatives ($MW-p = 0.011$, $Cohen's d = 0.49$).
Local participants also emphasize the importance of synthesizing and sharing local knowledge. For example, locals ($M = 0.306$) valued Awareness of local options and alternatives in their localness interpretation more than non-locals ($M = 0.113$), $MW-p = 0.018$, $Cohen's d = 0.462$.

In contrast, non-local participants view localness primarily through residence duration and birthplace status. This perspective dominates the Physical domain, where non-locals mentioned more ($M = 3.113$) than locals ($M = 1.537$), showing a mean difference of -1.575 ($MW-p = 0.001$, $Cohen's d = -0.656$). The Temporal dimension within this domain further reinforces this pattern, with non-locals ($M = 3.056$) scoring higher than locals ($M = 1.485$) ($MW-p = 0.001$, $Cohen's d = -0.663$). This emphasis on historical connection appears in components like Being Born/Native ($MW-p < 0.001$, $Cohen's d = -0.745$) and Formative Years ($MW-p = 0.002$, $Cohen's d = -0.571$), and in sub-components such as Early Life Experiences ($MW-p < 0.001$, $Cohen's d = -0.536$) and Lifelong Familiarity ($MW-p < 0.001$, $Cohen's d = -0.672$).

These findings reveal an insight: while ``local'' status logically depends on one's location, people's self-identification significantly influences how they define localness itself. Our results show that locals and nonlocals essentially create different mental models of what makes someone ``local.'' Self-identified locals emphasize what they actively do --- community participation and relationship-building --- while nonlocals emphasize historical factors like birthplace and long-term residence. 
The significance of this pattern emerges when considering that our nonlocal participants had lived in their communities (where they identified themselves not local at) for four years on average, a meaningful amount of time. This suggests that nonlocals aren’t drawing on their original hometowns when thinking about localness. Instead, they’re reflecting on their experience of slowly becoming part of a new place. In both groups, people seem to define localness in ways that reinforce their current sense of identity and belonging. 
This shows that localness isn’t a fixed label --- it evolves over time through people’s interactions with their environment. For community development practitioners, these findings suggest that effective integration programs should recognize the diverse ways people come to feel local. By doing so, communities can bridge the gap between long-time residents and newcomers, creating space for multiple pathways to local identity.

\begin{figure}[t!]
    \centering
    \begin{subfigure}[b]{0.49\linewidth}
        \centering
        \includegraphics[width=\linewidth]{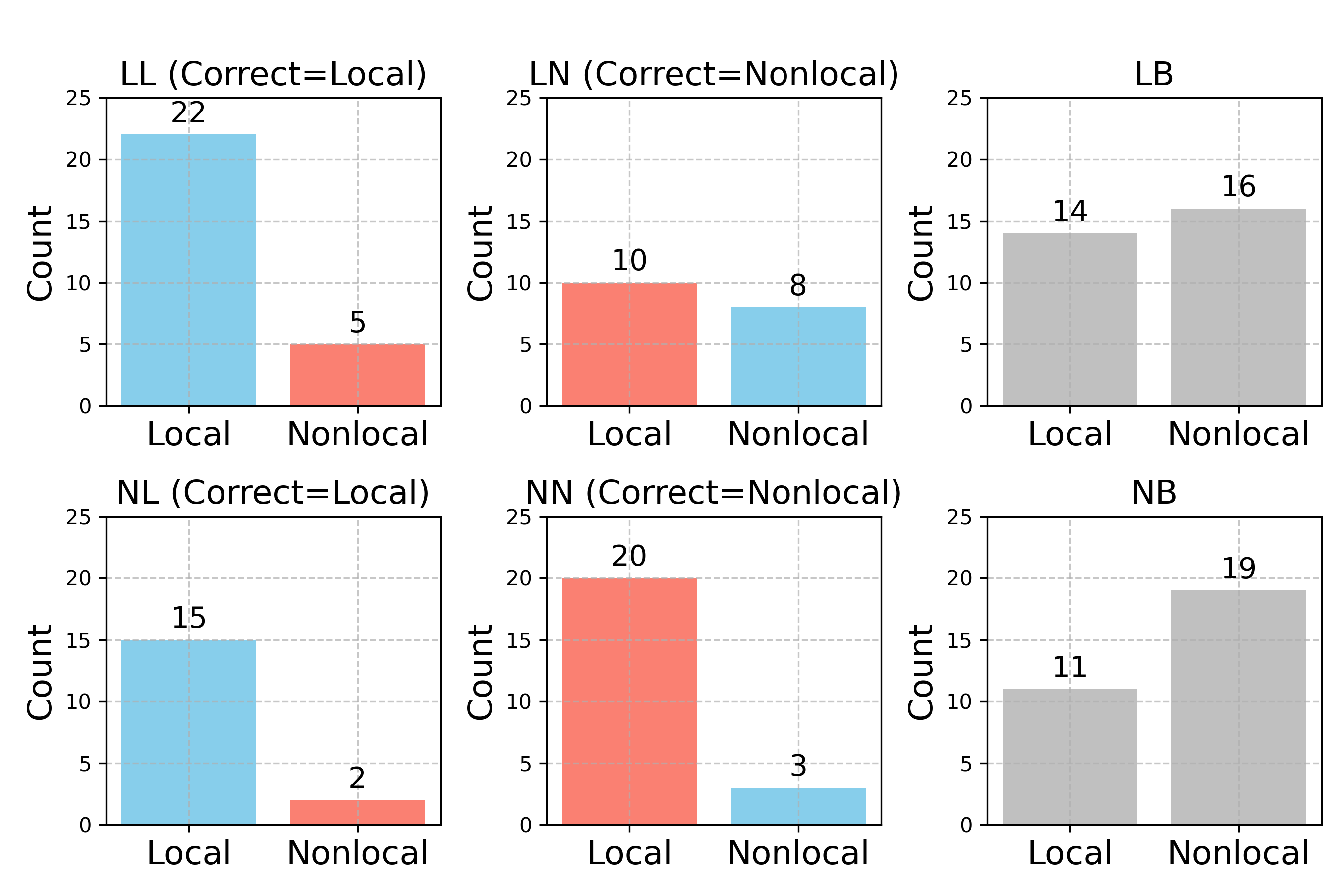}
        \caption{Comparison of LD\'s Local Judgment in Different Settings}
        \label{fig:ld-local-group}
    \end{subfigure}
    \hfill
    \begin{subfigure}[b]{0.49\linewidth}
        \centering
        \includegraphics[width=\linewidth]{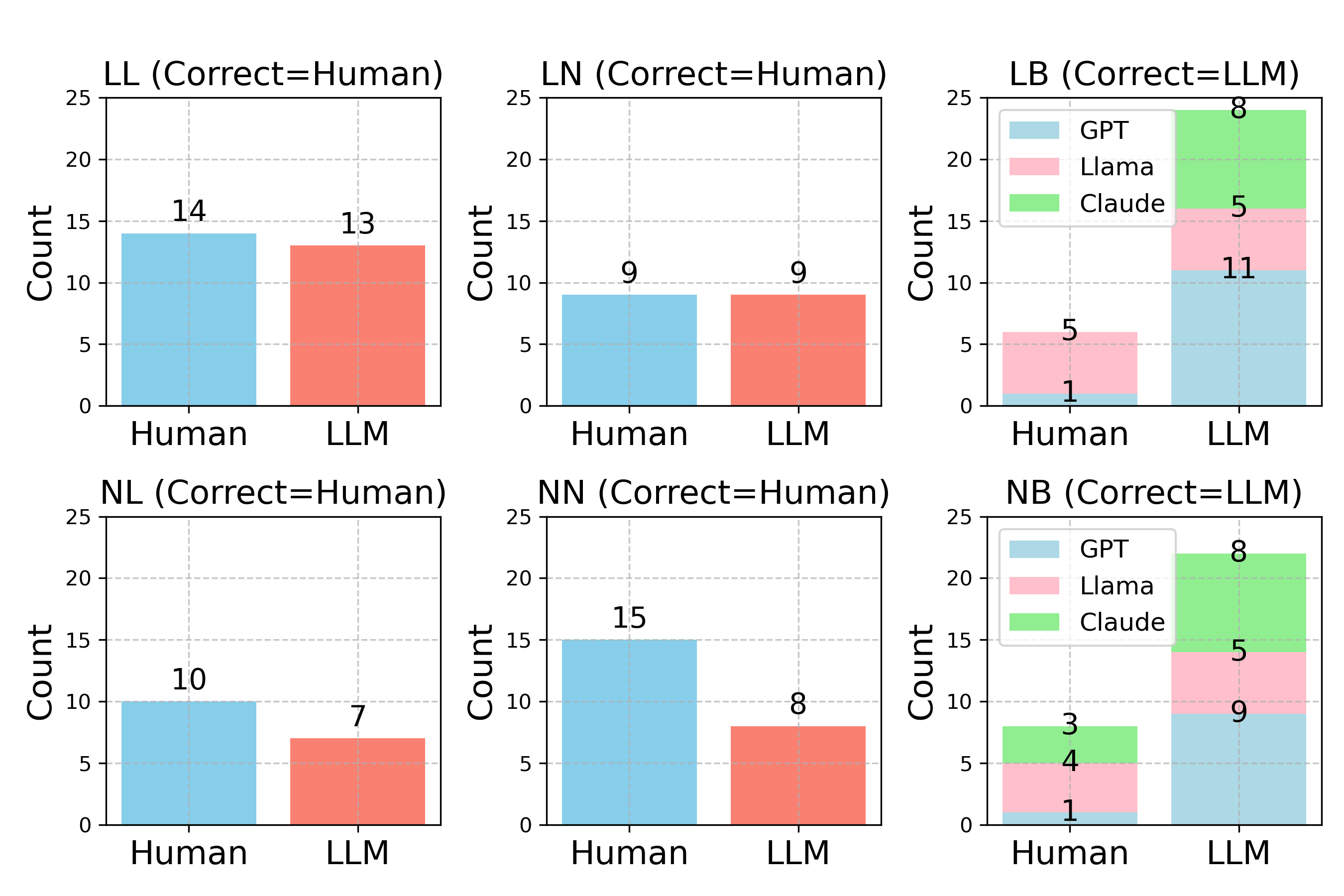}
        \caption{Comparison of LD\'s LLM Judgment in Different Settings}
        \label{fig:ld-llms-group}
    \end{subfigure}
    \caption{Comparison of LD\'s Judgments in Different Settings}
    \label{fig:ld-group-comparison}
\end{figure}

\subsection{Can people differentiate between local and nonlocal Chat Partners, and between human and LLM Chat Partners?}
\label{sec:rq2}

Our results in Section~\ref{sec:rq1} show that people construct local identity through a combination of cognitive knowledge, relational ties, and physical presence, which are interwoven concepts that create a rich, holistic characterization of a place. Importantly, however, people having an understanding of localness, conceptually, does not necessarily mean that people are able to accurately recognize if someone else is local. The multi-dimensional framework we uncovered in RQ1 outlines how people conceive of localness for themselves, but an open question remains: Are these ``localness signifiers'' actually actionable in deciding if someone is local?

Therefore, in RQ2, we investigate whether participants can accurately distinguish between local and nonlocal chat partners in conversation, as well as whether they can differentiate human partners from LLMs. By focusing on participants' accuracy patterns, we assess which aspects of localness can be reliably authenticated through dialogue and where participants encounter ambiguity or deception in their judgments.

\subsubsection{Differences in Localness Judgments}
Our analysis reveals that local LDs generally demonstrate greater accuracy in identifying chat partners' local status compared to nonlocal LDs  ($MW-p= 0.046$, $Cohen's~D = 0.442$). This advantage manifests differently depending on the chat partner's status.

In our results, when local LDs interacted with local chat partners, they achieved high accuracy, correctly identifying localness in 22 out of 27 conversations. However, their accuracy decreased notably when chatting with nonlocals, correctly identifying only 8 out of 18 nonlocal partners (Figure~\ref{fig:ld-local-group}). Statistical analysis confirms this pattern, showing significantly higher accuracy in LL group than LN group ($MW-p= 0.011$, $Cohen's~D = 0.832$). Further, nonlocal LDs also show remarkable accuracy in identifying local chat partners, correctly recognizing 15 out of 17 local participants. However, they struggle to identify fellow nonlocals, achieving correct identification in only 3 out of 23 conversations (Figure~\ref{fig:ld-local-group}). This contrast in accuracy is statistically significant as well ($MW-p< 0.001$, $Cohen's~D = 2.216$). 

Notably, these results suggest that both locals and nonlocals can \textit{positively} recognize if someone is a local with high rates of accuracy, but negatively recognizing someone as nonlocal is much more difficult. In particular, when the chat partner was not local, both locals and nonlocals frequently incorrectly judged their nonlocal chat partner to be local. While these results focus on human judgements, not an algorithmic decision, our results suggest that participants show ``high precision'' in identifying local chat partners, but their ``recall'' is not nearly as strong. In other words, they correctly identify the true positive chat partners who are local, but also have high false positive rates with chat partners who are not local. High false negative rates are not uncommon in domains where the judgement being made is focused on a complex and somewhat unclear concept, like our localness judgment task here \cite{thebault-spieker_diverse_2023, venkatagiri2019groundtruth}. Prior work has shown similar results in evaluating whether social media content is political \cite{thebault-spieker_diverse_2023}, or hate speech \cite{warner_detecting_2012}.

\subsubsection{Differences in LLM Judgments}

Beyond localness recognition, we examined participants' ability to distinguish between human and LLM chat partners. 
We find that local and nonlocal participants are generally not good at identifying their chat partner was human or not.
Local LDs correctly identified their chat partner as human in 14 out of 27 dyads and 9 out of 18 dyads in LL and LN groups respectively. Nonlocal LDs correctly identified their chat partner as human in 10 out of 17, and 15 out of 23 dyads in NL and NN groups respectively (Figure~\ref{fig:ld-llms-group}).

Nonlocal LDs demonstrated slightly better performance in identifying human partners, though we did not see significant differences between local and nonlocal LDs' accuracy in LLM Judgments ($MW-p= 0.296$, $Cohen's~D = 0.228$). 
Notably, local LDs maintained consistent accuracy rates regardless of whether their chat partners were local or nonlocal ($MW-p= 0.915$, $Cohen's~D = -0.129$), so as nonlocal LDs ($MW-p= 0.695$, $Cohen's~D = -0.036$), suggesting that local status does not significantly influence bot detection abilities.

\subsection{RQ3: When Do People Perceive LLMs as Local or as Human?}

Our analysis in RQ2 demonstrated that participants could reliably identify local individuals in conversation but struggled to correctly classify nonlocals. Furthermore, their ability to distinguish between human and LLM partners was notably weak, suggesting that conversational cues alone do not always provide clear signals of authenticity. These earlier questions, while potentially promising, point to a natural next question: if participants can recognize locals, but have difficulty differentiating between human and AI partners, does that mean LLMs can successfully mimic localness?

In RQ3, we turn our attention to how people perceive LLMs in conversation --- whether they mistake them for locals, whether they recognize their artificial nature, and how LLM-generated responses compare to those from human partners. By analyzing how participantsclassified LLMs  relative to human locals and nonlocals, we uncover whether AI can convincingly perform localness or if fundamental gaps in authentic local presence remain evident.

\subsubsection{Localness Recognition of LLMs Chat Partners}
Our analysis reveals that LLMs face significant challenges in being perceived as local, with human participants consistently identifying them as non-local. Using Kruskal-Wallis Test with Dunn's Post-hoc Test, we found that LL, LN and LB groups have significant differences of making ``local'' judgment ($KW-p= 0.024$). 

The pattern of results suggests that LLMs are particularly disadvantaged in appearing local. Local LDs were significantly more likely to recognize human local partners as "local" compared to LLM ($Dunn-p= 0.022$). Interestingly, LLMs were perceived similarly to nonlocal human partners, with no significant difference in local judgments between LN and LB groups ($Dunn-p= 1.00$). This suggests that for local LDs, LLMs' attempts at demonstrating local knowledge and connection fall short in ways similar to nonlocal human partners. Nonlocal LDs showed more consistent judgment patterns, with no significant differences in their local assessments across human and LLM interactions ($KW-p= 6.168$).

Participants demonstrated heightened ability to identify LLM partners compared to their performance in human-human interactions. While human-human interactions showed relatively balanced judgments between "human" and "LLM" classifications (with a slight tendency toward "human"), human-LLM interactions elicited stronger and more accurate "LLM" classifications (Figure~\ref{fig:ld-llms-group}).

This pattern held true across different participant groups. Among local LDs, we found significant differences in LLM recognition across LL, LN, and LB groups (KW-p = 0.026). Local LDs were significantly more likely to correctly identify LLM partners compared to their judgments in LL interactions (Dunn-p = 0.043), though their accuracy in LB and LN groups remained similar (Dunn-p = 0.12).

Nonlocal LDs showed similar patterns of enhanced LLM detection (KW-p = 0.012). They were significantly more accurate in identifying LLM partners compared to their judgments in NN interactions (Dunn-p = 0.017), with no significant difference between NB and NL groups (Dunn-p = 0.105). These results suggest that despite LLMs' sophisticated conversational abilities, participants maintain a robust capacity to distinguish them from human partners.

\subsubsection{Human Recognition of LLM Chat Partners}

Participants demonstrated heightened ability to identify LLM partners compared to their performance in human-human interactions. While human-human interactions showed relatively balanced judgments between ``human'' and ``LLM'' classifications (with a slight tendency toward ``human''), human-LLM interactions elicited stronger and more accurate ``LLM'' classifications (Figure~\ref{fig:ld-llms-group}).

This pattern held true across different participant groups. Among local LDs, we found significant differences in LLM recognition across LL, LN, and LB groups ($KW-p= 0.026$). LDs in LB groups are significantly more likely to judge their chat partners as ``LLM'' comparing with LDs in LL group ($Dunn-p= 0.043$). But LDs in LB groups and LN groups have similar probability of judging chat partners as ``LLM''  ($Dunn-p= 0.12$). This similarity in identifying both LLMs and nonlocal chat partners suggests that LLMs share common limitations with nonlocals in demonstrating local identity, failing to capture the subtle markers of authentic local experience that local LDs readily recognize.

Nonlocal LDs showed similar patterns of enhanced LLM detection. Nonlocal LDs have significant differences in the probability of judging chat partners as ``LLM'' ($KW-p= 0.012$). LDs in NB groups are significantly more likely to judge their chat partners as ``LLM'' comparing with LDs in NN group ($Dunn-p= 0.017$). But LDs in NB groups and NL groups have similar probability of judging chat partners as ``LLM'' ($Dunn-p= 0.105$). These findings indicate that even without hight local levels, nonlocal LDs can still detect patterns in LLM responses that distinguish them from human communication about place. 
\subsection{RQ4-1: Localness Sensemaking: What Information Do Participants Gather for Localness Judgments?}

\begin{figure*}[t]
\centering
\begin{subfigure}[b]{0.45\textwidth}
    \includegraphics[width=\textwidth]{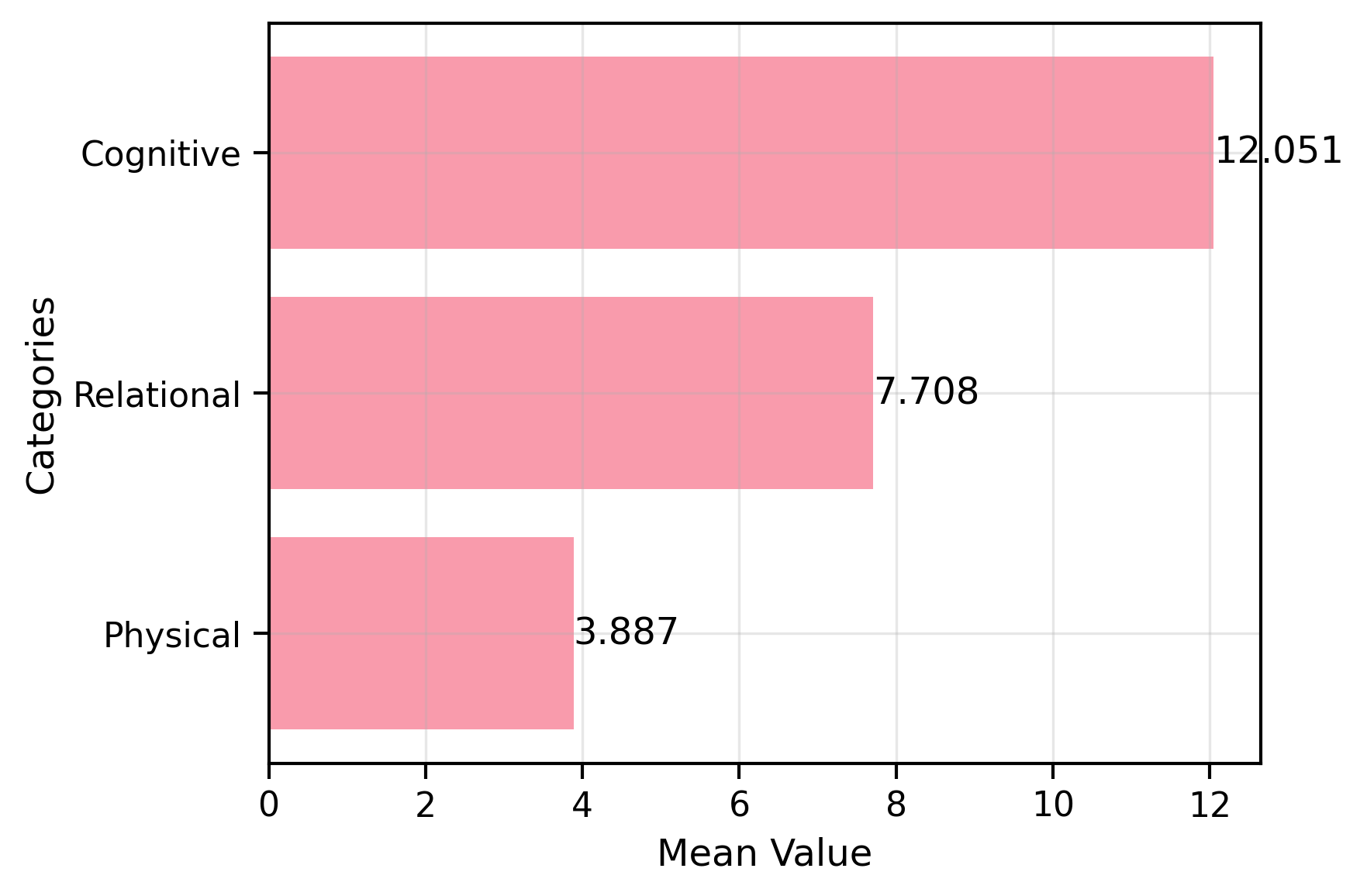}
    \caption{Distribution of Domain Categories}
    \label{fig:dom}
\end{subfigure}
\hfill
\begin{subfigure}[b]{0.45\textwidth}
    \includegraphics[width=\textwidth]{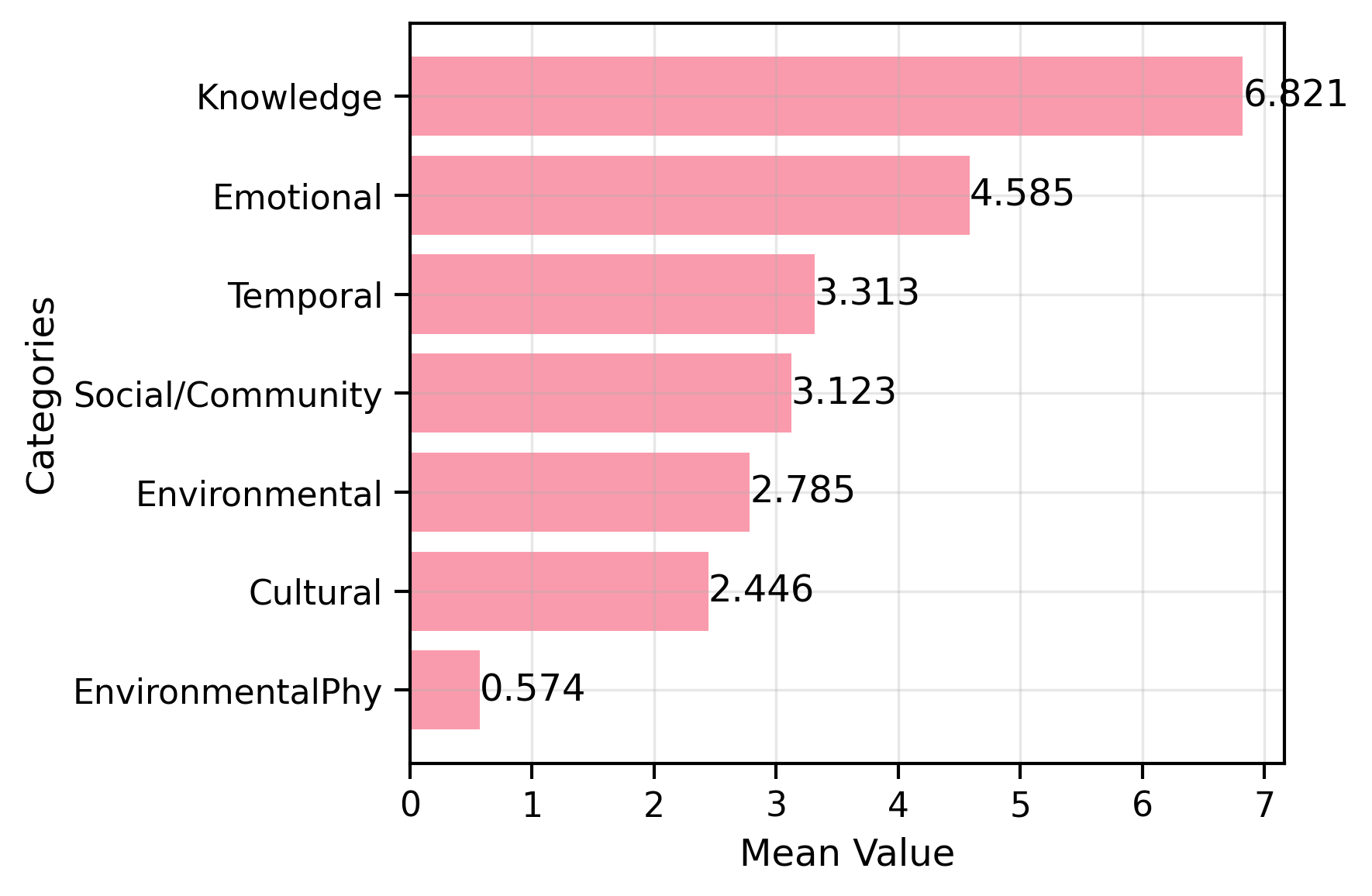}
    \caption{Distribution of Dimension Categories}
    \label{fig:dim}
\end{subfigure}

\begin{subfigure}[b]{0.45\textwidth}
    \includegraphics[width=\textwidth]{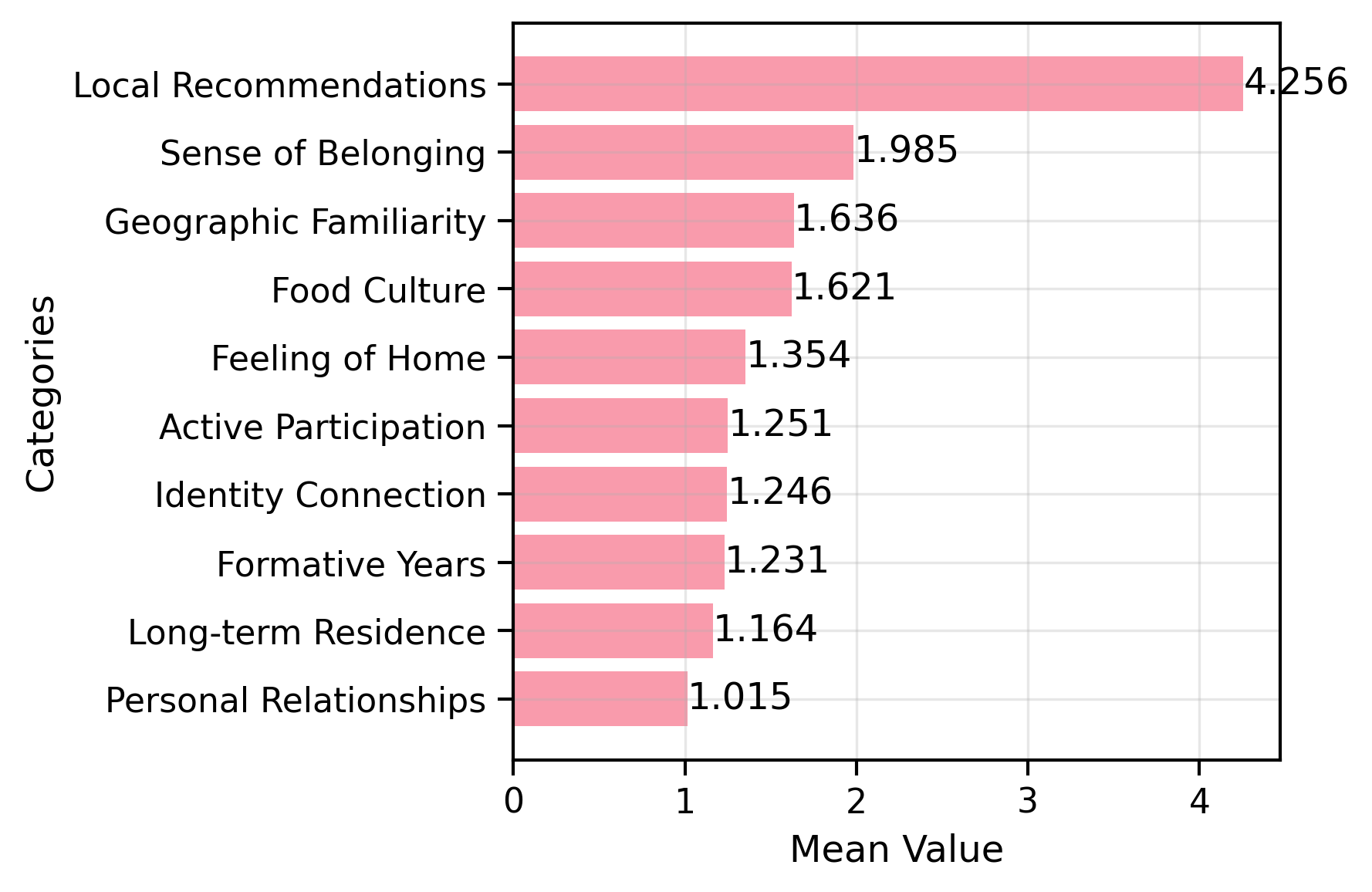}
    \caption{Distribution of Top 10 Components Categories}
    \label{fig:com}
\end{subfigure}
\hfill
\begin{subfigure}[b]{0.45\textwidth}
    \includegraphics[width=\textwidth]{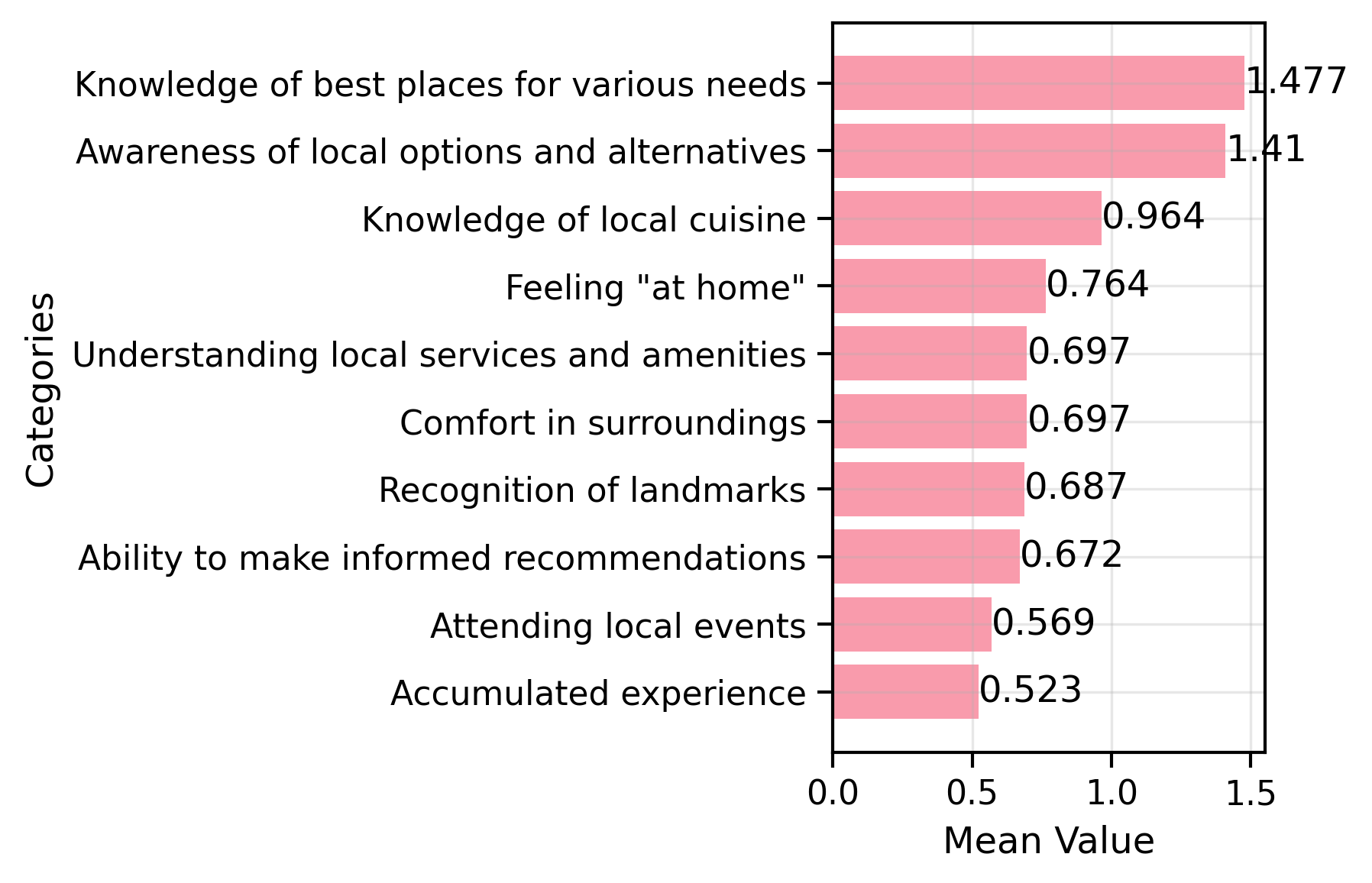}
    \caption{Distribution of Top 10 Sub-components Categories}
    \label{fig:sub}
\end{subfigure}
\caption{Different Category Distributions in LD's Questions during Conversations}
\label{fig:all}
\end{figure*}

Our findings in RQ3 revealed that participants were generally skeptical of LLMs' ability to authentically present as local. Despite LLMs’ growing proficiency in generating contextually appropriate responses, participants frequently identified them as nonlocal and as artificial, often judging them similarly to nonlocal human partners. This suggests that even when LLMs provide factually accurate information, their responses may lack the deeper markers of authentic local presence that participants unconsciously expect.

This raises a critical next question: what specific cues do participants rely on when evaluating localness? If participants struggle to identify nonlocals and LLMs in systematic ways, understanding their decision-making process becomes essential. In RQ4, we shift our focus from classification outcomes to the sensemaking strategies participants use to assess localness in conversation. How do they decide which information matters? What types of questions do they ask? And how do their approaches differ when making accurate versus inaccurate judgments?

To answer these questions, we examine the specific information participants seek when assessing a chat partner’s localness, providing insight into the foundations of localness authentication.

\subsubsection{What Information Do Participants Prioritize?}

Our analysis reveals that when evaluating a chat partner’s localness, participants overwhelmingly prioritized cognitive knowledge over other forms of local identity (Figure~\ref{fig:dom}). Questions related to factual and navigational knowledge of a place were far more frequent than those addressing emotional connection, social engagement, or environmental familiarity. On average, participants asked nearly twice as many questions about cognitive aspects of localness ($M=12.051$) as relational ($M=7.708$) and physical ($M=3.887$) dimensions, suggesting that participants associate local expertise with practical knowledge rather than personal experience or social integration.

Within this cognitive domain, the most frequently sought-after information fell under the knowledge dimension ($M=6.821$). Participants relied heavily on local recommendations ($M=4.256$) as an indicator of authentic local presence (Figure~\ref{fig:com}). This pattern suggests that participants viewed the ability to provide insider tips—such as recommending the best places to eat, shop, or visit—as a strong indicator of local familiarity. This emphasis is reflected in specific sub-components, where knowledge of best places for various needs ($M=1.447$) and awareness of local alternatives ($M=1.41$) were among the most common topics of inquiry (Figure~\ref{fig:sub}). Such practical, experience-based knowledge may have served as a heuristic for distinguishing deep familiarity from superficial researchable facts, as it reflects lived experiences that are difficult to acquire from external sources alone.

Beyond cognitive knowledge, participants also explored relational dimensions of localness, although with less frequency. Many sought evidence of emotional connection to place ($M=4.585$) or social and community engagement ($M=3.313$). This was particularly evident in questions about sense of belonging ($M=1.985$), which appeared in conversational cues about "feeling at home" ($M=0.764$) or comfort in surroundings ($M=0.697$). Some participants asked their chat partners how they personally identified with the place or what it meant to them, while others tested whether the partner could reference specific local traditions, seasonal events, or past historical changes. These inquiries suggest that participants were not only testing factual knowledge but also probing for a personal and emotional connection to the place, as deeper ties might indicate a stronger claim to local identity.

Although physical dimensions of localness were the least frequently addressed, temporal factors—such as length of residence—were still relatively common ($M=3.123$). This suggests that while participants did not strongly rely on birthplace or residency duration as primary indicators of localness, they still recognized time spent in a place as a potential proxy for familiarity. Meanwhile, environmental knowledge ($M=2.785$) and cultural aspects ($M=2.446$) were mentioned moderately often, while physical engagement with the natural environment received the least attention ($M=0.574$). This may indicate that participants viewed local identity as something constructed through social and cognitive interactions, rather than simply being present in a physical space.

Taken together, these findings suggest that participants approached localness assessments with a multi-faceted but knowledge-driven strategy, balancing practical expertise with personal connection to assess the authenticity of their chat partner's local identity.

\begin{table*}[t]
    \centering
    \caption{Results of Zero-inflation Component Using \texttt{beta\_accuracy\_count} and \texttt{beta\_round\_count} Parameters in Bayesian Mixed-effect ZINB Model for Conversation Comparison. Pink highlights indicate significant results where HDI (Highest Density Interval) bounds do not cross zero}
    \label{tab:zero_conv}
    \begin{tabular}{l l r r r  r r r}
        \toprule
        \multirow{2}{*}{Levels} & \multirow{2}{*}{Features Name} & \multicolumn{3}{c}{\texttt{beta\_accuracy\_zero}} & \multicolumn{3}{c}{\texttt{beta\_round\_zero}} \\
        \cmidrule(lr){3-5} \cmidrule(lr){6-8}
         & & Mean & HDI 3\% & HDI 97\% & Mean & HDI 3\% & HDI 97\% \\
        \midrule
        Domain       & Cognitive                & -0.02                & -0.32                &  0.27              & \cellcolor{pink}0.25                  & \cellcolor{pink}0.06                   &  \cellcolor{pink}0.43                   \\
        Domain       & Physical                 & -0.28                & -0.57                &  0.02              &  0.12              & -0.08              &  0.31              \\
        Domain       & Relational               & \cellcolor{pink}-0.31 & \cellcolor{pink}-0.62 & \cellcolor{pink}-0.03 &  0.05              & -0.14              &  0.24              \\
        Dimension    & Cultural                 & \cellcolor{pink}-0.31 & \cellcolor{pink}-0.61 & \cellcolor{pink}-0.01 & -0.16              & -0.34              &  0.03              \\
        Dimension    & Emotional                & \cellcolor{pink}-0.37 & \cellcolor{pink}-0.66 & \cellcolor{pink}-0.06 & -0.04              & -0.23              &  0.15              \\
        Dimension    & Environmental            & -0.29                & -0.59                &  0.02              &  0.14              & -0.05              &  0.34              \\
        Dimension    & Knowledge                & -0.05                & -0.34                &  0.25              &  0.19              &  0                 &  0.36              \\
        Dimension    & Social/Community         & \cellcolor{pink}-0.39 & \cellcolor{pink}-0.68 & \cellcolor{pink}-0.09 & -0                 & -0.20              &  0.19              \\
        Dimension    & Temporal                 & \cellcolor{pink}-0.31 & \cellcolor{pink}-0.61 & \cellcolor{pink}-0.01 &  0.05              & -0.14              &  0.26              \\
        Dimension    & EnvironmentalPhy         & \cellcolor{pink}-0.46 & \cellcolor{pink}-0.76 & \cellcolor{pink}-0.14 & -0.26              & -0.47              & -0.06              \\
        Component    & Active Participation     & \cellcolor{pink}-0.44 & \cellcolor{pink}-0.75 & \cellcolor{pink}-0.16 & \cellcolor{pink}-0.32 & \cellcolor{pink}-0.52 & \cellcolor{pink}-0.13 \\
        Component    & Being Born/Native        & \cellcolor{pink}-0.37 & \cellcolor{pink}-0.67 & \cellcolor{pink}-0.07 & -0.13              & -0.34              &  0.07              \\
        Component    & Change Awareness         & \cellcolor{pink}-0.36 & \cellcolor{pink}-0.67 & \cellcolor{pink}-0.06 & \cellcolor{pink}-0.33 & \cellcolor{pink}-0.53 & \cellcolor{pink}-0.13 \\
        Component    & Civic Engagement         & \cellcolor{pink}-0.48 & \cellcolor{pink}-0.79 & \cellcolor{pink}-0.17 & \cellcolor{pink}-0.41 & \cellcolor{pink}-0.62 & \cellcolor{pink}-0.21 \\
        Component    & Community Investment     & \cellcolor{pink}-0.46 & \cellcolor{pink}-0.78 & \cellcolor{pink}-0.18 & -0.16              & -0.36              &  0.04              \\
        Component    & Ecological Understanding & \cellcolor{pink}-0.48 & \cellcolor{pink}-0.77 & \cellcolor{pink}-0.16 & \cellcolor{pink}-0.24 & \cellcolor{pink}-0.44 & \cellcolor{pink}-0.03 \\
        Component    & Feeling of Home          & \cellcolor{pink}-0.44 & \cellcolor{pink}-0.74 & \cellcolor{pink}-0.14 & -0.18              & -0.38              &  0.02              \\
        Component    & Food Culture             & \cellcolor{pink}-0.29                & \cellcolor{pink}-0.59                & \cellcolor{pink}-0                & -0.16              & -0.33              &  0.04              \\
        Component    & Formative Years          & \cellcolor{pink}-0.33 & \cellcolor{pink}-0.62 & \cellcolor{pink}-0.02 & -0.11              & -0.31              &  0.10              \\
        Component    & Geographic Familiarity   & \cellcolor{pink}-0.31                & \cellcolor{pink}-0.61                & \cellcolor{pink}-0                &  0                 & -0.20              &  0.20              \\
        Component    & Hidden Gems              & \cellcolor{pink}-0.45 & \cellcolor{pink}-0.75 & \cellcolor{pink}-0.15 & \cellcolor{pink}-0.37 & \cellcolor{pink}-0.58 & \cellcolor{pink}-0.17 \\
        Component    & Historical Knowledge     & \cellcolor{pink}-0.44 & \cellcolor{pink}-0.72 & \cellcolor{pink}-0.13 & \cellcolor{pink}-0.29 & \cellcolor{pink}-0.48 & \cellcolor{pink}-0.09 \\
        Component    & Identity Connection      & \cellcolor{pink}-0.44 & \cellcolor{pink}-0.75 & \cellcolor{pink}-0.14 & \cellcolor{pink}-0.23 & \cellcolor{pink}-0.42 & \cellcolor{pink}-0.03 \\
        Component    & Language/Dialect         & \cellcolor{pink}-0.49 & \cellcolor{pink}-0.79 & \cellcolor{pink}-0.19 & \cellcolor{pink}-0.32 & \cellcolor{pink}-0.53 & \cellcolor{pink}-0.13 \\
        Component    & Local Customs/Norms      & \cellcolor{pink}-0.50 & \cellcolor{pink}-0.81 & \cellcolor{pink}-0.20 & \cellcolor{pink}-0.43 & \cellcolor{pink}-0.63 & \cellcolor{pink}-0.22 \\
        Component    & Local Recommendations    & -0.22                & -0.53                &  0.07              & -0.03              & -0.21              &  0.15              \\
        Component    & Long-term Residence      & \cellcolor{pink}-0.42 & \cellcolor{pink}-0.72 & \cellcolor{pink}-0.11 & -0.17              & -0.37              &  0.03              \\
        Component    & Natural Environment      & \cellcolor{pink}-0.47 & \cellcolor{pink}-0.78 & \cellcolor{pink}-0.18 & \cellcolor{pink}-0.23 & \cellcolor{pink}-0.44 & \cellcolor{pink}-0.03 \\
        Component    & Navigation/Wayfinding    & \cellcolor{pink}-0.37 & \cellcolor{pink}-0.68 & \cellcolor{pink}-0.07 & \cellcolor{pink}-0.22 & \cellcolor{pink}-0.42 & \cellcolor{pink}-0.02 \\
        Component    & Personal Relationships   & \cellcolor{pink}-0.49 & \cellcolor{pink}-0.81 & \cellcolor{pink}-0.20 & \cellcolor{pink}-0.21 & \cellcolor{pink}-0.42 & \cellcolor{pink}-0.02 \\
        Component    & Sense of Belonging       & \cellcolor{pink}-0.36 & \cellcolor{pink}-0.66 & \cellcolor{pink}-0.06 & -0.07              & -0.28              &  0.12              \\
        Component    & Geographic Familiarity (Phy)   & \cellcolor{pink}-0.49 & \cellcolor{pink}-0.79 & \cellcolor{pink}-0.18 & \cellcolor{pink}-0.37 & \cellcolor{pink}-0.58 & \cellcolor{pink}-0.17 \\
        Component    & Natural Environment (Phy)      & \cellcolor{pink}-0.46 & \cellcolor{pink}-0.76 & \cellcolor{pink}-0.15 & \cellcolor{pink}-0.29 & \cellcolor{pink}-0.49 & \cellcolor{pink}-0.08 \\
        Component    & Ecological Understanding (Phy) & \cellcolor{pink}-0.51 & \cellcolor{pink}-0.81 & \cellcolor{pink}-0.21 & \cellcolor{pink}-0.37 & \cellcolor{pink}-0.56 & \cellcolor{pink}-0.15 \\
        \bottomrule
    \end{tabular}
\end{table*}

\subsubsection{Differences Between Correct and Incorrect Localness Judgments}

To better understand why some participants were more successful than others in assessing localness, we examined differences in how correct and incorrect judges structured their information-seeking strategies. Using a Bayesian mixed-effects ZINB model, we identified patterns that distinguished ``effective'' from ``ineffective'' questioning behaviors.


One of the most striking findings was that participants who made correct localness judgments engaged more consistently with relational aspects of localness. Our analysis revealed a significant decrease in the likelihood of ignoring relational questions among correct judges ($\beta = -0.31$, HDI: $[-0.62, -0.03]$). This effect was particularly strong in the emotional ($\beta = -0.37$, HDI: $[-0.66, -0.06]$) and social/community ($\beta = -0.39$, HDI: $[-0.68, -0.09]$) dimensions, suggesting that correct judges were more likely to probe for social and emotional indicators of localness, whereas incorrect judges overlooked these cues.

This difference in questioning depth extended to specific topics. Participants who successfully identified localness tended to ask more about active community participation ($\beta = -0.44$, HDI: $[-0.75, -0.16]$) and hidden local knowledge ($\beta = -0.4$, HDI: $[-0.75, -0.15]$). This suggests that a strong local identity is often demonstrated through personal engagement in the community rather than just knowledge of locations. In contrast, incorrect judges disproportionately relied on cognitive knowledge alone, with no significant differences in their questioning frequency for knowledge-based topics compared to correct judges.

Interestingly, the ability to ask about cultural aspects of localness was a key predictor of successful judgment ($\beta = 0.35$, HDI: $[0.04, 0.64]$). Participants who incorporated cultural questions—such as references to local dialects, food culture, or historical events—were more likely to differentiate between truly local and nonlocal chat partners. This aligns with prior research on sensemaking, which suggests that combining multiple, diverse sources of evidence leads to more accurate judgments.

\subsubsection{Temporal Evolution in Questioning Strategies}

Beyond differences in what participants asked, we also examined how questioning strategies changed over time. The round effects in our Bayesian model revealed that as conversations progressed, participants asked fewer diverse types of questions. In particular, the likelihood of zero interactions (i.e., failing to ask about a specific category) increased significantly in the cognitive domain over time ($\beta = 0.25$, HDI: $[0.06, 0.43]$). This suggests that as participants became more confident in their assessments, they narrowed their focus, asking about fewer types of localness features.

Notably, participants became less likely to explore community engagement, cultural identity, and historical awareness over time. For example, the probability of engaging with civic engagement topics dropped significantly as conversations progressed ($\beta = -0.41$, HDI: $[-0.62, -0.21]$), as did questions about local customs ($\beta = -0.43$, HDI: $[-0.63, -0.22]$) and historical knowledge ($\beta = -0.29$, HDI: $[-0.48, -0.09]$). This narrowing of inquiry suggests that participants developed early impressions about localness and then stuck to a specific line of questioning, potentially missing important cues later in the conversation.

However, some localness features remained stable throughout interactions. Participants consistently asked about residency duration, geographic familiarity, and local recommendations from the beginning to the end of the conversation, with no significant decrease in questioning over time. This suggests that some indicators of localness—particularly those related to place-based knowledge—are seen as fundamental and remain central regardless of conversational progression.

\subsection{RQ4-2: Localness Sensemaking: How Do Participants Filter and Reframe Localness Information?}

While the previous section examined how participants gather information to assess localness, this section focuses on how they filter and prioritize that information to reach a final judgment. Simply collecting information is not enough—participants must decide which details are meaningful, which are unreliable, and which best demonstrate an authentic local presence. Drawing on sensemaking theory \cite{pirolli2005sensemaking}, we examine how participants refine their understanding of localness by selecting key conversational moments that they believe are the most diagnostic of their chat partner’s identity.

This filtering process is directly captured in our study by asking participants to select conversation rounds they considered most important in their localness assessments. By analyzing which conversational exchanges were chosen and comparing them to the broader information-gathering trends, we can understand how participants winnow down broad exploration into focused, actionable evidence.

\subsubsection{The Distribution of Localness Features in Selected Conversation}

Participants exhibited a clear pattern in how they refined their localness assessments. While the same general categories of localness appeared in both their full conversations and their selected key exchanges, the intensity of focus shifted.

At the highest level (Figure~\ref{fig:dom_selected}), participants continued to prioritize cognitive knowledge as the most informative domain, but with a notable reduction in its presence ($M=12.051$ in full conversations vs. $M=6.869$ in selected exchanges). Similarly, relational ($M=4.471$) and physical ($M=2.021$) aspects were retained but filtered down in magnitude. This suggests that participants did not disregard relational and physical dimensions but relied more selectively on key moments that demonstrated these aspects convincingly.

Drilling deeper into which aspects of localness were most retained, we see that knowledge-based indicators remained dominant, though they were used in a more targeted manner. The knowledge dimension ($M=3.932$, down from $M=6.821$) was still the most frequent, followed by emotional connection ($M=2.618$) and social/community engagement ($M=1.853$). This indicates that while factual knowledge remained the primary lens for assessing localness, participants still considered emotional and social connections important but relied on fewer, higher-quality examples to validate their judgments.

At the component level (Figure~\ref{fig:com_selected}), local recommendations ($M=2.372$) and sense of belonging ($M=1.110$) were the most frequently selected indicators. This pattern shows that participants valued practical knowledge and emotional attachment as key signals of authentic local presence. Notably, at the sub-component level, knowledge of best places ($M=0.796$) and awareness of local options ($M=0.785$) remained at the top, reinforcing the idea that the ability to provide specific, experience-based recommendations carried substantial weight in determining localness.

In contrast, experiential factors—such as comfort in surroundings ($M=0.361$) and accumulated experience ($M=0.293$)—were retained far less frequently. This suggests that while participants may have initially asked about lived experience, they ultimately placed greater trust in tangible knowledge rather than subjective self-reports when making their final decisions.

When selecting conversations for supporting localness judgment, participants maintained similar categorical priorities but with notably reduced magnitudes across all dimensions.

\subsubsection{Differences between Correct and Incorrect Localness Judgment}

To better understand how effective participants were at filtering relevant information, we compared the selected conversation rounds of those who correctly identified their partner’s local status to those who misjudged it. Using a Bayesian mixed-effects ZINB model, we examined whether correct and incorrect judges differed in what information they retained and how they structured their evidence selection.

Unlike in the information-gathering phase, where significant differences emerged between correct and incorrect judges, no clear patterns appeared in how participants filtered their final selections at the broad domain and dimension levels. The zero-inflation component ($\beta_{local_zero}$), which measures whether certain types of localness features were completely omitted from selection, showed no significant differences between the two groups (Table~\ref{tab:zero_selected}). This suggests that both correct and incorrect judges engaged with the same overall categories of localness when filtering information—they did not systematically exclude or overlook specific domains.

Similarly, the feature prevalence component (count model $\beta_{local_count}$), which measures how much weight participants placed on different localness categories, also yielded no significant effects (Table~\ref{tab:count_selected}). This means that while participants differed in their ability to correctly assess localness, they did not differ substantially in how much importance they assigned to different aspects of localness during their filtering process. In other words, both groups retained a similar mix of factual, relational, and experiential cues, but correct judges were more effective at interpreting them.

\subsubsection{Temporal Evolution in Filtering Strategies}

This finding contrasts with the information-gathering phase, where participants became more selective in their questioning as conversations progressed. In that phase, they gradually reduced their inquiry scope, asking about fewer but more targeted aspects of localness as they formed clearer impressions. However, this adaptive strategy did not carry over into filtering. Once participants had formed an idea of what they considered strong evidence of localness, they applied that interpretation consistently, rather than adjusting their selection criteria as they gained more experience.

This rigidity suggests that some participants may have prematurely locked into a particular model of localness, leading to misjudgments. Because incorrect judges were more likely to filter out relational and cultural elements, their fixed selection criteria may have reinforced a narrow, fact-based perspective that did not account for the full complexity of local identity (this finding is supported by Section~\ref{sec:incorrect}). Conversely, correct judges appeared to develop a broader and more integrative approach from the outset, allowing them to retain the right balance of cognitive, social, and cultural indicators.

\subsection{RQ4-3: Localness Sensemaking: How Do Participants Interpret and Use Information to Judge Localness?}

\begin{figure*}[t]
\centering
\begin{subfigure}[b]{0.45\textwidth}
    \includegraphics[width=\textwidth]{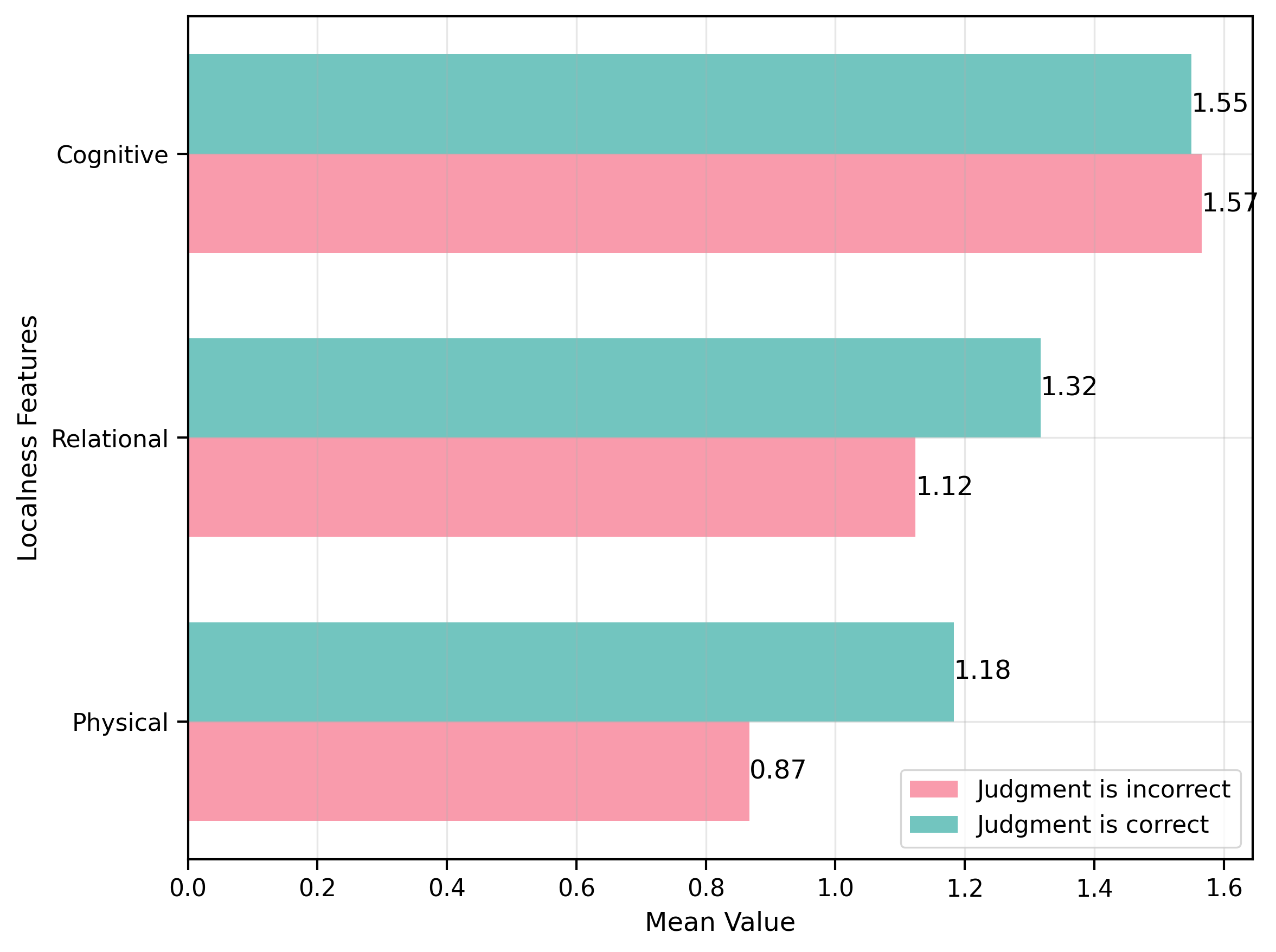}
    \caption{Distribution of Domain Categories}
    \label{fig:dom_reason}
\end{subfigure}
\hfill
\begin{subfigure}[b]{0.45\textwidth}
    \includegraphics[width=\textwidth]{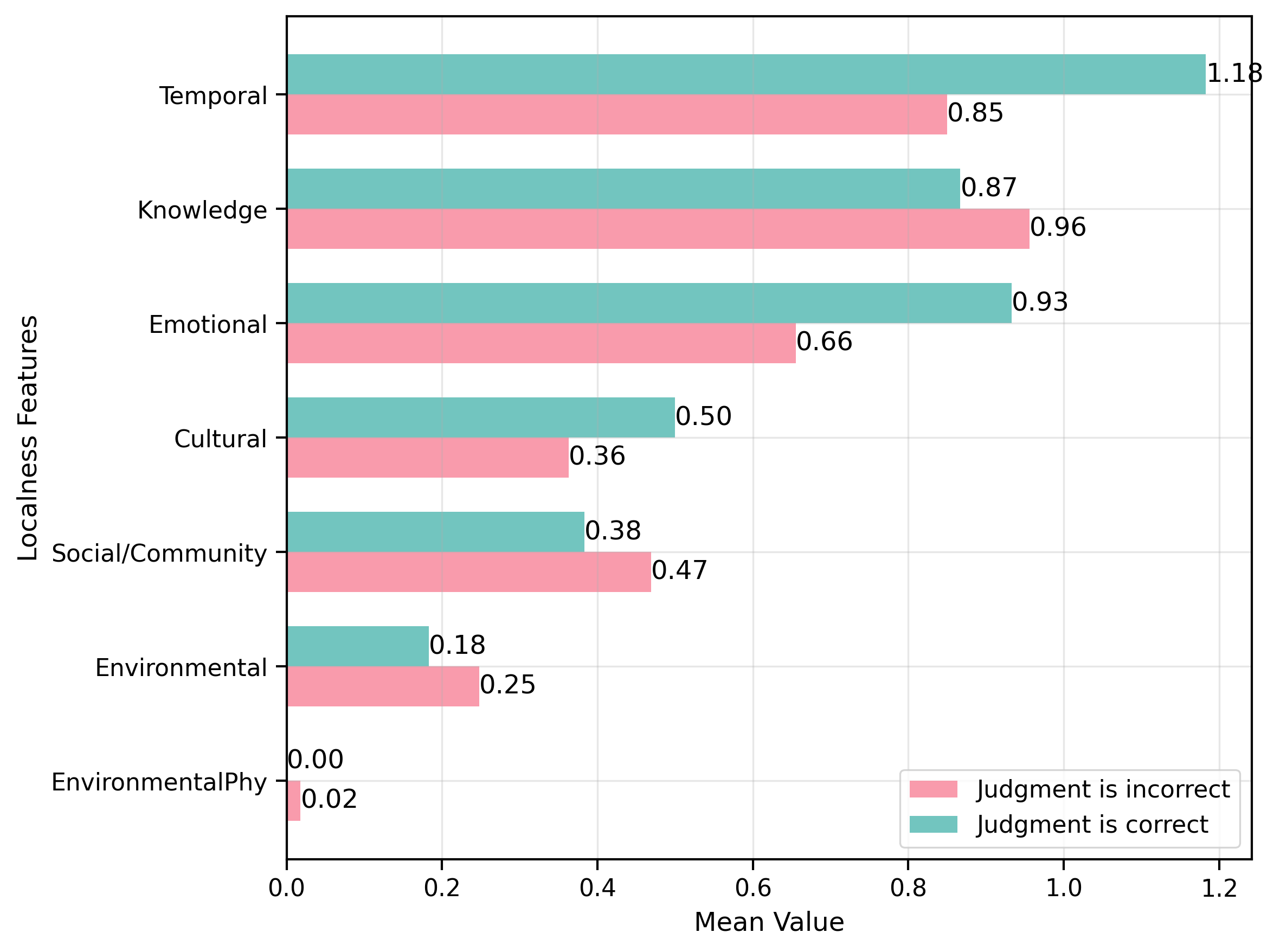}
    \caption{Distribution of Dimension Categories}
    \label{fig:dim_reason}
\end{subfigure}

\begin{subfigure}[b]{0.45\textwidth}
    \includegraphics[width=\textwidth]{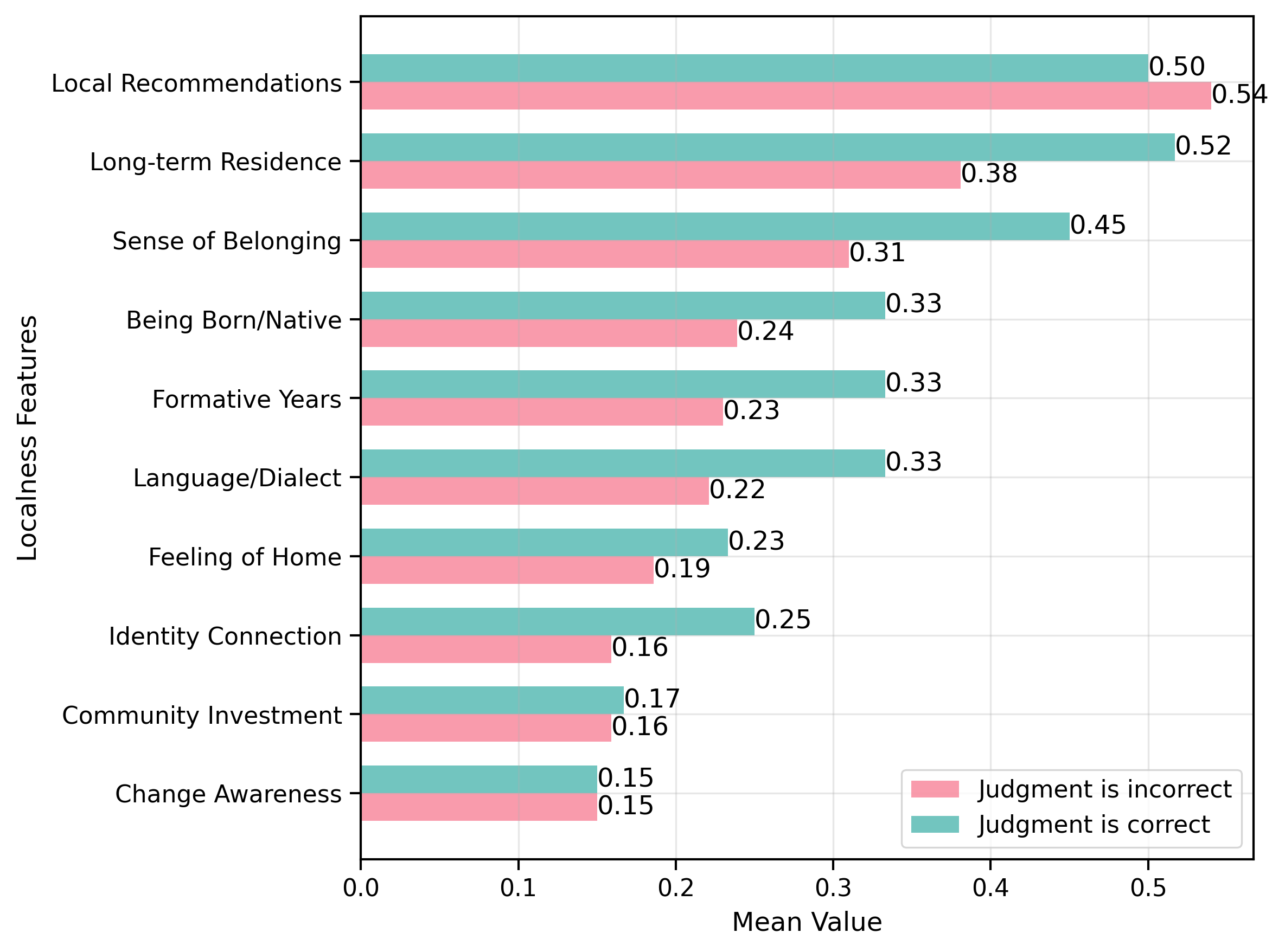}
    \caption{Distribution of Top 10 Components Categories}
    \label{fig:com_reason}
\end{subfigure}
\hfill
\begin{subfigure}[b]{0.45\textwidth}
    \includegraphics[width=\textwidth]{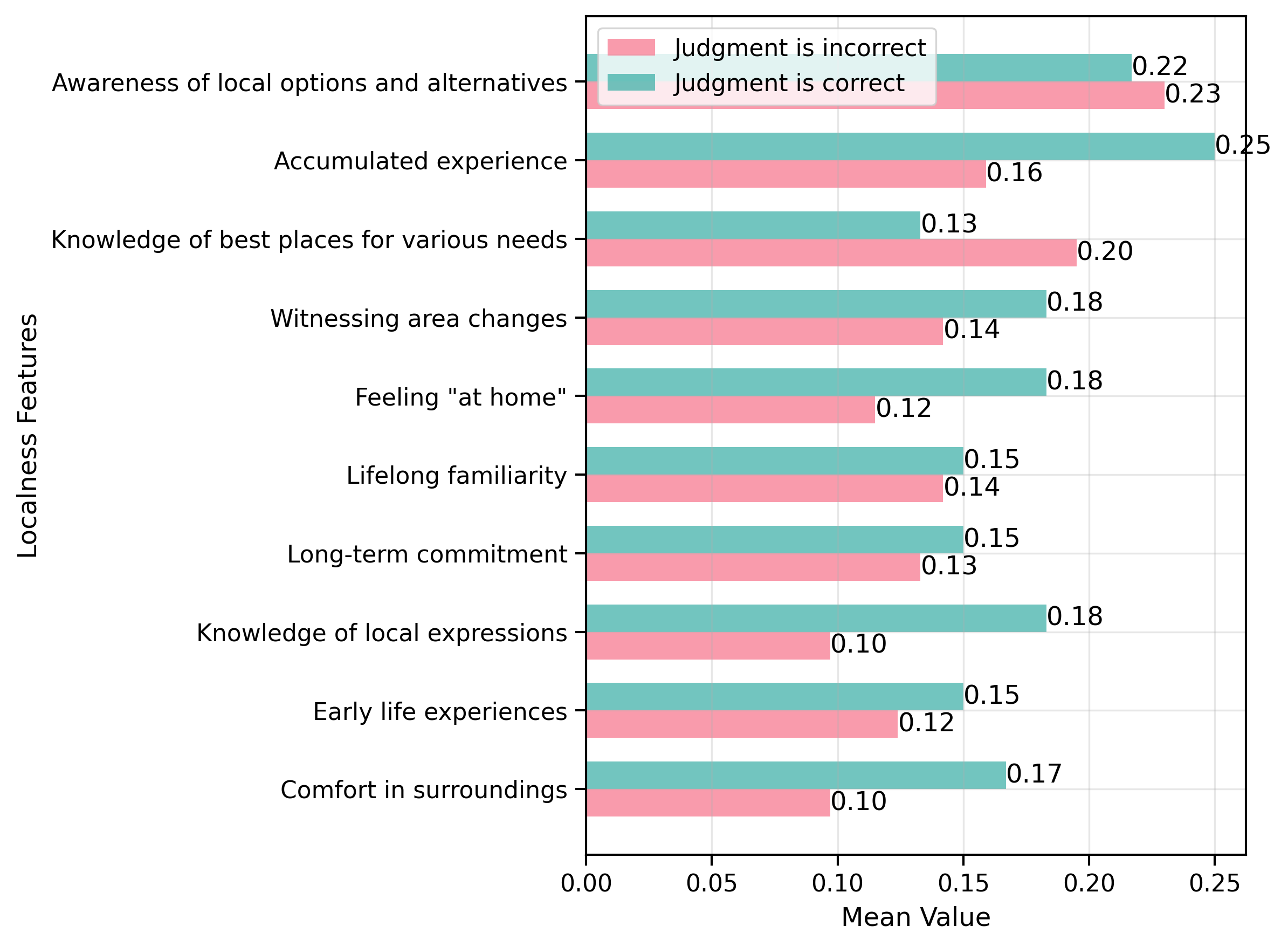}
    \caption{Distribution of Top 10 Sub-components Categories}
    \label{fig:sub_reason}
\end{subfigure}
\caption{Comparison of Category Distributions in LD's Localness Judgment Reason}
\label{fig:all_judgment_reason}
\end{figure*}

Our analysis of RQ4-2 revealed how participants filtered and prioritized the information they gathered, selecting conversation rounds that they believed were most diagnostic of localness. While this filtering process sheds light on what participants considered important, it does not yet explain how they arrived at their final judgments. Selecting a subset of information is only part of the sensemaking process—participants must also interpret that information, weighing different types of evidence, resolving inconsistencies, and forming structured reasoning about localness.

Thus, in this section, we shift from examining what information participants prioritized to how they made sense of it. What reasoning strategies led to accurate judgments? Where did incorrect assessments go wrong? And how did participants integrate different aspects of localness—such as factual knowledge, lived experience, and emotional connection—into their final determinations? By analyzing participants’ explanations for their decisions, we uncover the interpretative frameworks they used to transform filtered information into confident judgments about local authenticity.

\subsubsection{Correct Judgments: Effective Cross-Referencing and Detailed Analysis}

Participants who made correct judgments about whether their chat partner was local employed robust cross-referencing strategies and exhibited nuanced contextual analysis. A key strategy was drawing on multiple layers of evidence rather than relying on specific individual indicators. This is supported by Figure~\ref{fig:com_reason}, which shows that participants who made correct judgments emphasized features such as Local Recommendations ($mean = 0.54$) and Long-term Residence ($mean = 0.52$) more than those who made incorrect judgments. For instance, P7 synthesized personal history and local cultural knowledge to form an accurate judgment, reasoning that their chat partner was:

\begin{quote}
\textit{``not just a student, seems to have a really good understanding of [city name] culture and has experienced many seasons in [city name].''}
\end{quote}

When participants correctly judged their chat partner as non-local, they often detected gaps in specific knowledge or inconsistencies in behavior. As Figure~\ref{fig:dim_reason} indicates, Knowledge dimension ($mean = 0.87$ for correct judgments) was a strong predictor of accurate judgments. For example, P95 correctly concluded that their chat partner was not local, reasoning: 

\begin{quote}
\textit{``didn’t know what [city name] was known for. Said favorite restaurant was a deli on the east side.''}
\end{quote}

Participants also excelled in contextual and experiential analysis, detecting whether the chat partner had a deeper understanding of local events and activities. As Figure~\ref{fig:sub_reason} illustrates, correct judgments often included recognition of specific localness sub-components Awareness of Local Options and Alternatives ($mean = 0.23$) and Knowledge of Best Places for Various Needs ($mean = 0.20$). For example, P33 stated:

\begin{quote}
\textit{``They knew about a lot of local events and things to do, like [trail name] and [nature preserve name], that someone interested in the community would know about and do.''}
\end{quote}

\subsubsection{Incorrect Judgments: Over-Reliance on Single Factors and Narrow Definitions}
\label{sec:incorrect}

Participants who made incorrect judgments frequently relied on abstract or generic conceptions of localness or failed to cross-check their assumptions with sufficient evidence. Figure~\ref{fig:com_reason} shows that incorrect judgments were more often associated with an over-reliance on components like Being Born/Native ($mean = 0.24$) and Language/Dialect ($mean = 0.22$), suggesting a tendency to overvalue surface-level indicators. For example, P152 overemphasized self-identification as local without corroborating other evidence, stating:

\begin{quote}
\textit{``They said they were from [city name], and they had spent much of their adult life here.''}
\end{quote}

Similarly, participants who judged someone incorrectly often discounted cultural differences or communication styles. Figure~\ref{fig:dom_reason} highlights how incorrect judgments frequently underestimated the Relational dimension ($mean = 1.12$ for incorrect judgments compared to $1.32$ for correct judgments). For instance, P127 assumed their chat partner was non-local due to language patterns, stating:

\begin{quote}
\textit{``My sense is the person is a non-native English speaker, and I speculate a possible foreign student.''}
\end{quote}

Participants who made incorrect judgments also often prioritized perceived transience over active participation in the local community. For example, P122 doubted their chat partner’s localness, saying:

\begin{quote}
\textit{``I think they like [city name] but are not committed to it being their permanent home.''}
\end{quote}

As shown in Figure~\ref{fig:sub_reason}, this overemphasis on sub-components Long-term Commitment ($mean = 0.13$ for incorrect judgments) reflects a narrow interpretation of localness that undervalues current engagement and community integration.

The integrated analysis of Figures~\ref{fig:com_reason} through \ref{fig:sub_reason} reveals important distinctions in how participants approached localness judgments. Correct judgments were more likely to incorporate a broader range of features, such as practical knowledge, temporal understanding, and emotional engagement. In contrast, incorrect judgments relied more on isolated indicators like residency or language, often leading to overgeneralizations. These findings emphasize the detailed definitions of localness that account for diverse and dynamic community interactions.

\subsubsection{Judgments Involving LLMs: Detecting Localness through Pattern Recognition and Inconsistencies}

\begin{figure*}[t]
\centering
\begin{subfigure}[b]{0.45\textwidth}
    \includegraphics[width=\textwidth]{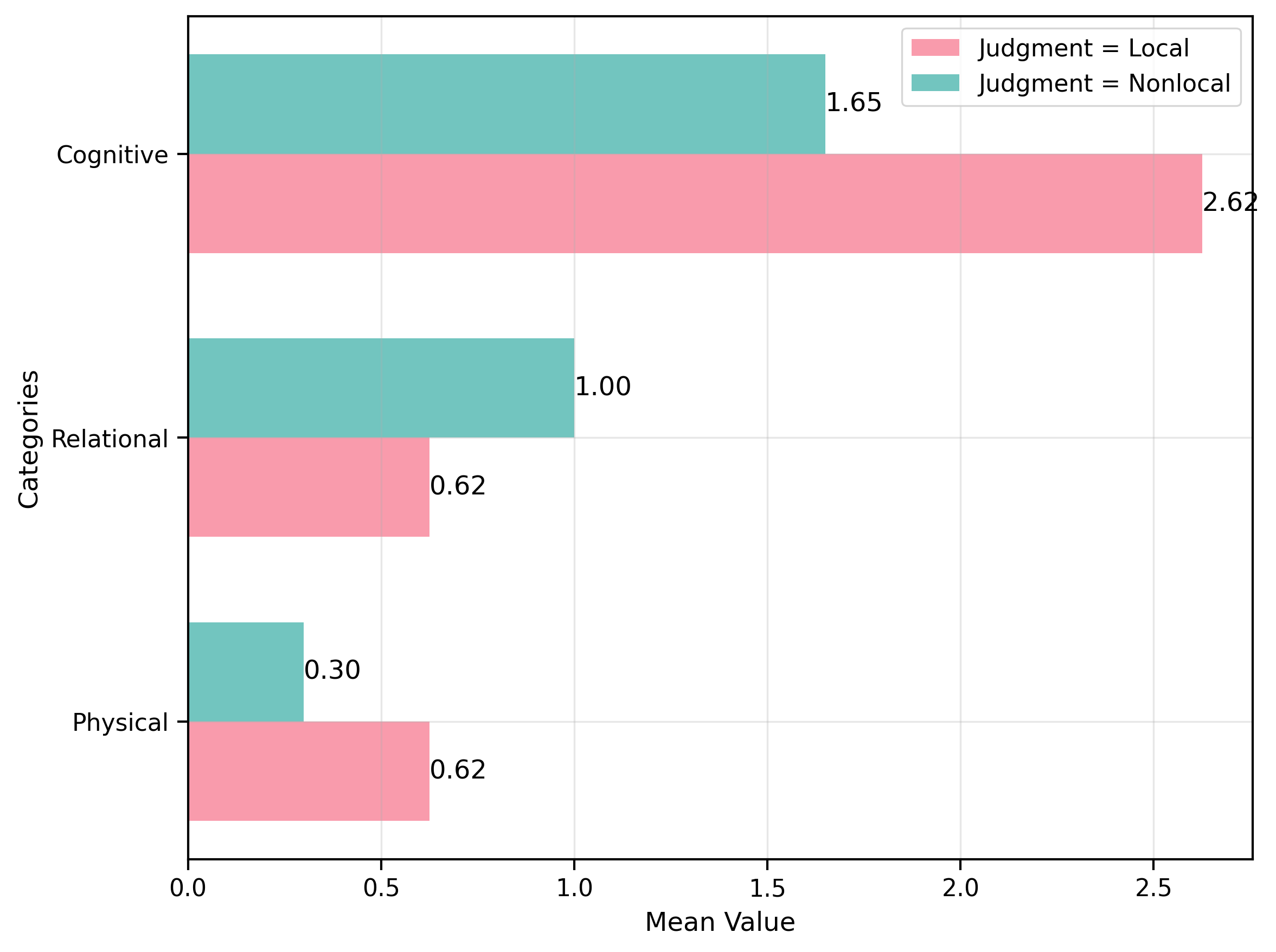}
    \caption{Distribution of Domain Categories}
    \label{fig:dom_llm}
\end{subfigure}
\hfill
\begin{subfigure}[b]{0.45\textwidth}
    \includegraphics[width=\textwidth]{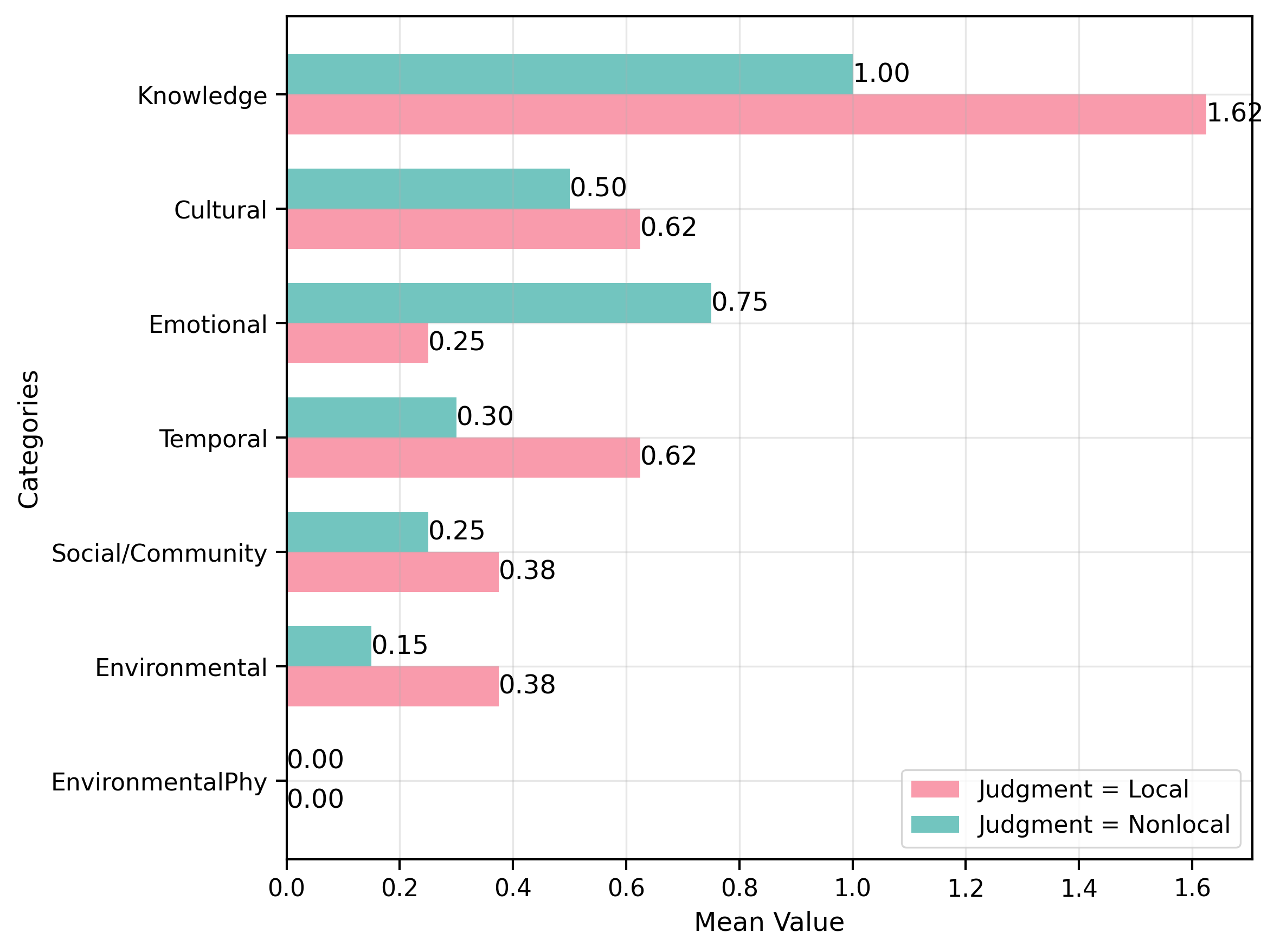}
    \caption{Distribution of Dimension Categories}
    \label{fig:dim_llm}
\end{subfigure}

\begin{subfigure}[b]{0.45\textwidth}
    \includegraphics[width=\textwidth]{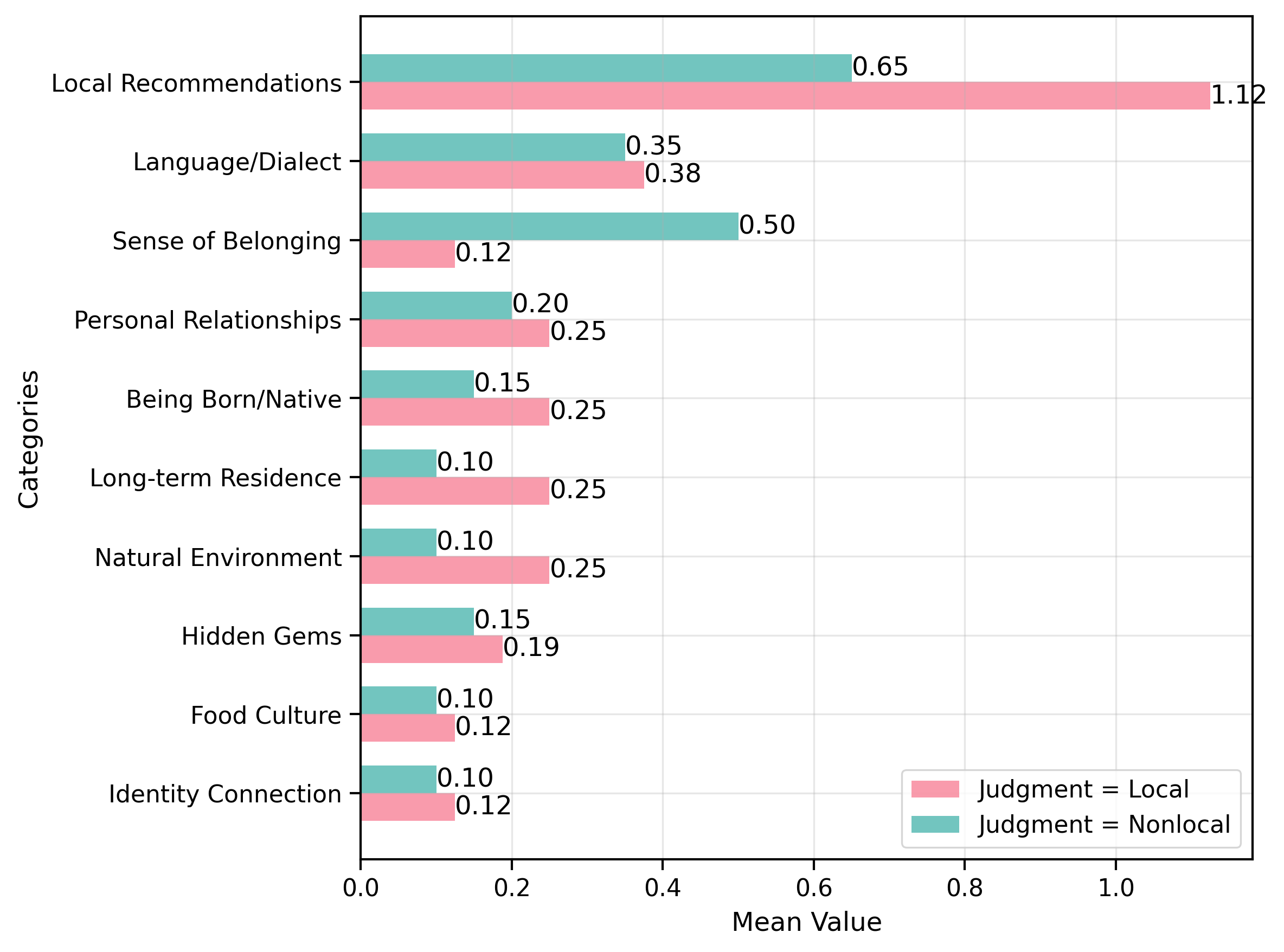}
    \caption{Distribution of Top 10 Components Categories}
    \label{fig:com_llm}
\end{subfigure}
\hfill
\begin{subfigure}[b]{0.45\textwidth}
    \includegraphics[width=\textwidth]{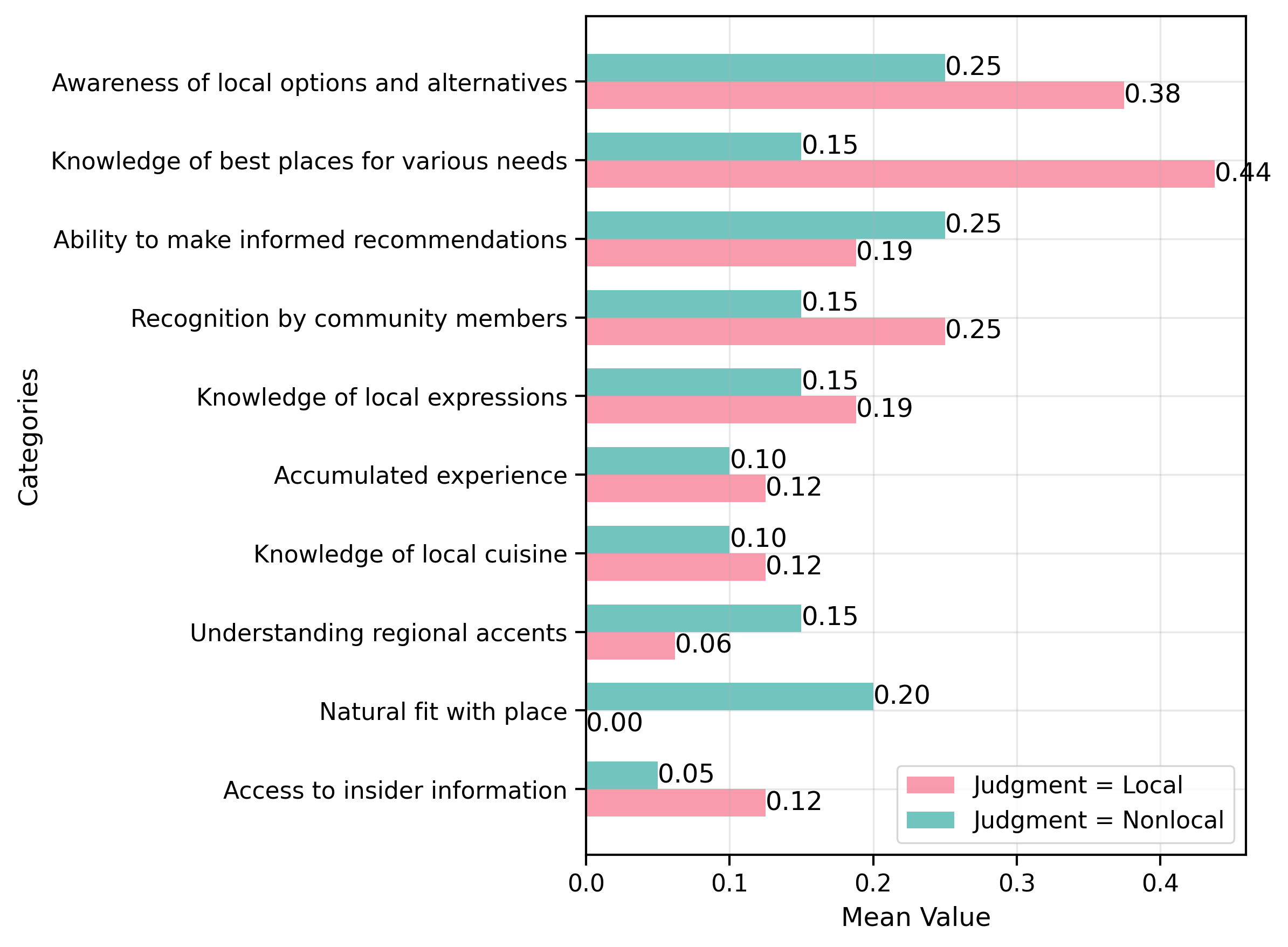}
    \caption{Distribution of Top 10 Sub-components Categories}
    \label{fig:sub_llm}
\end{subfigure}
\caption{Comparison of Category Distributions in LD's Reasons for Their Local and Nonlocal Localness Judgment}
\label{fig:all_llm}
\end{figure*}

The ability of LLMs to provide Local Recommendations emerged as the most valued feature in localness judgments, with the highest mean score ($1.125$), highlighting their strength in offering practical, contextually relevant suggestions. This aligns with participant reports emphasizing the importance of breadth and specificity in local knowledge. For instance, P114 noted:  

\begin{quote}
    ``[My chat partner] seems to know a variety of different restaurants and shops in the [city name] area. They seemed to know of different activities to do like their favorite walking trails.''  
\end{quote}

However, while LLMs were effective in recommending widely recognized businesses and locations, participants penalized them when responses lacked deep cultural or experiential nuance. The relatively lower mean score for Knowledge of Best Places for Various Needs ($0.438$) suggests that while LLMs provide general knowledge, they struggle to surface hidden gems—an important expectation for localness. This was echoed by P49, who remarked:

\begin{quote}
    ``It seemed like the person researched some popular businesses, places, and things to do in [city name] rather than knowing other hidden gems that a local person would know.''  
\end{quote}

This illustrates that true localness is often associated with knowledge beyond what is easily searchable, including personalized insights and community-specific information.


While Language/Dialect ($0.375$) ranked as a secondary factor, its impact on localness judgments was nuanced. Participants expected a local chat partner to use familiar linguistic expressions and region-specific vocabulary. However, minor inconsistencies or unnatural phrasing led to skepticism. Some participants even used linguistic inconsistencies as evidence of AI presence, as demonstrated by P153:

\begin{quote}
    ``He seems to know so much that I'm starting to doubt whether they're AI. But they use different ways of writing 'Midwesterner' and 'Mid-West.' That inconsistency seems like something AI would not do.''  
\end{quote}

This suggests that while LLMs can mimic local dialects to some extent, minor errors in consistency can inadvertently trigger doubts about authenticity.

Similarly, Sense of Belonging ($0.5$) was a key determinant of non-localness. Participants penalized LLMs when they failed to convey emotional or cultural connection to the region. This was evident when responses lacked personal anecdotes or displayed rigid factual correctness without deeper context. For instance, P72 noted:

\begin{quote}
    ``The answers were too vague, most likely from AI. While they were regionally accurate, they weren’t specific enough for me to believe the person was local to [city name].''  
\end{quote}

This suggests that emotional and relational components of localness remain critical gaps for LLMs, reinforcing the need for enhancements in these areas.


LLMs demonstrated moderate success in factual accuracy, as reflected in the Knowledge dimension (mean of $1.625$ for local judgments). However, inaccuracies significantly contributed to judgments of non-localness, with a penalized mean score of $1.0$. Participants frequently relied on contextual accuracy as a determinant of localness. For example, P32 flagged an incorrect time for a local event as evidence that their chat partner was non-local:

\begin{quote}
    ``It said it was going to the Farmers' Market at 12:30 when the Farmers' Market closes at 1.''  
\end{quote}

Similarly, users assessed whether LLM-generated responses aligned with realistic local behaviors. When responses suggested impractical activities, such as an unrealistic outdoor plan on a rainy day, participants were quick to judge them as non-local. P41 illustrated this by stating:

\begin{quote}
    ``No one would do the lake loop on a rainy day! Not enough time to finish it with today’s rain.''  
\end{quote}

These findings indicate that true localness requires not just factual correctness but also practical awareness of local norms, events, and contextual appropriateness—areas where LLMs still struggle.


A key area of concern for LLMs was the Relational dimension, where they scored lower for local judgments ($0.625$) compared to non-local judgments ($1.0$), highlighting the difficulty in achieving relational authenticity. Similarly, Understanding Community Values had the lowest mean score ($0.125$), emphasizing LLMs' struggle in conveying social and cultural integration. This is reflected in participant feedback penalizing responses that lacked a personal touch or relational depth.

The Emotional dimension also played a critical role in non-local judgments, with a higher mean score in non-local evaluations ($0.75$) compared to local ones ($0.25$). Participants expected a local to express attachment, nostalgia, or emotional resonance. When responses failed to exhibit these characteristics, LLMs were judged as non-local.

Our findings reveals that LLMs excel in cognitive and practical aspects of localness, particularly in providing recommendations and accessing factual information. However, emotional connection, relational depth, and hidden cultural knowledge remain significant challenges. Participants actively detected these weaknesses using contextual accuracy checks, linguistic cues, and expectations of realistic behaviors.

\subsection{RQ5: What Factors Predict Accurate Localness Judgments?}
\label{sec:predict}

\begin{figure}
    \centering
    \includegraphics[width=0.5\linewidth]{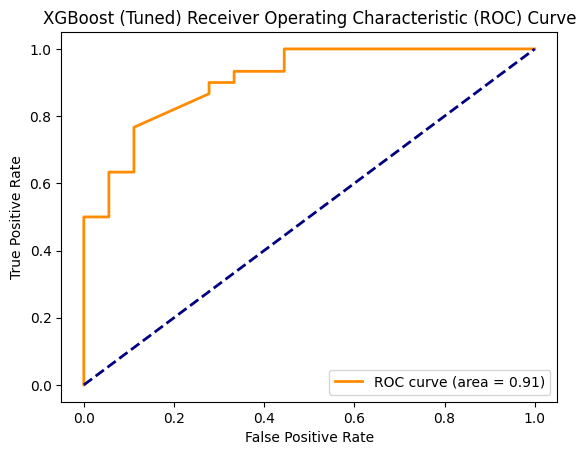}
    \caption{XGBoost ROC Curve}
    \label{fig:roc}
\end{figure}

\begin{figure*}[t]
    \centering
    \begin{subfigure}[t]{0.78\textwidth}
        \centering
        \includegraphics[width=\textwidth]{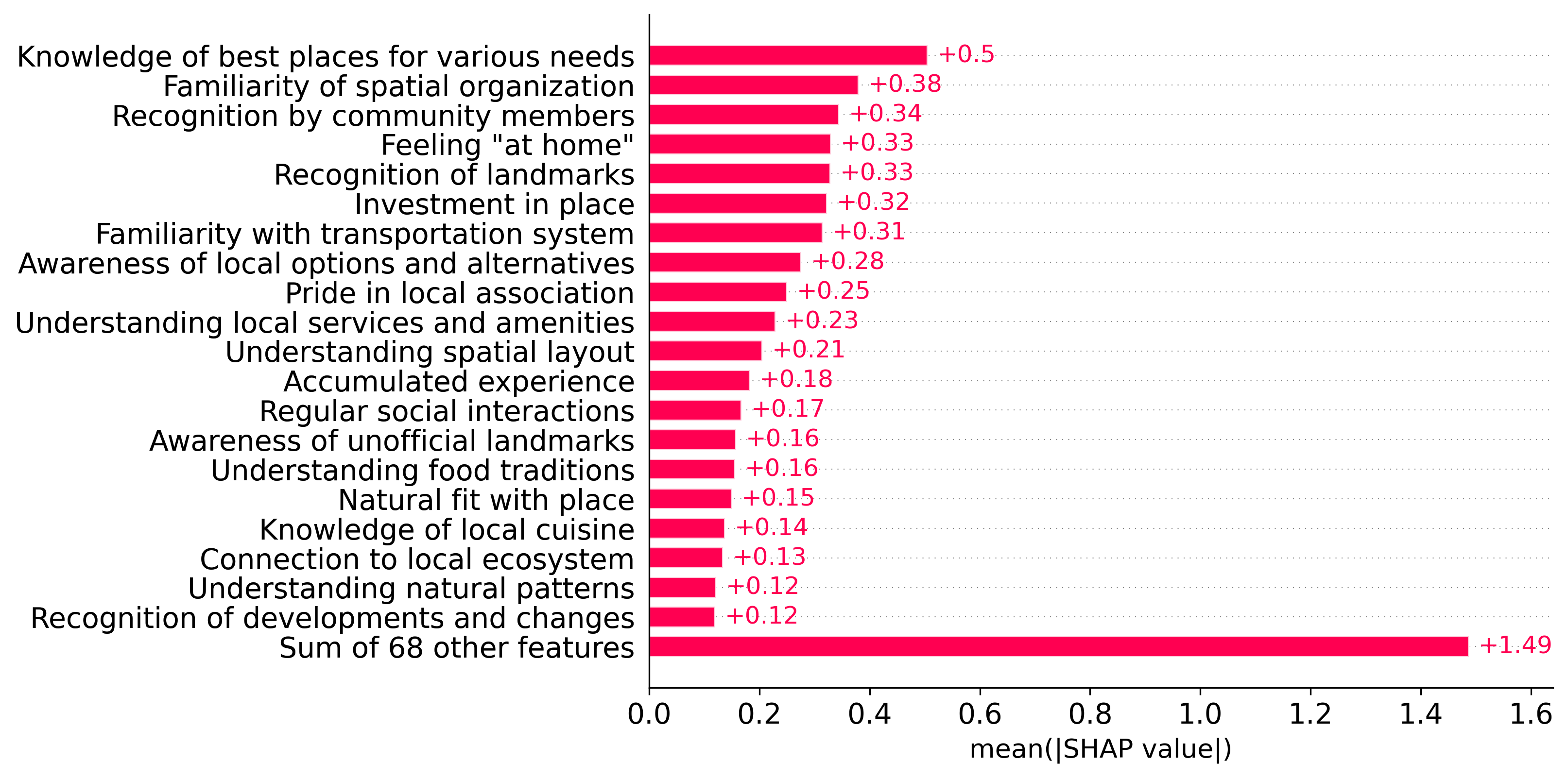}
        \caption{SHAP values of sub-components showing their relative importance in localness prediction}
        \label{fig:shap_subcom}
    \end{subfigure}

    \vspace{0.5em} 
    \begin{subfigure}[t]{0.58\textwidth}
        \centering
        \includegraphics[width=\textwidth]{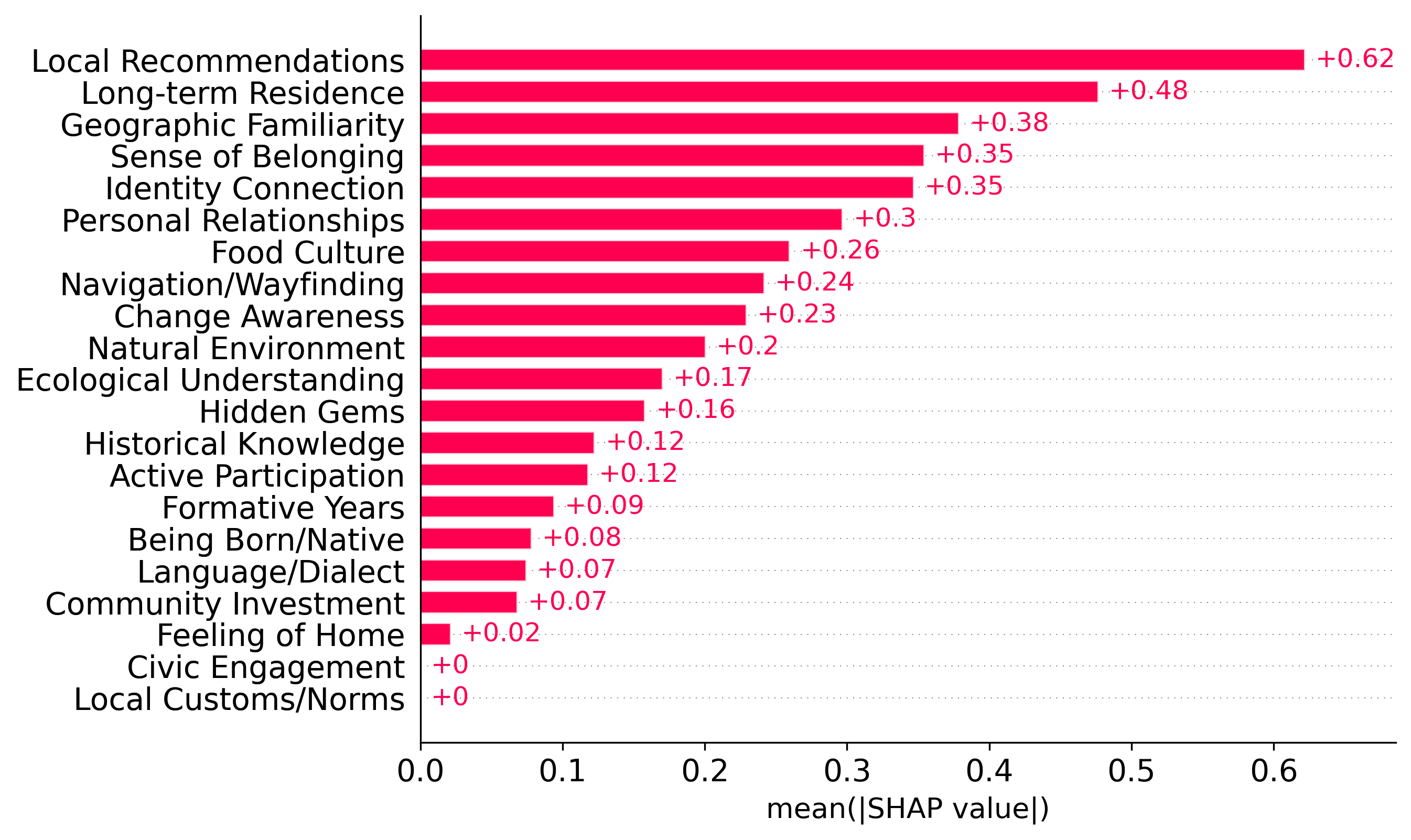}
        \caption{SHAP values aggregated by components}
        \label{fig:shap_com}
    \end{subfigure}
    \hfill 
    \begin{subfigure}[t]{0.38\textwidth}
        \centering
        \includegraphics[width=\textwidth]{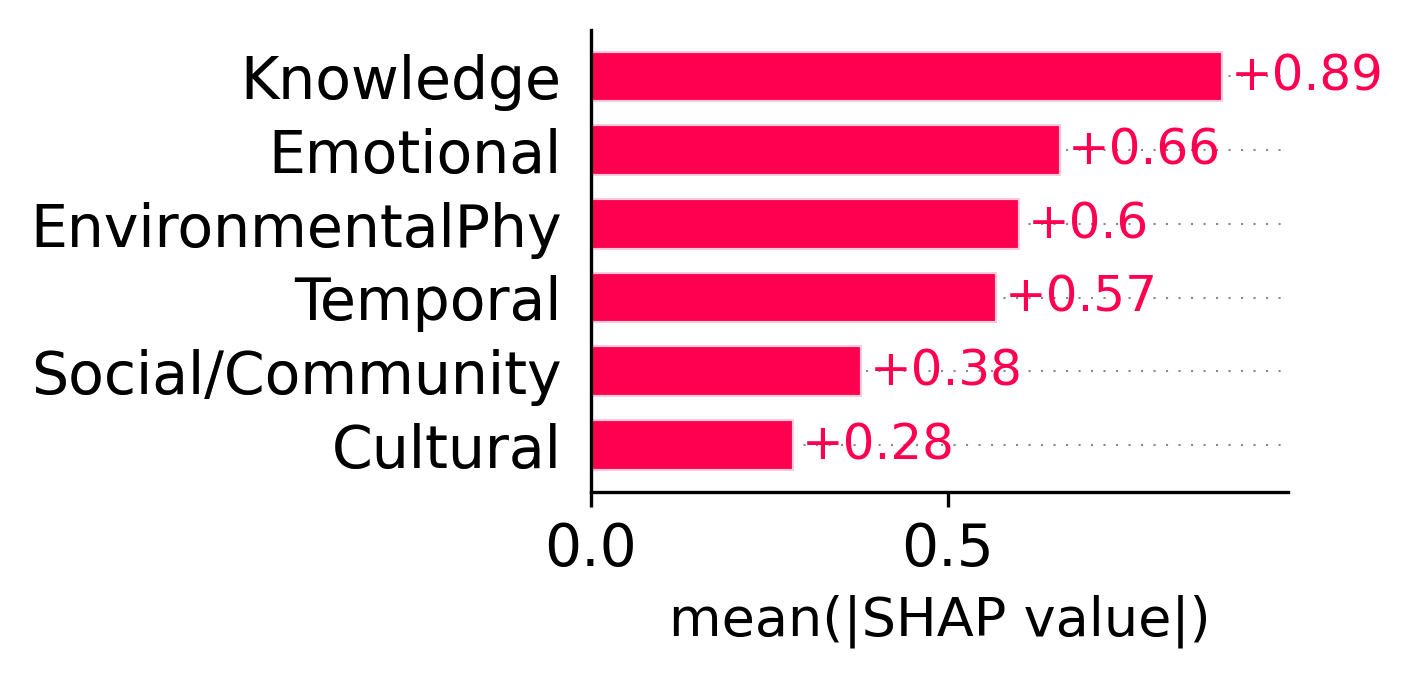}
        \caption{SHAP values aggregated by dimensions}
        \label{fig:shap_dim}
    \end{subfigure}

    \caption{SHAP analysis results showing the importance of different factors in localness prediction at three levels: (a) individual sub-components, (b) aggregated components, and (c) dimensions.
    mean(|SHAP value|) represents the average magnitude of a feature's contribution to the model's predictions across all data points. It measures how strongly a feature impacts predictions, regardless of whether the impact increases or decreases the predicted value. Higher values indicate greater importance of the feature in the model.}
    \label{fig:shap_analysis}
\end{figure*}

Our analysis of RQ4-3 uncovered how participants interpreted and synthesized information to make localness judgments, revealing key reasoning strategies that distinguished accurate assessments from inaccurate ones. Participants who made correct judgments engaged in cross-referencing multiple sources of evidence, detecting inconsistencies, and evaluating contextual appropriateness, while those who made incorrect judgments often relied on narrow definitions or single indicators, such as birthplace or self-identification.

However, while these qualitative insights reveal how participants reason about localness, they do not yet tell us which factors are most predictive of successful judgments. Were certain localness dimensions—such as cognitive knowledge, emotional attachment, or environmental familiarity—more influential in leading to accurate assessments? Did some aspects of local identity consistently mislead participants?

To answer these questions, RQ5 shifts from qualitative analysis to predictive modeling, employing machine learning to identify which conceptual elements most strongly predict judgment accuracy. This model quantifies the relative importance of identified localness components in practical classification tasks.

The XGBoost classifier demonstrated strong performance in predicting the accuracy of localness judgments. The model achieved an overall accuracy of 83\%, with balanced performance across both positive and negative classes. For accurate localness judgments, the model showed high precision (0.84) and recall (0.90), resulting in an F1-score of 0.87. For inaccurate judgments, the model maintained strong precision (0.81) with somewhat lower recall (0.72), yielding an F1-score of 0.76.The Receiver Operating Characteristic (ROC) curve shows an AUC of 0.91 (Figure~\ref{fig:roc}). This high AUC value indicates that the model effectively distinguishes between accurate and inaccurate localness judgments across different classification thresholds. The ROC curve's shape, with its sharp initial rise and early plateau, suggests that the model achieves high true positive rates while maintaining low false positive rates, indicating robust predictive performance.

We then employed SHAP analysis to elucidate how different factors contribute to model predictions of localness judgment accuracy. Figure~\ref{fig:shap_analysis} highlights the contributions at three levels of granularity.

The Knowledge dimension emerged as the most influential predictor, which underscores the significance of practical expertise and local familiarity ($mean(|SHAP~value|)=0.89$). Among its components, ``Local Recommendations'' holds the highest importance ($mean(|SHAP~value|)=0.62$). Sub-components such as ``Knowledge of best places for various needs'' ($mean(|SHAP~value|)=0.5$) and ``Familiarity with spatial organization'' ($mean(|SHAP~value|)=0.38$) are particularly impactful, signifying that the ability to navigate and recommend unique local spots distinguishes localness effectively. Additionally, ``Recognition of landmarks'' further strengthens this association, emphasizing the importance of geographic and spatial understanding in local identity ($mean(|SHAP~value|)=0.33$).

The Emotional dimension exhibited the second-highest impact, demonstrating the role of intrinsic and affective connections to a place ($mean(|SHAP~value|)=0.66$). ``Sense of Belonging'' ($mean(|SHAP~value|)=0.35$) and ``Identity Connection'' ($mean(|SHAP~value|)=0.35$) emerge as critical components, with sub-components like ``Feeling at home'' ($mean(|SHAP~value|)=0.33$) and ``Pride in local association'' ($mean(|SHAP~value|)=0.25$) showing substantial contributions. These findings highlight that emotional attachment and a deep sense of association with the community play a crucial role in influencing localness perceptions. The SHAP results also underscore the influence of ``Investment in place'' ($mean(|SHAP~value|)=0.32$) and ``Accumulated experience'' ($mean(|SHAP~value|)=0.18$), reinforcing the importance of time and effort in fostering a sense of local belonging.

The Physical-Environmental dimension reflects the significance of physical interaction with the surrounding environment ($mean(|SHAP~value|)=0.6$). Key components, such as ``Navigation/Wayfinding'' ($mean(|SHAP~value|)=0.24$) and ``Natural Environment'' ($mean(|SHAP~value|)=0.2$), showcase the importance of spatial awareness. Sub-components like ``Understanding spatial layout'' ($mean(|SHAP~value|)=0.21$) and ``Connection to local ecosystem'' ($mean(|SHAP~value|)=0.13$) rank prominently, indicating that familiarity with natural and built environments is vital for localness judgments.

The Temporal dimension, while showing lower overall impact ($mean(|SHAP~value|)=0.57$), sheds light on the longitudinal aspects of localness. ``Long-term Residence'' emerges as a key component ($mean(|SHAP~value|)=0.48$), with components like ``Formative Years'' ($mean(|SHAP~value|)=0.09$) and ``Being Born/Native'' 
($mean(|SHAP~value|)=0.08$) indicating the importance of historical ties to the area. This temporal connection is complemented by ``Change Awareness'' ($mean(|SHAP~value|)=0.23$), where an understanding of developments and changes over time adds depth to local identity.

Within the Social/Community dimension ($mean(|SHAP~value|)=0.38$), ``Personal Relationships'' 
($mean(|SHAP~value|)=0.3$) the importance of social ties. Sub-components like ``Recognition by community members'' ($mean(|SHAP~value|)=0.34$) and ``Regular social interactions'' ($mean(|SHAP~value|)=0.17$) underline that localness is not just about being present but also about being an integral part of the community's fabric. The SHAP values suggest that active involvement and visible contributions to the community are also key markers of local identity.

Finally, the Cultural dimension, while ranked lowest ($mean(|SHAP~value|)=0.28$), provided complementary predictive value through specific behavioral indicators. ``Food Culture'' ($mean(|SHAP~value|)=0.26$) and ``Language/Dialect'' ($mean(|SHAP~value|)=0.07$) stand out as influential components. Subcomponents such as ``Understanding food traditions'' ($mean(|SHAP~value|)=0.16$) and ``Knowledge of local expressions'' ($mean(|SHAP~value|)=0.06$) reveal the role of cultural practices and linguistic familiarity in distinguishing locals from non-locals.

\section{Discussion}

This study provides a structured examination of localness authentication by analyzing 932 conversation rounds from 230 participants across six experimental groups. 
Our findings reveal that participants’ conceptualization of localness is multi-dimensional, structured around three domains: Cognitive (e.g., local recommendations and historical knowledge), Physical (e.g., birthplace and long-term residence), and Relational (e.g., sense of belonging and community engagement). The analysis reveals that while both locals and nonlocals invoke cognitive and physical cues to define local identity, locals tend to emphasize active, ongoing engagement and dynamic knowledge sharing, while nonlocals rely more on static, historical indicators.

The second research question examines the accuracy with which participants can discern local versus nonlocal partners as well as distinguish human partners from LLMs. The findings indicate that local participants exhibit high accuracy when identifying local human partners; however, both local and nonlocal participants struggle to correctly classify nonlocals. Moreover, while participants are moderately successful in distinguishing human from LLM partners, the findings of the third research question consistently show that LLMs are perceived as nonlocal, largely due to their failure to exhibit deep relational and experiential nuances. These insights point to a persistent gap between the sophisticated conversational output of LLMs and the subtle, embodied markers of genuine local identity.

Furthermore, the analysis of participants’ sensemaking processes revealed that participants actively gather, filter, and reframe information during interactions, prioritizing cognitive aspects such as local recommendations alongside emotional and social cues that signify a “sense of belonging.” The results highlight that participants making correct judgments tend to integrate multiple layers of evidence, whereas those making incorrect judgments often rely on single, superficial indicators. This layered approach not only explains the observed patterns in judgment accuracy but also provides actionable insights for designing systems aimed at authenticating local presence.

Finally, predictive modeling using an XGBoost classifier (83\% accuracy, AUC 0.91) confirmed that the Knowledge and Emotional dimensions are the most influential predictors of correct localness judgments. These core findings suggest that both robust cognitive markers and affective ties are critical for accurately perceiving and authenticating local identity in interactive systems.

\subsection{Rethinking Localness with a Richer Definition}

Our study fundamentally extends traditional conceptualizations of localness by demonstrating that authentic local presence is not merely a matter of physical proximity or static geographic markers but rather a dynamic construct built on cognitive, relational, and physical dimensions. Drawing inspiration from Heidegger’s notion of dwelling \cite{heidegger1975poetry}, we found that individuals articulate localness through a rich, multi-layered framework.
This work builds upon foundational theories of space, place, and localness, which emphasize that space is not merely a physical location but is transformed into place through social relationships and emotional connections \cite{tuan1977space,relph1976place}. Our findings align with and extend these groundings, revealing that participants defined localness through nuanced expressions of local knowledge, active community engagement, and emotional bonds. This theoretical shift underscores that local identity emerges from a complex interplay of lived experiences, ranging from practical know-how (e.g., local recommendations and navigation skills) to deep-seated feelings of belonging, thereby providing a more holistic lens for examining human-place relationships in digital contexts.

Our findings further reveal that localness is fundamentally more complex and nuanced than current computational approaches suggest. In particular, rather than relying solely on conventional proxies --- such as location check-ins \cite{sanchez2022travelers}, zip codes \cite{bandy2021errors}, search logs \cite{white2012characterizing}, or self-reported profiles \cite{wu2011mining} --- participants emphasized dimensions that are rarely captured computationally: in-depth knowledge of community dynamics, emotional connections and a sense of belonging, and active social engagement. This emphasis calls for grappling with the sociotechnical nature of localness — moving beyond readily available geographic or temporal indicators toward understanding how people actively participate in and emotionally connect with their communities.

Moreover, our results reveal a meaningful divergence in how locals and nonlocals define localness. This pattern is consistent with Social Identity Theory, which suggests that people tend to define group membership in ways that highlight qualities they already possess \cite{tajfel1979integrative, hogg2016social}. In our study, locals tended to define localness through what they actively do—such as participating in community life—while nonlocals relied more on what they have, like birthplace or the number of years they’ve lived in the area. This difference reflects principles from Self-Categorization Theory, which suggests that people highlight aspects of their identity that best support their position in a given social context \cite{turner2012self, reicher2010social}. For example, nonlocals may emphasize time spent in a place because active community ties --- more common among locals --- may feel less accessible or legitimate to them. Local and nonlocal people internalize their experiences with the physical environment as part of their self-concept \cite{proshansky1983place}, thus emphasizing different criteria of localness to justify their right to belong \cite{low1992place}. More broadly, our findings extend objectivist approaches to localness often used in policy and planning, which emphasize factors like birthplace or length of stay as key indicators of local identity \cite{kariryaa2018defining}. Our data complement this perspective by highlighting how localness is also socially constructed—shaped through evolving relationships and lived experiences with place. For community development and integration efforts, this suggests a need to move beyond rigid definitions. Programs should consider both objective markers and subjective experiences of belonging that develop over time. Recognizing these multiple pathways to local identity can help bridge the divide between long-time residents and newcomers, creating more inclusive and resilient communities \cite{ mcmillan1986sense, baumeister2017need,manzo2006finding}.

In addition to advancing theoretical understanding, our work suggests significant opportunities and challenges for system design. By integrating insights from both traditional sense of place research and recent advances in location-based services \cite{almohamed2016designing,cranshaw2016journeys,sun2017movemeant}, we propose that digital platforms should evolve to capture the sociotechnical nuances of localness. 
For example, location-based services and locally aware platforms could benefit from integrating personalized local information, facilitating user-generated content that reflects local traditions and cultural practices, and enabling community engagement features that encourage both online and offline interactions. By bridging the gap between a rich theoretical understanding of localness and practical system design, our findings lay the groundwork for developing more context-aware platforms that foster authentic local participation and support newcomers in adapting to new environments.

Together, these contributions advocate for a paradigm shift—from reductionist geographic proxies to a comprehensive, multi-dimensional understanding of localness that accounts for the cognitive, relational, and physical nuances of dwelling. This enriched framework not only advances academic discourse on sense of place and digital placemaking but also provides concrete guidance for designing systems that more accurately capture and support genuine local engagement in our increasingly digital and mobile world.

\subsection{The Challenges of Human Localness Judgment—Accuracy, Biases, and Sensemaking}
Our findings reveal several fundamental challenges in how humans assess localness authenticity, highlighting systematic asymmetries, biases, and cognitive limitations in localness authentication. These insights offers several implications for HCI design, platform trust models, and hybrid human-AI collaboration in localness verification.

\subsubsection{Humans Are Better at Identifying Locals than Nonlocals}
Our results revealed that participants were significantly more accurate in identifying locals than nonlocals, suggesting that confirming authentic local presence is easier than detecting nonlocal status. This asymmetry likely arises because localness is signaled through a diverse and consistent set of cues—such as geographic familiarity, cultural awareness, and social engagement—whereas there are fewer definitive indicators of nonlocal status. As a result, people rely on affirmative signals of localness rather than trying to detect nonlocals based on their absence of local markers. 

\textbf{Practical Consideration about Identifying ``Outsiders'' in Location-Based Services: }
Efforts to identify and exclude outsiders are present in some location-based services. For example, platforms like Nextdoor implement verification systems that restrict participation to verified local residents, typically through postal addresses or phone number verification. While these mechanisms help affirm localness, they also serve as a gatekeeping tool, potentially excluding those who do not meet rigid verification criteria. However, in most cases, local verification mechanisms function by proving the presence of localness rather than actively detecting nonlocalness. This mirrors broader authentication practices in security, where verification relies on proving a positive identity (e.g., logging in with credentials) rather than proving the absence of an imposter. Similarly, localness must be actively demonstrated and socially affirmed rather than determined through exclusionary means.

\textbf{Theoretical Consideration about Localness as an Affirmative Social Identity: }
Our findings align with the notion that localness is not simply the absence of nonlocalness, but rather an actively conferred and socially recognized identity. This challenges a binary perspective in which people are either local or nonlocal by default. Instead, localness operates as an affirmative grouping that requires validation, recognition, and participation. Being local is not purely an accumulation of knowledge—such as knowing the best restaurants or historical facts about a place—but also a social and relational process. It involves belonging, engagement, and acknowledgment from the community. The inability to identify nonlocals suggests that localness is something that is bestowed rather than withheld, meaning that an individual may not be seen as local simply by virtue of lacking ties to the community. Instead, they must establish a presence, contribute meaningfully, and be recognized by others as part of the local fabric.

\textbf{Implications for Supporting People Becoming Local: } 
If localness is an affirmative status rather than a default one, this suggests new ways digital platforms and communities can support individuals in integrating into a place. First, platforms could promote engagement with existing locals by encouraging new residents to participate in community events, discussions, and local traditions. This could be facilitated through design features that connect newcomers with established community members in shared activities. Second, rather than defining localness strictly through knowledge or time spent in a place, systems could incorporate social validation mechanisms, allowing established locals to ``vouch'' for others based on their presence and contributions. This could make the authentication of localness more dynamic and inclusive. Finally, instead of relying solely on static metrics such as GPS history or address verification, localness could be inferred through patterns of social interaction, participation in community discussions, and active contributions to local knowledge-sharing platforms.

\subsubsection{Cognitive Biases in Localness Judgments}
Despite participants’ relative success in identifying local partners, our study also reveals biases in how humans assess localness authenticity. Specifically, participants tended to over-rely on single factors, such as birthplace, residency duration, or surface-level geographic knowledge. This heuristic approach led to incorrect classifications, particularly when nonlocals displayed some surface familiarity or when locals failed to explicitly signal key local identity markers.

These biases align with cognitive science research on heuristics and expertise-driven decision-making \cite{foong2017novice, venkatagiri2019groundtruth}. Experts (locals) relied on pattern recognition and multi-dimensional contextual cues, whereas non-experts (nonlocals) gravitated toward simplistic, rule-based heuristics—such as assuming long-term residency equates to deeper local knowledge. This suggests that nonlocal individuals evaluating localness may be particularly susceptible to errors when encountering surface-level knowledge without deeper contextual cues. These findings have two key implications for developing computational approaches to localness verification:  (1) our study shows that humans struggle to verify localness when relying on simple static indicators, often misjudging partial familiarity as true local expertise. Computational assessment models must incorporate multi-dimensional assessments --- as humans naturally do when evaluating local authenticity over extended interactions. (2) Our findings indicate that nonlocals can easily pass as locals if they exhibit limited but convincing knowledge of a place. This suggests that computational models may also be vulnerable to adversarial localness attempts, where individuals—or AI systems—strategically deploy superficial markers of localness to appear credible. To mitigate this, localness inference models should assess patterns of interaction rather than isolated knowledge points. For example, LLMs trained for localness evaluation should incorporate response coherence over time, testing for depth and cross-domain consistency rather than simply checking for the presence of a few well-known local references.

\subsubsection{Optimal Representation Methods for Localness Sensemaking}
Our findings align with sensemaking theory, which posits that authenticity judgments are iterative, multi-cue processes where individuals actively gather, filter, and synthesize different signals before reaching conclusions. While current computational localness assessments tend to be static, one-shot classifications, our results suggest that computational localness authenticity assessments should better model the human sensemaking process, adapting dynamically to evolving inputs. Rather than treating localness as a fixed binary attribute, computational models should reflect its fluid and context-dependent nature, incorporating adaptive learning mechanisms that allow AI to refine assessments over time based on behavioral signals, evolving community knowledge, and longitudinal engagement patterns.

Beyond accuracy and adaptability, fairness remains a critical concern in AI-based localness assessments. Prior research has demonstrated that computational models trained on user-generated content tend to be biased toward urban areas and dominant cultural narratives \cite{johnson2016not}, potentially marginalizing underrepresented communities. Ensuring equitable localness assessments requires developing methods for bias correction, fair representation of diverse community perspectives, and strategies to prevent AI-driven gatekeeping that disproportionately favors certain demographic groups. Future research should explore techniques for sourcing geographically and culturally diverse training data, as well as incorporating community-driven participatory approaches to local knowledge verification.

\subsection{Human vs. LLM Performance in Localness Authenticity Assessment}

Our analysis reveals distinct patterns in how humans and LLMs perform in localness autheticity assessment tasks, highlighting both the capabilities and limitations of current AI approaches. Most notably, we found an asymmetric pattern in human authentication: both local and nonlocal participants showed high accuracy in identifying local chat partners but struggled significantly with identifying nonlocal partners. This asymmetry suggests that markers of authentic local presence are more recognizable and consistent than indicators of nonlocal status, reinforcing the idea that localness is actively constructed through engagement rather than defined by a lack of knowledge or connection. LLMs demonstrated pronounced limitations in conveying authentic localness. While they excelled at retrieving and presenting factual local information, participants consistently identified them as nonlocal, often rating them even lower than nonlocal human partners in conveying authentic local presence. This performance gap manifested particularly in three key areas:

First, LLMs struggled with conveying emotional connections and a sense of belonging, which emerged as critical dimensions in human assessments of localness. While LLMs could accurately list local places and events, they failed to demonstrate the kind of personal investment and emotional resonance that participants associated with genuine local presence. This aligns with prior research in place attachment and sense of place, which highlights how local identity is deeply embedded in social relationships, lived experiences, and feelings of belonging \cite{tuan1977space,relph1976place,jorgensen2001sense}.

Second, participants frequently identified inconsistencies in LLMs' practical knowledge, such as suggesting unrealistic activities given local conditions (e.g., ``doing the lake loop on a rainy day'') or misunderstanding local timing patterns (e.g., incorrect market hours). These errors reveal a crucial gap between having access to information about a place and understanding how that information applies in real-world contexts. This aligns with prior work suggesting that LLMs struggle with integrating commonsense reasoning into local or cultural contexts, as their knowledge is often high-level and lacks specificity \cite{sap2019atomic, bosselut2019comet, petroni2019language}.

Third, LLMs struggled to convey experiential depth, particularly in surfacing ``hidden gems'' or localized knowledge that locals often take for granted. Participants noted that LLM responses often felt like ``researched'' knowledge rather than lived experience, lacking the personal anecdotes and deep contextual awareness that characterized authentic local discourse. This aligns with findings from human sensemaking research, where localness assessments rely heavily on cumulative, interactive, and dynamically evolving knowledge processes rather than isolated factual retrieval \cite{pirolli2005sensemaking}.

Existing work on geo-contextual LLM knowledge \cite{acharya2020towards,yin2022geomlama} suggests that these limitations stem from the lack of fine-grained local data in LLM training corpora. While LLMs encode vast amounts of general world knowledge, they lack access to high-resolution, context-specific, and culturally embedded local information. Our findings indicate that effectively supporting localness in LLMs necessitates more than just expanding factual knowledge—it requires incorporating social, contextual, and experiential dimensions that define true local expertise.

These findings have important implications for the design of more locally-aware AI systems. Rather than focusing solely on expanding factual databases, future LLMs should integrate mechanisms to validate the contextual appropriateness of suggestions, develop models for grounding AI-generated responses in real-world local experiences, and incorporate broader indicators of local authenticity beyond factual accuracy. Furthermore, the clear detectability of LLM-generated content suggests that hybrid systems—where LLMs perform initial information retrieval while human local experts validate and enrich the data—could be a promising avenue for more context-aware and socially embedded AI applications. Future work should explore how to effectively combine the complementary strengths of AI-driven information retrieval with human expertise in relational and experiential knowledge while maintaining authentic local voice.

\subsection{LLMs as an Avenue Of Localness Representation}

Given the limitations identified in LLM performance, a key question emerges: how can AI systems better represent localness? To address this gap, we propose two critical areas of focus: (1) enhancing high-resolution local data collection and (2) developing formalized evaluation benchmarks for localness in LLMs.

\subsubsection{Expanding Localness Representation Through Better Data Collection}

Our results suggest that existing LLMs rely on large-scale internet data, which disproportionately represents urban centers and lacks the nuanced, community-driven perspectives that characterize true localness. Studies in location-based social networks and geo-contextual AI suggest that local knowledge is often underrepresented in digital platforms, particularly in rural or lower-income regions \cite{wu2011mining, bandy2021errors, hardy2019design}. To improve localness representation, we propose the development of comprehensive localness datasets designed to capture the multifaceted aspects of localness:

First, data collection methods should focus on gathering geo-social content from various local sources. This could involve using social media platforms, local forums, and community events to capture real-time, authentic local content. Ensuring the data includes various dimensions of localness ---such as historical knowledge, cultural practices, and local landmarks --- will provide a richer dataset for training LLMs. 
Second, similar to prior work in crowdsourced knowledge systems (e.g., Wikipedia, OpenStreetMap) \cite{jenkins2016crowdsourcing}, new participatory frameworks should enable local residents to contribute insights, experiences, and cultural narratives to help ensure the data is representative and comprehensive. Of course, this may be very difficult --- for instance, some have argued that current peer production modalities cannot support rural communities \cite{hardy2019design}, which means that it may be necessary to develop novel alternative approaches to data production. 
Third, labeling local information at varying levels of specificity (e.g., widely known landmarks vs. niche community spaces) will improve AI’s ability to distinguish between general vs. hyperlocal knowledge.

However, prior research has identified structural challenges in digital participation—for instance, crowdsourced knowledge production disproportionately benefits urban, high-income communities, while rural or marginalized groups often lack representation \cite{hardy2019design, johnson2016not}. Developing alternative participatory mechanisms—such as incentivized local knowledge sharing, direct collaborations with community organizations, or integrating AI-assisted knowledge validation—may be necessary to address these disparities and ensure equitable representation.

\subsubsection{Establishing Localness Evaluation Benchmarks for LLMs}

A second major challenge is the lack of standardized benchmarks for evaluating LLM performance in localness tasks. Current LLM benchmarks emphasize general knowledge retrieval, linguistic coherence, and factual correctness, but they do not assess AI’s ability to capture localized and culturally specific knowledge \cite{yin2022geomlama}. Our findings underscore the need for new evaluation metrics that reflect the core dimensions of localness, including: (1) \textit{geographical specificity} (does the LLM demonstrate deep, granular knowledge about local spaces and cultural practices?) (2) \textit{contextual appropriateness} (can the model provide locally relevant responses without inconsistencies or factual errors?) (3) \textit{relational and experiential depth} (does the AI’s output capture the emotional, social, and lived experience aspects of localness?). We explored these metrics further in Subsection~\ref{sec:metrics}

While benchmarking localness in LLMs is a non-trivial challenge, prior work in commonsense reasoning benchmarks (e.g., ATOMIC, COMET) provides useful methodologies \cite{sap2019atomic, bosselut2019comet}. By adapting these techniques—e.g., using human-in-the-loop evaluations, structured verification pipelines, and region-specific task datasets—we can build more robust assessment frameworks for locally-aware AI.

Beyond benchmarking, structural biases in AI training data must also be addressed. LLMs are, generally, trained on user-generated content scraped from the internet. Prior work shows that user-generated content disproportionately reflects non-local perspectives—for example, many travel and restaurant reviews are written by tourists rather than residents \cite{johnson2016not,thebault2018distance,sen2015turkers}, and user-generated content is biased towards urban areas \cite{thebault2018distance, thebault2018geographic, johnson2016not}. Thus, achieving effective localness representation will require rethinking how AI training data is collected, curated, and weighted. Future research in HCI, AI ethics, and computational social science must explore novel approaches to ensuring parity in whose perspectives are reflected in AI-generated local knowledge.

\subsection{Computational Localness Authenticity Assessment}
\label{sec:metrics}

Our findings reveal that localness is a multi-dimensional construct, encompassing cognitive, physical, and relational aspects. Human judgments of localness rely on a nuanced combination of these dimensions, whereas traditional computational approaches primarily depend on spatial and temporal signals. 

\subsubsection{Beyond Spatial-Temporal Metrics in Localness Detection}
Existing computational approaches focus on geospatial data to determine localness. \texttt{nDays} and \texttt{Plurality} classify users based on the temporal spread and concentration of location-based activity, while \texttt{LocationField} and \texttt{GeometricMedian} rely on explicit user-provided locations and median geographic positioning \cite{kariryaa2018defining}. While effective in detecting physical presence, these methods fail to capture how humans assess local authenticity.

Our sensemaking analysis (RQ4) demonstrated that participants relied on a diverse range of indicators beyond spatial-temporal presence, such as deep local knowledge, emotional belonging, and community engagement. Furthermore, our SHAP analysis (RQ5) highlighted that factors such as \texttt{Local Recommendations, Sense of Belonging}, and \texttt{Personal Relationships} were the strongest predictors of accurate localness judgments. This suggests that computational models should incorporate cognitive and relational metrics alongside spatial-temporal metrics.

\subsubsection{Knowledge-Based Authentication Enhancement}

Our findings indicate that knowledge-based authentication can significantly improve localness assessment. The SHAP analysis showed that the \texttt{Knowledge} dimension has a strong predictive value, underscoring the importance of detailed local expertise in distinguishing genuine locals from transient visitors. Current spatial-temporal models could be enhanced by analyzing not just the frequency of location mentions but also their depth and contextual relevance.

One approach is to develop a \texttt{Local Knowledge Index}, which evaluates the specificity, diversity, and appropriateness of location-based references in user-generated content. Computational methods such as NLP can assess whether users refer to commonly known places versus insider landmarks, historical sites, or seasonal events. The high importance of ``Local Recommendations'' suggests that authentication systems should evaluate not just the quantity but also the contextual appropriateness of location references. This could include analyzing whether users demonstrate awareness of local changes over time, seasonal accessibility of places, or evolving cultural norms. 

Furthermore, \texttt{GeometricMedian} could be expanded beyond a pure spatial aggregation metric to assess users' ability to articulate and contextualize spatial relationships. Features such as ``Geographic Familiarity'' indicate that authentic locals possess a nuanced understanding of their surroundings, including neighborhood boundaries, travel patterns, and ecological landmarks. Computational assessment of navigation descriptions and mental maps could enhance existing geospatial models, allowing authentication systems to differentiate between true locals and external users with surface-level geographic familiarity.

\subsubsection{Relational Authentication Integration}

The second major area for enhancement involves relational authentication, supported by the high predictive value of the \texttt{Emotional} dimension. Traditional computational methods overlook the importance of social integration in establishing local identity. Our results suggest that relational indicators, such as sustained interactions with local organizations and embeddedness in community networks, are crucial for localness authentication.

A \texttt{Relational Connectivity Score} could be developed using social network analysis to assess the depth and consistency of a user's local interactions. This score would measure not just the number of connections but also the nature and longevity of those relationships. ``Personal Relationships'' emerged as a strong predictor in our analysis, indicating that authentic locals maintain sustained and meaningful engagements with others in their community. Computational models could incorporate interaction frequency, diversity of social ties, and participation in community discussions to refine relational authentication. 

\subsubsection{Enhanced Temporal-Physical Assessment}

Temporal metrics, while foundational to existing approaches, require refinement to align with human authentication strategies. The traditional \texttt{nDays} approach determines localness based on repeated presence over time, but our results suggest that authentic locals also demonstrate deep awareness of temporal patterns unique to their community. ``Change Awareness'' ($mean(|SHAP~value|) = 0.23$) significantly contributes to accurate localness assessments, highlighting the need for more nuanced temporal analysis.

A refined \texttt{Temporal Knowledge Metric} could assess whether users demonstrate awareness of local seasonal variations, historical transformations, and community events. NLP models could analyze content for references to past and upcoming events, distinguishing between users who merely frequent a location and those who have internalized its temporal rhythms. This approach would allow authentication systems to integrate temporal context beyond mere presence detection, improving differentiation between residents and recurrent visitors.

\subsubsection{Implementation Framework Considerations}

Implementing these enhanced metrics requires a multi-modal authentication approach that integrates spatial, cognitive, and relational signals in a weighted scoring model. Rather than treating each metric independently, systems should adopt a probabilistic model that considers how different localness indicators interact. Natural language processing can assess the depth of local knowledge, while machine learning models can analyze network engagement patterns. 

Pattern matching techniques could be employed to validate user-provided information against verified local content sources. For example, if a user claims familiarity with a specific neighborhood or local tradition, computational systems could cross-reference this against historical data, social media discussions, and geotagged community events. The high predictive value of integrated knowledge markers in our SHAP analysis suggests that authentic localness manifests through the consistent demonstration of multiple types of local understanding.

A key challenge in developing these enhanced metrics is ensuring adaptability across different geographic and cultural contexts. Our findings on the differences between local and nonlocal conceptualizations of localness suggest that validation approaches must be flexible enough to accommodate diverse expressions of local identity. Dynamic assessment models that adjust to regional variations while maintaining robust assessment standards are necessary for practical deployment.

Moreover, the increasing sophistication of language models raises concerns about their ability to simulate local knowledge convincingly. Our human-LLM interaction analysis revealed that modern LLMs can generate well-researched but ultimately inauthentic representations of local identity. This underscores the need for ongoing refinement of metrics to distinguish between learned knowledge and genuine lived experience. Future authentication systems must be increasingly nuanced in detecting subtle inconsistencies in local knowledge depth, emotional attachment, and community integration.

\subsection{Limitations and Future Directions}
While our study provides insights into human localness assessment and its implications for AI, several limitations point to future research directions. First, our reliance on conversation-based assessment may not fully capture behavioral, spatial, and community-based indicators of localness. Future work should explore multi-modal approaches incorporating geospatial activity, social network participation, and real-world engagement to develop more holistic localness models. Second, our participant recruitment from a single university community in the upper Midwest limits generalizability. Our findings may be context-dependent, as perceptions of localness vary across urban vs. rural settings, cultural backgrounds, and mobility patterns. Expanding research across diverse geographic and demographic groups will help refine AI models for broader applicability. Third, our study offers a static snapshot of localness judgments, whereas local identity evolves over time. Longitudinal research should investigate how people develop local ties and how AI can track dynamic engagement patterns, moving beyond rigid classification toward adaptive, time-sensitive models.

Our findings also highlight AI limitations in local knowledge representation. LLMs struggle with contextual adaptation, emotional connection, and experiential depth, often providing factually correct but impersonal responses. Future work should explore training AI on richer, locally sourced data and developing evaluation benchmarks to systematically assess AI’s ability to convey authentic localness.

Bias mitigation remains a challenge, as current computational approaches tend to overrepresent well-documented urban areas while marginalizing rural and underrepresented communities. Future research should develop fairer data collection methods to ensure inclusivity in AI-driven localness verification.

Finally, explainability in AI localness assessments is crucial. Participants often relied on implicit heuristics they struggled to articulate, suggesting that AI should provide interpretable justifications for its assessments. Future systems could incorporate human-in-the-loop verification, allowing communities to refine and validate AI-generated judgments for greater transparency and trust.
\section{conclusion}

This study highlights the fundamental role of localness in shaping meaningful interactions between individuals and their communities. As digital platforms increasingly mediate our engagement with place, understanding how people conceptualize and recognize localness is essential for improving location-based services, community-driven information systems, and AI-driven local content generation.

Our study designed an experimental chat system inspired by Turing’s Imitation Game and Von Ahn’s Games With A Purpose \cite{von2006games}, and sensemaking theory \cite{pirolli2005sensemaking}, to investigate localness assessment. This system facilitate structured conversational interactions between locals, non-locals, and LLMs. Through this framework, we identified the multi-dimensional nature of localness, revealing how different groups emphasize distinct aspects of local identity. Our findings demonstrate that both locals and non-locals effectively differentiate local and non-local content, underscoring the cognitive and social mechanisms people use to assess authenticity in digital environments.

Despite the growing sophistication of LLMs, our results indicate that these models struggle to convey localness at a level comparable to humans. The primary limitations stem from their insufficient integration of deep local knowledge, an inability to capture lived experiences, and a lack of social and cultural awareness that underpins authentic local engagement. This gap suggests that current AI-driven location-based services may struggle to generate truly localized, contextually rich content that resonates with real community dynamics.

To address these shortcomings, we advocate for the development of computational localness authenticity assessment methods that integrate high-resolution local data into LLM training processes. Future work should focus on curating comprehensive datasets that capture not only factual local knowledge but also social interactions, community involvement, and cultural nuances. By enhancing the ability of AI systems to dwell --- to understand and engage with local identity beyond mere geographic presence --- we can design digital tools that foster deeper, more inclusive, and socially embedded experiences in location-based interactions.

\begin{acks}
\end{acks}

\bibliographystyle{ACM-Reference-Format}

\begin{thebibliography}{106}


\ifx \showCODEN    \undefined \def \showCODEN     #1{\unskip}     \fi
\ifx \showDOI      \undefined \def \showDOI       #1{#1}\fi
\ifx \showISBNx    \undefined \def \showISBNx     #1{\unskip}     \fi
\ifx \showISBNxiii \undefined \def \showISBNxiii  #1{\unskip}     \fi
\ifx \showISSN     \undefined \def \showISSN      #1{\unskip}     \fi
\ifx \showLCCN     \undefined \def \showLCCN      #1{\unskip}     \fi
\ifx \shownote     \undefined \def \shownote      #1{#1}          \fi
\ifx \showarticletitle \undefined \def \showarticletitle #1{#1}   \fi
\ifx \showURL      \undefined \def \showURL       {\relax}        \fi
\providecommand\bibfield[2]{#2}
\providecommand\bibinfo[2]{#2}
\providecommand\natexlab[1]{#1}
\providecommand\showeprint[2][]{arXiv:#2}

\bibitem[Acharya et~al\mbox{.}(2020)]%
        {acharya2020towards}
\bibfield{author}{\bibinfo{person}{Anurag Acharya}, \bibinfo{person}{Kartik Talamadupula}, {and} \bibinfo{person}{Mark~A Finlayson}.} \bibinfo{year}{2020}\natexlab{}.
\newblock \showarticletitle{Towards an atlas of cultural commonsense for machine reasoning}.
\newblock \bibinfo{journal}{\emph{arXiv preprint arXiv:2009.05664}} (\bibinfo{year}{2020}).
\newblock


\bibitem[Aiello et~al\mbox{.}(2016)]%
        {aiello2016chatty}
\bibfield{author}{\bibinfo{person}{Luca~Maria Aiello}, \bibinfo{person}{Rossano Schifanella}, \bibinfo{person}{Daniele Quercia}, {and} \bibinfo{person}{Francesco Aletta}.} \bibinfo{year}{2016}\natexlab{}.
\newblock \showarticletitle{Chatty maps: constructing sound maps of urban areas from social media data}.
\newblock \bibinfo{journal}{\emph{Royal Society open science}} \bibinfo{volume}{3}, \bibinfo{number}{3} (\bibinfo{year}{2016}), \bibinfo{pages}{150690}.
\newblock


\bibitem[Almohamed and Vyas(2016)]%
        {almohamed2016designing}
\bibfield{author}{\bibinfo{person}{Asam Almohamed} {and} \bibinfo{person}{Dhaval Vyas}.} \bibinfo{year}{2016}\natexlab{}.
\newblock \showarticletitle{Designing for the Marginalized: A step towards understanding the lives of refugees and asylum seekers}. In \bibinfo{booktitle}{\emph{Proceedings of the 2016 acm conference companion publication on designing interactive systems}}. \bibinfo{pages}{165--168}.
\newblock


\bibitem[Alvarado~Garcia et~al\mbox{.}(2020)]%
        {alvarado2020fostering}
\bibfield{author}{\bibinfo{person}{Adriana Alvarado~Garcia}, \bibinfo{person}{Karla Badillo-Urquiola}, \bibinfo{person}{Mayra~D Barrera~Machuca}, \bibinfo{person}{Franceli~L Cibrian}, \bibinfo{person}{Marianela Ciolfi~Felice}, \bibinfo{person}{Laura~S Gayt{\'a}n-Lugo}, \bibinfo{person}{Diego G{\'o}mez-Zar{\'a}}, \bibinfo{person}{Carla~F Griggio}, \bibinfo{person}{Monica Perusquia-Hernandez}, \bibinfo{person}{Soraia Silva-Prietch}, {et~al\mbox{.}}} \bibinfo{year}{2020}\natexlab{}.
\newblock \showarticletitle{Fostering HCI Research in, by, and for Latin America}. In \bibinfo{booktitle}{\emph{Extended Abstracts of the 2020 CHI Conference on Human Factors in Computing Systems}}. \bibinfo{pages}{1--4}.
\newblock


\bibitem[Balestrini et~al\mbox{.}(2017)]%
        {balestrini2017city}
\bibfield{author}{\bibinfo{person}{Mara Balestrini}, \bibinfo{person}{Yvonne Rogers}, \bibinfo{person}{Carolyn Hassan}, \bibinfo{person}{Javi Creus}, \bibinfo{person}{Martha King}, {and} \bibinfo{person}{Paul Marshall}.} \bibinfo{year}{2017}\natexlab{}.
\newblock \showarticletitle{A city in common: a framework to orchestrate large-scale citizen engagement around urban issues}. In \bibinfo{booktitle}{\emph{Proceedings of the 2017 CHI conference on human factors in computing systems}}. \bibinfo{pages}{2282--2294}.
\newblock


\bibitem[Bandy and Hecht(2021)]%
        {bandy2021errors}
\bibfield{author}{\bibinfo{person}{Jack Bandy} {and} \bibinfo{person}{Brent Hecht}.} \bibinfo{year}{2021}\natexlab{}.
\newblock \showarticletitle{Errors in Geotargeted Display Advertising: Good News for Local Journalism?}
\newblock \bibinfo{journal}{\emph{Proceedings of the ACM on Human-Computer Interaction}} \bibinfo{volume}{5}, \bibinfo{number}{CSCW1} (\bibinfo{year}{2021}), \bibinfo{pages}{1--19}.
\newblock


\bibitem[Baumeister and Leary(2017)]%
        {baumeister2017need}
\bibfield{author}{\bibinfo{person}{Roy~F Baumeister} {and} \bibinfo{person}{Mark~R Leary}.} \bibinfo{year}{2017}\natexlab{}.
\newblock \showarticletitle{The need to belong: Desire for interpersonal attachments as a fundamental human motivation}.
\newblock \bibinfo{journal}{\emph{Interpersonal development}} (\bibinfo{year}{2017}), \bibinfo{pages}{57--89}.
\newblock


\bibitem[Bender et~al\mbox{.}(2021)]%
        {bender2021dangers}
\bibfield{author}{\bibinfo{person}{Emily~M Bender}, \bibinfo{person}{Timnit Gebru}, \bibinfo{person}{Angelina McMillan-Major}, {and} \bibinfo{person}{Shmargaret Shmitchell}.} \bibinfo{year}{2021}\natexlab{}.
\newblock \showarticletitle{On the dangers of stochastic parrots: Can language models be too big?}. In \bibinfo{booktitle}{\emph{Proceedings of the 2021 ACM conference on fairness, accountability, and transparency}}. \bibinfo{pages}{610--623}.
\newblock


\bibitem[Bosselut et~al\mbox{.}(2019)]%
        {bosselut2019comet}
\bibfield{author}{\bibinfo{person}{Antoine Bosselut}, \bibinfo{person}{Hannah Rashkin}, \bibinfo{person}{Maarten Sap}, \bibinfo{person}{Chaitanya Malaviya}, \bibinfo{person}{Asli Celikyilmaz}, {and} \bibinfo{person}{Yejin Choi}.} \bibinfo{year}{2019}\natexlab{}.
\newblock \showarticletitle{COMET: Commonsense transformers for automatic knowledge graph construction}.
\newblock \bibinfo{journal}{\emph{arXiv preprint arXiv:1906.05317}} (\bibinfo{year}{2019}).
\newblock


\bibitem[Bradel et~al\mbox{.}(2014)]%
        {bradel2014multi}
\bibfield{author}{\bibinfo{person}{Lauren Bradel}, \bibinfo{person}{Chris North}, {and} \bibinfo{person}{Leanna House}.} \bibinfo{year}{2014}\natexlab{}.
\newblock \showarticletitle{Multi-model semantic interaction for text analytics}. In \bibinfo{booktitle}{\emph{2014 IEEE Conference on Visual Analytics Science and Technology (VAST)}}. IEEE, \bibinfo{pages}{163--172}.
\newblock


\bibitem[Chang et~al\mbox{.}(2023)]%
        {chang2023citesee}
\bibfield{author}{\bibinfo{person}{Joseph~Chee Chang}, \bibinfo{person}{Amy~X Zhang}, \bibinfo{person}{Jonathan Bragg}, \bibinfo{person}{Andrew Head}, \bibinfo{person}{Kyle Lo}, \bibinfo{person}{Doug Downey}, {and} \bibinfo{person}{Daniel~S Weld}.} \bibinfo{year}{2023}\natexlab{}.
\newblock \showarticletitle{Citesee: Augmenting citations in scientific papers with persistent and personalized historical context}. In \bibinfo{booktitle}{\emph{Proceedings of the 2023 CHI Conference on Human Factors in Computing Systems}}. \bibinfo{pages}{1--15}.
\newblock


\bibitem[Choi et~al\mbox{.}(2018)]%
        {choi2018quac}
\bibfield{author}{\bibinfo{person}{Eunsol Choi}, \bibinfo{person}{He He}, \bibinfo{person}{Mohit Iyyer}, \bibinfo{person}{Mark Yatskar}, \bibinfo{person}{Wen-tau Yih}, \bibinfo{person}{Yejin Choi}, \bibinfo{person}{Percy Liang}, {and} \bibinfo{person}{Luke Zettlemoyer}.} \bibinfo{year}{2018}\natexlab{}.
\newblock \showarticletitle{QuAC: Question answering in context}.
\newblock \bibinfo{journal}{\emph{arXiv preprint arXiv:1808.07036}} (\bibinfo{year}{2018}).
\newblock


\bibitem[Colley et~al\mbox{.}(2017)]%
        {colley2017geography}
\bibfield{author}{\bibinfo{person}{Ashley Colley}, \bibinfo{person}{Jacob Thebault-Spieker}, \bibinfo{person}{Allen~Yilun Lin}, \bibinfo{person}{Donald Degraen}, \bibinfo{person}{Benjamin Fischman}, \bibinfo{person}{Jonna H{\"a}kkil{\"a}}, \bibinfo{person}{Kate Kuehl}, \bibinfo{person}{Valentina Nisi}, \bibinfo{person}{Nuno~Jardim Nunes}, \bibinfo{person}{Nina Wenig}, {et~al\mbox{.}}} \bibinfo{year}{2017}\natexlab{}.
\newblock \showarticletitle{The geography of Pok{\'e}mon GO: beneficial and problematic effects on places and movement}. In \bibinfo{booktitle}{\emph{Proceedings of the 2017 CHI conference on human factors in computing systems}}. \bibinfo{pages}{1179--1192}.
\newblock


\bibitem[Cranshaw et~al\mbox{.}(2016)]%
        {cranshaw2016journeys}
\bibfield{author}{\bibinfo{person}{Justin Cranshaw}, \bibinfo{person}{Andr{\'e}s Monroy-Hern{\'a}ndez}, {and} \bibinfo{person}{SA Needham}.} \bibinfo{year}{2016}\natexlab{}.
\newblock \showarticletitle{Journeys \& notes: Designing social computing for non-places}. In \bibinfo{booktitle}{\emph{Proceedings of the 2016 CHI Conference on Human Factors in Computing Systems}}. \bibinfo{pages}{4722--4733}.
\newblock


\bibitem[Cranshaw et~al\mbox{.}(2012)]%
        {cranshaw2012livehoods}
\bibfield{author}{\bibinfo{person}{Justin Cranshaw}, \bibinfo{person}{Raz Schwartz}, \bibinfo{person}{Jason Hong}, {and} \bibinfo{person}{Norman Sadeh}.} \bibinfo{year}{2012}\natexlab{}.
\newblock \showarticletitle{The livehoods project: Utilizing social media to understand the dynamics of a city}. In \bibinfo{booktitle}{\emph{Proceedings of the international AAAI conference on web and social media}}, Vol.~\bibinfo{volume}{6}. \bibinfo{pages}{58--65}.
\newblock


\bibitem[Dervin(1983)]%
        {dervin1983overview}
\bibfield{author}{\bibinfo{person}{Brenda Dervin}.} \bibinfo{year}{1983}\natexlab{}.
\newblock \showarticletitle{AN OVERVIEW OF SENSE-MAKING RESEARCH: CONCEPTS, METHODS AND RESULTS TO DATE}. In \bibinfo{booktitle}{\emph{the Annual Meeting of the International Communication Association}}.
\newblock


\bibitem[Dervin(1992)]%
        {dervin1992mind}
\bibfield{author}{\bibinfo{person}{Brenda Dervin}.} \bibinfo{year}{1992}\natexlab{}.
\newblock \showarticletitle{From the mind’s eye of the user: The sense-making qualitative-quantitative methodology}.
\newblock \bibinfo{journal}{\emph{Qualitative research in information management}} \bibinfo{volume}{9}, \bibinfo{number}{1} (\bibinfo{year}{1992}), \bibinfo{pages}{61--84}.
\newblock


\bibitem[Deshpande et~al\mbox{.}(2005)]%
        {deshpande2005building}
\bibfield{author}{\bibinfo{person}{Nishchal Deshpande}, \bibinfo{person}{Bauke de Vries}, {and} \bibinfo{person}{Jos~P van Leeuwen}.} \bibinfo{year}{2005}\natexlab{}.
\newblock \showarticletitle{Building and supporting shared understanding in collaborative problem-solving}. In \bibinfo{booktitle}{\emph{Ninth International Conference on Information Visualisation (IV'05)}}. IEEE, \bibinfo{pages}{737--742}.
\newblock


\bibitem[Ei{\ss}feldt(2019)]%
        {eissfeldt2019supporting}
\bibfield{author}{\bibinfo{person}{Hinnerk Ei{\ss}feldt}.} \bibinfo{year}{2019}\natexlab{}.
\newblock \showarticletitle{Supporting urban air mobility with citizen participatory noise sensing: A concept}. In \bibinfo{booktitle}{\emph{Companion Proceedings of The 2019 World Wide Web Conference}}. \bibinfo{pages}{93--95}.
\newblock


\bibitem[Ference et~al\mbox{.}(2013)]%
        {ference2013location}
\bibfield{author}{\bibinfo{person}{Gregory Ference}, \bibinfo{person}{Mao Ye}, {and} \bibinfo{person}{Wang-Chien Lee}.} \bibinfo{year}{2013}\natexlab{}.
\newblock \showarticletitle{Location recommendation for out-of-town users in location-based social networks}. In \bibinfo{booktitle}{\emph{Proceedings of the 22nd ACM international conference on Information \& Knowledge Management}}. \bibinfo{pages}{721--726}.
\newblock


\bibitem[Foong et~al\mbox{.}(2017)]%
        {foong2017novice}
\bibfield{author}{\bibinfo{person}{Eureka Foong}, \bibinfo{person}{Darren Gergle}, {and} \bibinfo{person}{Elizabeth~M Gerber}.} \bibinfo{year}{2017}\natexlab{}.
\newblock \showarticletitle{Novice and expert sensemaking of crowdsourced design feedback}.
\newblock \bibinfo{journal}{\emph{Proceedings of the ACM on Human-Computer Interaction}} \bibinfo{volume}{1}, \bibinfo{number}{CSCW} (\bibinfo{year}{2017}), \bibinfo{pages}{1--18}.
\newblock


\bibitem[Gao et~al\mbox{.}(2024)]%
        {gao2024journeying}
\bibfield{author}{\bibinfo{person}{Zihan Gao}, \bibinfo{person}{Justin Cranshaw}, {and} \bibinfo{person}{Jacob Thebault-Spieker}.} \bibinfo{year}{2024}\natexlab{}.
\newblock \showarticletitle{Journeying Through Sense of Place with Mental Maps: Characterizing Changing Spatial Understanding and Sense of Place During Migration for Work}.
\newblock \bibinfo{journal}{\emph{Proceedings of the ACM on Human-Computer Interaction}} \bibinfo{volume}{8}, \bibinfo{number}{CSCW2} (\bibinfo{year}{2024}), \bibinfo{pages}{1--31}.
\newblock


\bibitem[Garbett et~al\mbox{.}(2016)]%
        {garbett2016app}
\bibfield{author}{\bibinfo{person}{Andrew Garbett}, \bibinfo{person}{Rob Comber}, \bibinfo{person}{Edward Jenkins}, {and} \bibinfo{person}{Patrick Olivier}.} \bibinfo{year}{2016}\natexlab{}.
\newblock \showarticletitle{App movement: A platform for community commissioning of mobile applications}. In \bibinfo{booktitle}{\emph{Proceedings of the 2016 CHI conference on human factors in computing systems}}. \bibinfo{pages}{26--37}.
\newblock


\bibitem[Goodchild(2007)]%
        {goodchild2007citizens}
\bibfield{author}{\bibinfo{person}{Michael~F Goodchild}.} \bibinfo{year}{2007}\natexlab{}.
\newblock \showarticletitle{Citizens as sensors: the world of volunteered geography}.
\newblock \bibinfo{journal}{\emph{GeoJournal}}  \bibinfo{volume}{69} (\bibinfo{year}{2007}), \bibinfo{pages}{211--221}.
\newblock


\bibitem[Goyal et~al\mbox{.}(2013a)]%
        {goyal2013effects}
\bibfield{author}{\bibinfo{person}{Nitesh Goyal}, \bibinfo{person}{Gilly Leshed}, {and} \bibinfo{person}{Susan~R Fussell}.} \bibinfo{year}{2013}\natexlab{a}.
\newblock \showarticletitle{Effects of Visualization and Note-taking on Sensemaking and Analysis}. In \bibinfo{booktitle}{\emph{Proceedings of the SIGCHI Conference on Human Factors in Computing Systems}}. \bibinfo{pages}{2721--2724}.
\newblock


\bibitem[Goyal et~al\mbox{.}(2013b)]%
        {goyal2013leveraging}
\bibfield{author}{\bibinfo{person}{Nitesh Goyal}, \bibinfo{person}{Gilly Leshed}, {and} \bibinfo{person}{Susan~R Fussell}.} \bibinfo{year}{2013}\natexlab{b}.
\newblock \showarticletitle{Leveraging partner's insights for distributed collaborative sensemaking}. In \bibinfo{booktitle}{\emph{Proceedings of the 2013 conference on Computer supported cooperative work companion}}. \bibinfo{pages}{15--18}.
\newblock


\bibitem[Graham and Dittus(2022)]%
        {graham2022geographies}
\bibfield{author}{\bibinfo{person}{Mark Graham} {and} \bibinfo{person}{Martin Dittus}.} \bibinfo{year}{2022}\natexlab{}.
\newblock \showarticletitle{Geographies of digital exclusion: Data and inequality}.
\newblock \bibinfo{journal}{\emph{(No Title)}} (\bibinfo{year}{2022}).
\newblock


\bibitem[Hardy(2019)]%
        {hardy2019design}
\bibfield{author}{\bibinfo{person}{Jean Hardy}.} \bibinfo{year}{2019}\natexlab{}.
\newblock \showarticletitle{How the design of social technology fails rural America}. In \bibinfo{booktitle}{\emph{Companion Publication of the 2019 on Designing Interactive Systems Conference 2019 Companion}}. \bibinfo{pages}{189--193}.
\newblock


\bibitem[Harrison and Dourish(1996)]%
        {harrison1996re}
\bibfield{author}{\bibinfo{person}{Steve Harrison} {and} \bibinfo{person}{Paul Dourish}.} \bibinfo{year}{1996}\natexlab{}.
\newblock \showarticletitle{Re-place-ing space: the roles of place and space in collaborative systems}. In \bibinfo{booktitle}{\emph{Proceedings of the 1996 ACM conference on Computer supported cooperative work}}. \bibinfo{pages}{67--76}.
\newblock


\bibitem[Hecht and Gergle(2010a)]%
        {hecht2010tower}
\bibfield{author}{\bibinfo{person}{Brent Hecht} {and} \bibinfo{person}{Darren Gergle}.} \bibinfo{year}{2010}\natexlab{a}.
\newblock \showarticletitle{The tower of Babel meets web 2.0: user-generated content and its applications in a multilingual context}. In \bibinfo{booktitle}{\emph{Proceedings of the SIGCHI conference on human factors in computing systems}}. \bibinfo{pages}{291--300}.
\newblock


\bibitem[Hecht et~al\mbox{.}(2011)]%
        {hecht2011tweets}
\bibfield{author}{\bibinfo{person}{Brent Hecht}, \bibinfo{person}{Lichan Hong}, \bibinfo{person}{Bongwon Suh}, {and} \bibinfo{person}{Ed~H Chi}.} \bibinfo{year}{2011}\natexlab{}.
\newblock \showarticletitle{Tweets from Justin Bieber's heart: the dynamics of the location field in user profiles}. In \bibinfo{booktitle}{\emph{Proceedings of the SIGCHI conference on human factors in computing systems}}. \bibinfo{pages}{237--246}.
\newblock


\bibitem[Hecht and Gergle(2010b)]%
        {hecht2010localness}
\bibfield{author}{\bibinfo{person}{Brent~J Hecht} {and} \bibinfo{person}{Darren Gergle}.} \bibinfo{year}{2010}\natexlab{b}.
\newblock \showarticletitle{On the" localness" of user-generated content}. In \bibinfo{booktitle}{\emph{Proceedings of the 2010 ACM conference on Computer supported cooperative work}}. \bibinfo{pages}{229--232}.
\newblock


\bibitem[Heidegger et~al\mbox{.}(1971)]%
        {heidegger1971building}
\bibfield{author}{\bibinfo{person}{Martin Heidegger} {et~al\mbox{.}}} \bibinfo{year}{1971}\natexlab{}.
\newblock \showarticletitle{Building dwelling thinking}.
\newblock \bibinfo{journal}{\emph{Poetry, language, thought}}  \bibinfo{volume}{154} (\bibinfo{year}{1971}), \bibinfo{pages}{1--26}.
\newblock


\bibitem[Heidegger et~al\mbox{.}(1975)]%
        {heidegger1975poetry}
\bibfield{author}{\bibinfo{person}{Martin Heidegger}, \bibinfo{person}{Albert Hofstadter}, {et~al\mbox{.}}} \bibinfo{year}{1975}\natexlab{}.
\newblock \bibinfo{booktitle}{\emph{Poetry, language, thought}}.
\newblock \bibinfo{publisher}{Harper \& Row New York}.
\newblock


\bibitem[Hogg(2016)]%
        {hogg2016social}
\bibfield{author}{\bibinfo{person}{Michael~A Hogg}.} \bibinfo{year}{2016}\natexlab{}.
\newblock \bibinfo{booktitle}{\emph{Social identity theory}}.
\newblock \bibinfo{publisher}{Springer}.
\newblock


\bibitem[Hsu et~al\mbox{.}(2019)]%
        {hsu2019smell}
\bibfield{author}{\bibinfo{person}{Yen-Chia Hsu}, \bibinfo{person}{Jennifer Cross}, \bibinfo{person}{Paul Dille}, \bibinfo{person}{Michael Tasota}, \bibinfo{person}{Beatrice Dias}, \bibinfo{person}{Randy Sargent}, \bibinfo{person}{Ting-Hao Huang}, {and} \bibinfo{person}{Illah Nourbakhsh}.} \bibinfo{year}{2019}\natexlab{}.
\newblock \showarticletitle{Smell Pittsburgh: Community-empowered mobile smell reporting system}. In \bibinfo{booktitle}{\emph{Proceedings of the 24th International Conference on Intelligent User Interfaces}}. \bibinfo{pages}{65--79}.
\newblock


\bibitem[Hu et~al\mbox{.}(2013)]%
        {hu2013whoo}
\bibfield{author}{\bibinfo{person}{Yuheng Hu}, \bibinfo{person}{Shelly~D Farnham}, {and} \bibinfo{person}{Andr{\'e}s Monroy-Hern{\'a}ndez}.} \bibinfo{year}{2013}\natexlab{}.
\newblock \showarticletitle{Whoo. ly: Facilitating information seeking for hyperlocal communities using social media}. In \bibinfo{booktitle}{\emph{Proceedings of the SIGCHI Conference on Human Factors in Computing Systems}}. \bibinfo{pages}{3481--3490}.
\newblock


\bibitem[Jenkins et~al\mbox{.}(2016)]%
        {jenkins2016crowdsourcing}
\bibfield{author}{\bibinfo{person}{Andrew Jenkins}, \bibinfo{person}{Arie Croitoru}, \bibinfo{person}{Andrew~T Crooks}, {and} \bibinfo{person}{Anthony Stefanidis}.} \bibinfo{year}{2016}\natexlab{}.
\newblock \showarticletitle{Crowdsourcing a collective sense of place}.
\newblock \bibinfo{journal}{\emph{PloS one}} \bibinfo{volume}{11}, \bibinfo{number}{4} (\bibinfo{year}{2016}), \bibinfo{pages}{e0152932}.
\newblock


\bibitem[Jiang et~al\mbox{.}(2023)]%
        {jiang2023graphologue}
\bibfield{author}{\bibinfo{person}{Peiling Jiang}, \bibinfo{person}{Jude Rayan}, \bibinfo{person}{Steven~P Dow}, {and} \bibinfo{person}{Haijun Xia}.} \bibinfo{year}{2023}\natexlab{}.
\newblock \showarticletitle{Graphologue: Exploring large language model responses with interactive diagrams}. In \bibinfo{booktitle}{\emph{Proceedings of the 36th Annual ACM Symposium on User Interface Software and Technology}}. \bibinfo{pages}{1--20}.
\newblock


\bibitem[Johnson et~al\mbox{.}(2016a)]%
        {johnson2016not}
\bibfield{author}{\bibinfo{person}{Isaac~L Johnson}, \bibinfo{person}{Yilun Lin}, \bibinfo{person}{Toby Jia-Jun Li}, \bibinfo{person}{Andrew Hall}, \bibinfo{person}{Aaron Halfaker}, \bibinfo{person}{Johannes Sch{\"o}ning}, {and} \bibinfo{person}{Brent Hecht}.} \bibinfo{year}{2016}\natexlab{a}.
\newblock \showarticletitle{Not at home on the range: Peer production and the urban/rural divide}. In \bibinfo{booktitle}{\emph{Proceedings of the 2016 CHI conference on Human Factors in Computing Systems}}. \bibinfo{pages}{13--25}.
\newblock


\bibitem[Johnson et~al\mbox{.}(2016b)]%
        {johnson2016geography}
\bibfield{author}{\bibinfo{person}{Isaac~L Johnson}, \bibinfo{person}{Subhasree Sengupta}, \bibinfo{person}{Johannes Sch{\"o}ning}, {and} \bibinfo{person}{Brent Hecht}.} \bibinfo{year}{2016}\natexlab{b}.
\newblock \showarticletitle{The geography and importance of localness in geotagged social media}. In \bibinfo{booktitle}{\emph{Proceedings of the 2016 CHI Conference on Human Factors in Computing Systems}}. \bibinfo{pages}{515--526}.
\newblock


\bibitem[Jorgensen and Stedman(2001)]%
        {jorgensen2001sense}
\bibfield{author}{\bibinfo{person}{Bradley~S Jorgensen} {and} \bibinfo{person}{Richard~C Stedman}.} \bibinfo{year}{2001}\natexlab{}.
\newblock \showarticletitle{Sense of place as an attitude: Lakeshore owners attitudes toward their properties}.
\newblock \bibinfo{journal}{\emph{Journal of environmental psychology}} \bibinfo{volume}{21}, \bibinfo{number}{3} (\bibinfo{year}{2001}), \bibinfo{pages}{233--248}.
\newblock


\bibitem[Jurgens et~al\mbox{.}(2015)]%
        {jurgens2015geolocation}
\bibfield{author}{\bibinfo{person}{David Jurgens}, \bibinfo{person}{Tyler Finethy}, \bibinfo{person}{James McCorriston}, \bibinfo{person}{Yi Xu}, {and} \bibinfo{person}{Derek Ruths}.} \bibinfo{year}{2015}\natexlab{}.
\newblock \showarticletitle{Geolocation prediction in twitter using social networks: A critical analysis and review of current practice}. In \bibinfo{booktitle}{\emph{Proceedings of the international AAAI conference on web and social media}}, Vol.~\bibinfo{volume}{9}. \bibinfo{pages}{188--197}.
\newblock


\bibitem[Kang et~al\mbox{.}(2023)]%
        {kang2023synergi}
\bibfield{author}{\bibinfo{person}{Hyeonsu~B Kang}, \bibinfo{person}{Tongshuang Wu}, \bibinfo{person}{Joseph~Chee Chang}, {and} \bibinfo{person}{Aniket Kittur}.} \bibinfo{year}{2023}\natexlab{}.
\newblock \showarticletitle{Synergi: A Mixed-Initiative System for Scholarly Synthesis and Sensemaking}. In \bibinfo{booktitle}{\emph{Proceedings of the 36th Annual ACM Symposium on User Interface Software and Technology}}. \bibinfo{pages}{1--19}.
\newblock


\bibitem[Kariryaa et~al\mbox{.}(2018)]%
        {kariryaa2018defining}
\bibfield{author}{\bibinfo{person}{Ankit Kariryaa}, \bibinfo{person}{Isaac Johnson}, \bibinfo{person}{Johannes Sch{\"o}ning}, {and} \bibinfo{person}{Brent Hecht}.} \bibinfo{year}{2018}\natexlab{}.
\newblock \showarticletitle{Defining and predicting the localness of volunteered geographic information using ground truth data}. In \bibinfo{booktitle}{\emph{Proceedings of the 2018 CHI conference on human factors in computing systems}}. \bibinfo{pages}{1--12}.
\newblock


\bibitem[Kumar and Shah(2018)]%
        {kumar2018false}
\bibfield{author}{\bibinfo{person}{Srijan Kumar} {and} \bibinfo{person}{Neil Shah}.} \bibinfo{year}{2018}\natexlab{}.
\newblock \showarticletitle{False information on web and social media: A survey}.
\newblock \bibinfo{journal}{\emph{arXiv preprint arXiv:1804.08559}} (\bibinfo{year}{2018}).
\newblock


\bibitem[Lentini and Decortis(2010)]%
        {lentini2010space}
\bibfield{author}{\bibinfo{person}{Laura Lentini} {and} \bibinfo{person}{Fran{\c{c}}oise Decortis}.} \bibinfo{year}{2010}\natexlab{}.
\newblock \showarticletitle{Space and places: when interacting with and in physical space becomes a meaningful experience}.
\newblock \bibinfo{journal}{\emph{Personal and Ubiquitous Computing}}  \bibinfo{volume}{14} (\bibinfo{year}{2010}), \bibinfo{pages}{407--415}.
\newblock


\bibitem[Li et~al\mbox{.}(2013)]%
        {li2013spatial}
\bibfield{author}{\bibinfo{person}{Linna Li}, \bibinfo{person}{Michael~F Goodchild}, {and} \bibinfo{person}{Bo Xu}.} \bibinfo{year}{2013}\natexlab{}.
\newblock \showarticletitle{Spatial, temporal, and socioeconomic patterns in the use of Twitter and Flickr}.
\newblock \bibinfo{journal}{\emph{Cartography and geographic information science}} \bibinfo{volume}{40}, \bibinfo{number}{2} (\bibinfo{year}{2013}), \bibinfo{pages}{61--77}.
\newblock


\bibitem[Liu et~al\mbox{.}(2021)]%
        {liu2021visually}
\bibfield{author}{\bibinfo{person}{Fangyu Liu}, \bibinfo{person}{Emanuele Bugliarello}, \bibinfo{person}{Edoardo~Maria Ponti}, \bibinfo{person}{Siva Reddy}, \bibinfo{person}{Nigel Collier}, {and} \bibinfo{person}{Desmond Elliott}.} \bibinfo{year}{2021}\natexlab{}.
\newblock \showarticletitle{Visually grounded reasoning across languages and cultures}.
\newblock \bibinfo{journal}{\emph{arXiv preprint arXiv:2109.13238}} (\bibinfo{year}{2021}).
\newblock


\bibitem[Long(2023)]%
        {long2023large}
\bibfield{author}{\bibinfo{person}{Jieyi Long}.} \bibinfo{year}{2023}\natexlab{}.
\newblock \showarticletitle{Large language model guided tree-of-thought}.
\newblock \bibinfo{journal}{\emph{arXiv preprint arXiv:2305.08291}} (\bibinfo{year}{2023}).
\newblock


\bibitem[Low and Altman(1992)]%
        {low1992place}
\bibfield{author}{\bibinfo{person}{Setha~M Low} {and} \bibinfo{person}{Irwin Altman}.} \bibinfo{year}{1992}\natexlab{}.
\newblock \showarticletitle{Place attachment: A conceptual inquiry}.
\newblock In \bibinfo{booktitle}{\emph{Place attachment}}. \bibinfo{publisher}{Springer}, \bibinfo{pages}{1--12}.
\newblock


\bibitem[Ludford et~al\mbox{.}(2007)]%
        {ludford2007capturing}
\bibfield{author}{\bibinfo{person}{Pamela~J Ludford}, \bibinfo{person}{Reid Priedhorsky}, \bibinfo{person}{Ken Reily}, {and} \bibinfo{person}{Loren Terveen}.} \bibinfo{year}{2007}\natexlab{}.
\newblock \showarticletitle{Capturing, sharing, and using local place information}. In \bibinfo{booktitle}{\emph{Proceedings of the SIGCHI conference on Human factors in computing systems}}. \bibinfo{pages}{1235--1244}.
\newblock


\bibitem[Ma et~al\mbox{.}(2024)]%
        {ma2024beyond}
\bibfield{author}{\bibinfo{person}{Xiao Ma}, \bibinfo{person}{Swaroop Mishra}, \bibinfo{person}{Ariel Liu}, \bibinfo{person}{Sophie~Ying Su}, \bibinfo{person}{Jilin Chen}, \bibinfo{person}{Chinmay Kulkarni}, \bibinfo{person}{Heng-Tze Cheng}, \bibinfo{person}{Quoc Le}, {and} \bibinfo{person}{Ed Chi}.} \bibinfo{year}{2024}\natexlab{}.
\newblock \showarticletitle{Beyond chatbots: Explorellm for structured thoughts and personalized model responses}. In \bibinfo{booktitle}{\emph{Extended Abstracts of the CHI Conference on Human Factors in Computing Systems}}. \bibinfo{pages}{1--12}.
\newblock


\bibitem[Malmborg et~al\mbox{.}(2015)]%
        {malmborg2015designing}
\bibfield{author}{\bibinfo{person}{Lone Malmborg}, \bibinfo{person}{Ann Light}, \bibinfo{person}{Geraldine Fitzpatrick}, \bibinfo{person}{Victoria Bellotti}, {and} \bibinfo{person}{Margot Brereton}.} \bibinfo{year}{2015}\natexlab{}.
\newblock \showarticletitle{Designing for sharing in local communities}. In \bibinfo{booktitle}{\emph{Proceedings of the 33rd Annual ACM Conference Extended Abstracts on Human Factors in Computing Systems}}. \bibinfo{pages}{2357--2360}.
\newblock


\bibitem[Manzo(2005)]%
        {manzo2005better}
\bibfield{author}{\bibinfo{person}{Lynne~C Manzo}.} \bibinfo{year}{2005}\natexlab{}.
\newblock \showarticletitle{For better or worse: Exploring multiple dimensions of place meaning}.
\newblock \bibinfo{journal}{\emph{Journal of environmental psychology}} \bibinfo{volume}{25}, \bibinfo{number}{1} (\bibinfo{year}{2005}), \bibinfo{pages}{67--86}.
\newblock


\bibitem[Manzo and Perkins(2006)]%
        {manzo2006finding}
\bibfield{author}{\bibinfo{person}{Lynne~C Manzo} {and} \bibinfo{person}{Douglas~D Perkins}.} \bibinfo{year}{2006}\natexlab{}.
\newblock \showarticletitle{Finding common ground: The importance of place attachment to community participation and planning}.
\newblock \bibinfo{journal}{\emph{Journal of planning literature}} \bibinfo{volume}{20}, \bibinfo{number}{4} (\bibinfo{year}{2006}), \bibinfo{pages}{335--350}.
\newblock


\bibitem[McMillan and Chavis(1986)]%
        {mcmillan1986sense}
\bibfield{author}{\bibinfo{person}{David~W McMillan} {and} \bibinfo{person}{David~M Chavis}.} \bibinfo{year}{1986}\natexlab{}.
\newblock \showarticletitle{Sense of community: A definition and theory}.
\newblock \bibinfo{journal}{\emph{Journal of community psychology}} \bibinfo{volume}{14}, \bibinfo{number}{1} (\bibinfo{year}{1986}), \bibinfo{pages}{6--23}.
\newblock


\bibitem[Mohanty et~al\mbox{.}(2019)]%
        {mohanty2019photo}
\bibfield{author}{\bibinfo{person}{Vikram Mohanty}, \bibinfo{person}{David Thames}, \bibinfo{person}{Sneha Mehta}, {and} \bibinfo{person}{Kurt Luther}.} \bibinfo{year}{2019}\natexlab{}.
\newblock \showarticletitle{Photo sleuth: Combining human expertise and face recognition to identify historical portraits}. In \bibinfo{booktitle}{\emph{Proceedings of the 24th International Conference on Intelligent User Interfaces}}. \bibinfo{pages}{547--557}.
\newblock


\bibitem[{OpenStreetMap Contributors}(nda)]%
        {osm_apple}
\bibfield{author}{\bibinfo{person}{{OpenStreetMap Contributors}}.} \bibinfo{year}{n.d.}\natexlab{a}.
\newblock \bibinfo{title}{Apple - OpenStreetMap Wiki}.
\newblock \bibinfo{howpublished}{\url{https://wiki.openstreetmap.org/wiki/Apple}}.
\newblock
\newblock
\shownote{Accessed: 2024-09-12}.


\bibitem[{OpenStreetMap Contributors}(ndb)]%
        {osm_lyft}
\bibfield{author}{\bibinfo{person}{{OpenStreetMap Contributors}}.} \bibinfo{year}{n.d.}\natexlab{b}.
\newblock \bibinfo{title}{Lyft - OpenStreetMap Wiki}.
\newblock \bibinfo{howpublished}{\url{https://wiki.openstreetmap.org/wiki/Lyft}}.
\newblock
\newblock
\shownote{Accessed: 2024-09-12}.


\bibitem[Paananen et~al\mbox{.}(2021)]%
        {paananen2021investigating}
\bibfield{author}{\bibinfo{person}{Ville Paananen}, \bibinfo{person}{Jonas Oppenlaender}, \bibinfo{person}{Jorge Goncalves}, \bibinfo{person}{Danula Hettiachchi}, {and} \bibinfo{person}{Simo Hosio}.} \bibinfo{year}{2021}\natexlab{}.
\newblock \showarticletitle{Investigating human scale spatial experience}.
\newblock \bibinfo{journal}{\emph{Proceedings of the ACM on Human-Computer Interaction}} \bibinfo{volume}{5}, \bibinfo{number}{ISS} (\bibinfo{year}{2021}), \bibinfo{pages}{1--18}.
\newblock


\bibitem[Panciera et~al\mbox{.}(2013)]%
        {panciera2013soil}
\bibfield{author}{\bibinfo{person}{Rocco Panciera}, \bibinfo{person}{Jeffrey~P Walker}, \bibinfo{person}{Thomas~J Jackson}, \bibinfo{person}{Douglas~A Gray}, \bibinfo{person}{Mihai~A Tanase}, \bibinfo{person}{Dongryeol Ryu}, \bibinfo{person}{Alessandra Monerris}, \bibinfo{person}{Heath Yardley}, \bibinfo{person}{Christoph R{\"u}diger}, \bibinfo{person}{Xiaoling Wu}, {et~al\mbox{.}}} \bibinfo{year}{2013}\natexlab{}.
\newblock \showarticletitle{The soil moisture active passive experiments (SMAPEx): Toward soil moisture retrieval from the SMAP mission}.
\newblock \bibinfo{journal}{\emph{IEEE transactions on geoscience and remote sensing}} \bibinfo{volume}{52}, \bibinfo{number}{1} (\bibinfo{year}{2013}), \bibinfo{pages}{490--507}.
\newblock


\bibitem[Petroni et~al\mbox{.}(2019)]%
        {petroni2019language}
\bibfield{author}{\bibinfo{person}{Fabio Petroni}, \bibinfo{person}{Tim Rockt{\"a}schel}, \bibinfo{person}{Patrick Lewis}, \bibinfo{person}{Anton Bakhtin}, \bibinfo{person}{Yuxiang Wu}, \bibinfo{person}{Alexander~H Miller}, {and} \bibinfo{person}{Sebastian Riedel}.} \bibinfo{year}{2019}\natexlab{}.
\newblock \showarticletitle{Language models as knowledge bases?}
\newblock \bibinfo{journal}{\emph{arXiv preprint arXiv:1909.01066}} (\bibinfo{year}{2019}).
\newblock


\bibitem[Pirolli and Card(2005)]%
        {pirolli2005sensemaking}
\bibfield{author}{\bibinfo{person}{Peter Pirolli} {and} \bibinfo{person}{Stuart Card}.} \bibinfo{year}{2005}\natexlab{}.
\newblock \showarticletitle{The sensemaking process and leverage points for analyst technology as identified through cognitive task analysis}. In \bibinfo{booktitle}{\emph{Proceedings of international conference on intelligence analysis}}, Vol.~\bibinfo{volume}{5}. McLean, VA, USA, \bibinfo{pages}{2--4}.
\newblock


\bibitem[Popescu and Grefenstette(2010)]%
        {popescu2010mining}
\bibfield{author}{\bibinfo{person}{Adrian Popescu} {and} \bibinfo{person}{Gregory Grefenstette}.} \bibinfo{year}{2010}\natexlab{}.
\newblock \showarticletitle{Mining user home location and gender from flickr tags}. In \bibinfo{booktitle}{\emph{Proceedings of the International AAAI Conference on Web and Social Media}}, Vol.~\bibinfo{volume}{4}. \bibinfo{pages}{307--310}.
\newblock


\bibitem[Proshansky et~al\mbox{.}(1983)]%
        {proshansky1983place}
\bibfield{author}{\bibinfo{person}{Harold~M Proshansky}, \bibinfo{person}{Abbe~K Fabian}, {and} \bibinfo{person}{Robert Kaminoff}.} \bibinfo{year}{1983}\natexlab{}.
\newblock \showarticletitle{Place-identity: Physical world socialization of the self}.
\newblock \bibinfo{journal}{\emph{Journal of Environmental Psychology}} \bibinfo{volume}{3}, \bibinfo{number}{1} (\bibinfo{year}{1983}), \bibinfo{pages}{57--83}.
\newblock


\bibitem[Puccetti et~al\mbox{.}(2024)]%
        {puccetti2024ai}
\bibfield{author}{\bibinfo{person}{Giovanni Puccetti}, \bibinfo{person}{Anna Rogers}, \bibinfo{person}{Chiara Alzetta}, \bibinfo{person}{Felice Dell'Orletta}, {and} \bibinfo{person}{Andrea Esuli}.} \bibinfo{year}{2024}\natexlab{}.
\newblock \showarticletitle{AI" News" Content Farms Are Easy to Make and Hard to Detect: A Case Study in Italian}.
\newblock \bibinfo{journal}{\emph{arXiv preprint arXiv:2406.12128}} (\bibinfo{year}{2024}).
\newblock


\bibitem[Quercia et~al\mbox{.}(2014)]%
        {quercia2014shortest}
\bibfield{author}{\bibinfo{person}{Daniele Quercia}, \bibinfo{person}{Rossano Schifanella}, {and} \bibinfo{person}{Luca~Maria Aiello}.} \bibinfo{year}{2014}\natexlab{}.
\newblock \showarticletitle{The shortest path to happiness: Recommending beautiful, quiet, and happy routes in the city}. In \bibinfo{booktitle}{\emph{Proceedings of the 25th ACM conference on Hypertext and social media}}. \bibinfo{pages}{116--125}.
\newblock


\bibitem[Quercia et~al\mbox{.}(2015)]%
        {quercia2015smelly}
\bibfield{author}{\bibinfo{person}{Daniele Quercia}, \bibinfo{person}{Rossano Schifanella}, \bibinfo{person}{Luca~Maria Aiello}, {and} \bibinfo{person}{Kate McLean}.} \bibinfo{year}{2015}\natexlab{}.
\newblock \showarticletitle{Smelly maps: the digital life of urban smellscapes}. In \bibinfo{booktitle}{\emph{Proceedings of the International AAAI conference on Web and Social Media}}, Vol.~\bibinfo{volume}{9}. \bibinfo{pages}{327--336}.
\newblock


\bibitem[Raymond et~al\mbox{.}(2010)]%
        {raymond2010measurement}
\bibfield{author}{\bibinfo{person}{Christopher~M Raymond}, \bibinfo{person}{Gregory Brown}, {and} \bibinfo{person}{Delene Weber}.} \bibinfo{year}{2010}\natexlab{}.
\newblock \showarticletitle{The measurement of place attachment: Personal, community, and environmental connections}.
\newblock \bibinfo{journal}{\emph{Journal of environmental psychology}} \bibinfo{volume}{30}, \bibinfo{number}{4} (\bibinfo{year}{2010}), \bibinfo{pages}{422--434}.
\newblock


\bibitem[Reicher et~al\mbox{.}(2010)]%
        {reicher2010social}
\bibfield{author}{\bibinfo{person}{Stephen Reicher}, \bibinfo{person}{Russell Spears}, {and} \bibinfo{person}{S~Alexander Haslam}.} \bibinfo{year}{2010}\natexlab{}.
\newblock \showarticletitle{The social identity approach in social psychology}.
\newblock \bibinfo{journal}{\emph{Sage identities handbook}} (\bibinfo{year}{2010}), \bibinfo{pages}{45--62}.
\newblock


\bibitem[Relph(1976)]%
        {relph1976place}
\bibfield{author}{\bibinfo{person}{Edward Relph}.} \bibinfo{year}{1976}\natexlab{}.
\newblock \bibinfo{booktitle}{\emph{Place and placelessness}}. Vol.~\bibinfo{volume}{67}.
\newblock \bibinfo{publisher}{Pion London}.
\newblock


\bibitem[Russell et~al\mbox{.}(1993)]%
        {russell1993cost}
\bibfield{author}{\bibinfo{person}{Daniel~M Russell}, \bibinfo{person}{Mark~J Stefik}, \bibinfo{person}{Peter Pirolli}, {and} \bibinfo{person}{Stuart~K Card}.} \bibinfo{year}{1993}\natexlab{}.
\newblock \showarticletitle{The cost structure of sensemaking}. In \bibinfo{booktitle}{\emph{Proceedings of the INTERACT'93 and CHI'93 conference on Human factors in computing systems}}. \bibinfo{pages}{269--276}.
\newblock


\bibitem[Salewski et~al\mbox{.}(2023)]%
        {salewski2023context}
\bibfield{author}{\bibinfo{person}{Leonard Salewski}, \bibinfo{person}{Stephan Alaniz}, \bibinfo{person}{Isabel Rio-Torto}, \bibinfo{person}{Eric Schulz}, {and} \bibinfo{person}{Zeynep Akata}.} \bibinfo{year}{2023}\natexlab{}.
\newblock \showarticletitle{In-context impersonation reveals Large Language Models' strengths and biases}.
\newblock \bibinfo{journal}{\emph{Advances in neural information processing systems}}  \bibinfo{volume}{36} (\bibinfo{year}{2023}), \bibinfo{pages}{72044--72057}.
\newblock


\bibitem[Sanchez and Dietz(2022)]%
        {sanchez2022travelers}
\bibfield{author}{\bibinfo{person}{Pablo Sanchez} {and} \bibinfo{person}{Linus~W Dietz}.} \bibinfo{year}{2022}\natexlab{}.
\newblock \showarticletitle{Travelers vs. Locals: The Effect of Cluster Analysis in Point-of-Interest Recommendation}. In \bibinfo{booktitle}{\emph{Proceedings of the 30th ACM Conference on User Modeling, Adaptation and Personalization}}. \bibinfo{pages}{132--142}.
\newblock


\bibitem[Sap et~al\mbox{.}(2019)]%
        {sap2019atomic}
\bibfield{author}{\bibinfo{person}{Maarten Sap}, \bibinfo{person}{Ronan Le~Bras}, \bibinfo{person}{Emily Allaway}, \bibinfo{person}{Chandra Bhagavatula}, \bibinfo{person}{Nicholas Lourie}, \bibinfo{person}{Hannah Rashkin}, \bibinfo{person}{Brendan Roof}, \bibinfo{person}{Noah~A Smith}, {and} \bibinfo{person}{Yejin Choi}.} \bibinfo{year}{2019}\natexlab{}.
\newblock \showarticletitle{Atomic: An atlas of machine commonsense for if-then reasoning}. In \bibinfo{booktitle}{\emph{Proceedings of the AAAI conference on artificial intelligence}}, Vol.~\bibinfo{volume}{33}. \bibinfo{pages}{3027--3035}.
\newblock


\bibitem[Sen et~al\mbox{.}(2015)]%
        {sen2015turkers}
\bibfield{author}{\bibinfo{person}{Shilad Sen}, \bibinfo{person}{Margaret~E. Giesel}, \bibinfo{person}{Rebecca Gold}, \bibinfo{person}{Benjamin Hillmann}, \bibinfo{person}{Matt Lesicko}, \bibinfo{person}{Samuel Naden}, \bibinfo{person}{Jesse Russell}, \bibinfo{person}{Zixiao~(Ken) Wang}, {and} \bibinfo{person}{Brent Hecht}.} \bibinfo{year}{2015}\natexlab{}.
\newblock \showarticletitle{Turkers, Scholars, "Arafat" and "Peace": Cultural Communities and Algorithmic Gold Standards}. In \bibinfo{booktitle}{\emph{Proceedings of the 18th ACM Conference on Computer Supported Cooperative Work \& Social Computing}} (Vancouver, BC, Canada) \emph{(\bibinfo{series}{CSCW '15})}. \bibinfo{publisher}{Association for Computing Machinery}, \bibinfo{address}{New York, NY, USA}, \bibinfo{pages}{826–838}.
\newblock
\showISBNx{9781450329224}
\urldef\tempurl%
\url{https://doi.org/10.1145/2675133.2675285}
\showDOI{\tempurl}


\bibitem[Shwartz(2022)]%
        {shwartz2022good}
\bibfield{author}{\bibinfo{person}{Vered Shwartz}.} \bibinfo{year}{2022}\natexlab{}.
\newblock \showarticletitle{Good night at 4 pm?! time expressions in different cultures}. In \bibinfo{booktitle}{\emph{Findings of the Association for Computational Linguistics: ACL 2022}}. \bibinfo{pages}{2842--2853}.
\newblock


\bibitem[Silva et~al\mbox{.}(2019)]%
        {silva2019urban}
\bibfield{author}{\bibinfo{person}{Thiago~H Silva}, \bibinfo{person}{Aline~Carneiro Viana}, \bibinfo{person}{Fabr{\'\i}cio Benevenuto}, \bibinfo{person}{Leandro Villas}, \bibinfo{person}{Juliana Salles}, \bibinfo{person}{Antonio Loureiro}, {and} \bibinfo{person}{Daniele Quercia}.} \bibinfo{year}{2019}\natexlab{}.
\newblock \showarticletitle{Urban computing leveraging location-based social network data: a survey}.
\newblock \bibinfo{journal}{\emph{ACM Computing Surveys (CSUR)}} \bibinfo{volume}{52}, \bibinfo{number}{1} (\bibinfo{year}{2019}), \bibinfo{pages}{1--39}.
\newblock


\bibitem[Starbird et~al\mbox{.}(2019)]%
        {starbird2019disinformation}
\bibfield{author}{\bibinfo{person}{Kate Starbird}, \bibinfo{person}{Ahmer Arif}, {and} \bibinfo{person}{Tom Wilson}.} \bibinfo{year}{2019}\natexlab{}.
\newblock \showarticletitle{Disinformation as collaborative work: Surfacing the participatory nature of strategic information operations}.
\newblock \bibinfo{journal}{\emph{Proceedings of the ACM on Human-Computer Interaction}} \bibinfo{volume}{3}, \bibinfo{number}{CSCW} (\bibinfo{year}{2019}), \bibinfo{pages}{1--26}.
\newblock


\bibitem[Stedman(2008)]%
        {stedman2008we}
\bibfield{author}{\bibinfo{person}{RC Stedman}.} \bibinfo{year}{2008}\natexlab{}.
\newblock \showarticletitle{What do we" mean" by place meanings? Implications of place meanings for managers and practitioners. Vols}.
\newblock \bibinfo{journal}{\emph{General Technical Report PNW-GTR-744. In LE Kruger, TE Hall, \& MC Stiefel (Eds.), Understanding concepts of place in recreation research and management. Portland, OR: US Department of Agriculture, Forest Service, Pacific Northwest Research Station}} (\bibinfo{year}{2008}).
\newblock


\bibitem[Suh et~al\mbox{.}(2023)]%
        {suh2023sensecape}
\bibfield{author}{\bibinfo{person}{Sangho Suh}, \bibinfo{person}{Bryan Min}, \bibinfo{person}{Srishti Palani}, {and} \bibinfo{person}{Haijun Xia}.} \bibinfo{year}{2023}\natexlab{}.
\newblock \showarticletitle{Sensecape: Enabling multilevel exploration and sensemaking with large language models}. In \bibinfo{booktitle}{\emph{Proceedings of the 36th Annual ACM Symposium on User Interface Software and Technology}}. \bibinfo{pages}{1--18}.
\newblock


\bibitem[Sun(2015)]%
        {sun2015importance}
\bibfield{author}{\bibinfo{person}{Emily Sun}.} \bibinfo{year}{2015}\natexlab{}.
\newblock \showarticletitle{The importance of play in digital placemaking}. In \bibinfo{booktitle}{\emph{Proceedings of the International AAAI Conference on Web and Social Media}}, Vol.~\bibinfo{volume}{9}. \bibinfo{pages}{23--25}.
\newblock


\bibitem[Sun et~al\mbox{.}(2017)]%
        {sun2017movemeant}
\bibfield{author}{\bibinfo{person}{Emily Sun}, \bibinfo{person}{Ross McLachlan}, {and} \bibinfo{person}{Mor Naaman}.} \bibinfo{year}{2017}\natexlab{}.
\newblock \showarticletitle{Movemeant: anonymously building community through shared location histories}. In \bibinfo{booktitle}{\emph{Proceedings of the 2017 CHI Conference on Human Factors in Computing Systems}}. \bibinfo{pages}{4284--4289}.
\newblock


\bibitem[Sun and Naaman(2018)]%
        {sun2018multi}
\bibfield{author}{\bibinfo{person}{Emily Sun} {and} \bibinfo{person}{Mor Naaman}.} \bibinfo{year}{2018}\natexlab{}.
\newblock \showarticletitle{A multi-site investigation of community awareness through passive location sharing}. In \bibinfo{booktitle}{\emph{Proceedings of the 2018 CHI Conference on Human Factors in Computing Systems}}. \bibinfo{pages}{1--13}.
\newblock


\bibitem[Tajfel et~al\mbox{.}(1979)]%
        {tajfel1979integrative}
\bibfield{author}{\bibinfo{person}{Henri Tajfel}, \bibinfo{person}{John~C Turner}, \bibinfo{person}{William~G Austin}, {and} \bibinfo{person}{Stephen Worchel}.} \bibinfo{year}{1979}\natexlab{}.
\newblock \showarticletitle{An integrative theory of intergroup conflict}.
\newblock \bibinfo{journal}{\emph{Organizational identity: A reader}} \bibinfo{volume}{56}, \bibinfo{number}{65} (\bibinfo{year}{1979}), \bibinfo{pages}{9780203505984--16}.
\newblock


\bibitem[Taylor et~al\mbox{.}(2015)]%
        {taylor2015data}
\bibfield{author}{\bibinfo{person}{Alex~S Taylor}, \bibinfo{person}{Si{\^a}n Lindley}, \bibinfo{person}{Tim Regan}, \bibinfo{person}{David Sweeney}, \bibinfo{person}{Vasillis Vlachokyriakos}, \bibinfo{person}{Lillie Grainger}, {and} \bibinfo{person}{Jessica Lingel}.} \bibinfo{year}{2015}\natexlab{}.
\newblock \showarticletitle{Data-in-place: Thinking through the relations between data and community}. In \bibinfo{booktitle}{\emph{Proceedings of the 33rd Annual ACM Conference on Human Factors in Computing Systems}}. \bibinfo{pages}{2863--2872}.
\newblock


\bibitem[Thebault-Spieker et~al\mbox{.}(2018a)]%
        {thebault2018distance}
\bibfield{author}{\bibinfo{person}{Jacob Thebault-Spieker}, \bibinfo{person}{Aaron Halfaker}, \bibinfo{person}{Loren~G Terveen}, {and} \bibinfo{person}{Brent Hecht}.} \bibinfo{year}{2018}\natexlab{a}.
\newblock \showarticletitle{Distance and attraction: Gravity models for geographic content production}. In \bibinfo{booktitle}{\emph{Proceedings of the 2018 CHI conference on human factors in computing systems}}. \bibinfo{pages}{1--13}.
\newblock


\bibitem[Thebault-Spieker et~al\mbox{.}(2018b)]%
        {thebault2018geographic}
\bibfield{author}{\bibinfo{person}{Jacob Thebault-Spieker}, \bibinfo{person}{Brent Hecht}, {and} \bibinfo{person}{Loren Terveen}.} \bibinfo{year}{2018}\natexlab{b}.
\newblock \showarticletitle{Geographic Biases are'Born, not Made' Exploring Contributors' Spatiotemporal Behavior in OpenStreetMap}. In \bibinfo{booktitle}{\emph{Proceedings of the 2018 ACM International Conference on Supporting Group Work}}. \bibinfo{pages}{71--82}.
\newblock


\bibitem[Tuan(1977)]%
        {tuan1977space}
\bibfield{author}{\bibinfo{person}{Yi-Fu Tuan}.} \bibinfo{year}{1977}\natexlab{}.
\newblock \bibinfo{booktitle}{\emph{Space and place: The perspective of experience}}.
\newblock \bibinfo{publisher}{U of Minnesota Press}.
\newblock


\bibitem[Turing(1950)]%
        {turing1950}
\bibfield{author}{\bibinfo{person}{A. Turing}.} \bibinfo{year}{1950}\natexlab{}.
\newblock \showarticletitle{Computing Machinery and Intelligence}.
\newblock \bibinfo{journal}{\emph{Mind}} \bibinfo{volume}{59}, \bibinfo{number}{236} (\bibinfo{year}{1950}), \bibinfo{pages}{433--460}.
\newblock


\bibitem[Turner and Reynolds(2012)]%
        {turner2012self}
\bibfield{author}{\bibinfo{person}{John~C Turner} {and} \bibinfo{person}{Katherine~J Reynolds}.} \bibinfo{year}{2012}\natexlab{}.
\newblock \showarticletitle{Self-categorization theory}.
\newblock \bibinfo{journal}{\emph{Handbook of theories of social psychology}}  \bibinfo{volume}{2} (\bibinfo{year}{2012}), \bibinfo{pages}{399--417}.
\newblock


\bibitem[Venkatagiri et~al\mbox{.}(2019)]%
        {venkatagiri2019groundtruth}
\bibfield{author}{\bibinfo{person}{Sukrit Venkatagiri}, \bibinfo{person}{Jacob Thebault-Spieker}, \bibinfo{person}{Rachel Kohler}, \bibinfo{person}{John Purviance}, \bibinfo{person}{Rifat~Sabbir Mansur}, {and} \bibinfo{person}{Kurt Luther}.} \bibinfo{year}{2019}\natexlab{}.
\newblock \showarticletitle{GroundTruth: Augmenting expert image geolocation with crowdsourcing and shared representations}.
\newblock \bibinfo{journal}{\emph{Proceedings of the ACM on Human-Computer Interaction}} \bibinfo{volume}{3}, \bibinfo{number}{CSCW} (\bibinfo{year}{2019}), \bibinfo{pages}{1--30}.
\newblock


\bibitem[Vlachokyriakos et~al\mbox{.}(2014)]%
        {vlachokyriakos2014postervote}
\bibfield{author}{\bibinfo{person}{Vasilis Vlachokyriakos}, \bibinfo{person}{Rob Comber}, \bibinfo{person}{Karim Ladha}, \bibinfo{person}{Nick Taylor}, \bibinfo{person}{Paul Dunphy}, \bibinfo{person}{Patrick McCorry}, {and} \bibinfo{person}{Patrick Olivier}.} \bibinfo{year}{2014}\natexlab{}.
\newblock \showarticletitle{PosterVote: expanding the action repertoire for local political activism}. In \bibinfo{booktitle}{\emph{Proceedings of the 2014 conference on Designing interactive systems}}. \bibinfo{pages}{795--804}.
\newblock


\bibitem[Von~Ahn(2006)]%
        {von2006games}
\bibfield{author}{\bibinfo{person}{Luis Von~Ahn}.} \bibinfo{year}{2006}\natexlab{}.
\newblock \showarticletitle{Games with a purpose}.
\newblock \bibinfo{journal}{\emph{Computer}} \bibinfo{volume}{39}, \bibinfo{number}{6} (\bibinfo{year}{2006}), \bibinfo{pages}{92--94}.
\newblock


\bibitem[Von~Ahn and Dabbish(2004)]%
        {von2004labeling}
\bibfield{author}{\bibinfo{person}{Luis Von~Ahn} {and} \bibinfo{person}{Laura Dabbish}.} \bibinfo{year}{2004}\natexlab{}.
\newblock \showarticletitle{Labeling images with a computer game}. In \bibinfo{booktitle}{\emph{Proceedings of the SIGCHI conference on Human factors in computing systems}}. \bibinfo{pages}{319--326}.
\newblock


\bibitem[Von~Ahn and Dabbish(2008)]%
        {von2008designing}
\bibfield{author}{\bibinfo{person}{Luis Von~Ahn} {and} \bibinfo{person}{Laura Dabbish}.} \bibinfo{year}{2008}\natexlab{}.
\newblock \showarticletitle{Designing games with a purpose}.
\newblock \bibinfo{journal}{\emph{Commun. ACM}} \bibinfo{volume}{51}, \bibinfo{number}{8} (\bibinfo{year}{2008}), \bibinfo{pages}{58--67}.
\newblock


\bibitem[White and Buscher(2012)]%
        {white2012characterizing}
\bibfield{author}{\bibinfo{person}{Ryen White} {and} \bibinfo{person}{Georg Buscher}.} \bibinfo{year}{2012}\natexlab{}.
\newblock \showarticletitle{Characterizing local interests and local knowledge}. In \bibinfo{booktitle}{\emph{Proceedings of the SIGCHI Conference on Human Factors in Computing Systems}}. \bibinfo{pages}{1607--1610}.
\newblock


\bibitem[Wu et~al\mbox{.}(2011)]%
        {wu2011mining}
\bibfield{author}{\bibinfo{person}{Shaomei Wu}, \bibinfo{person}{Shenwei Liu}, \bibinfo{person}{Dan Cosley}, {and} \bibinfo{person}{Michael Macy}.} \bibinfo{year}{2011}\natexlab{}.
\newblock \showarticletitle{Mining collective local knowledge from google mymaps}. In \bibinfo{booktitle}{\emph{Proceedings of the 20th international conference companion on World wide web}}. \bibinfo{pages}{151--152}.
\newblock


\bibitem[Yao et~al\mbox{.}(2024)]%
        {yao2024tree}
\bibfield{author}{\bibinfo{person}{Shunyu Yao}, \bibinfo{person}{Dian Yu}, \bibinfo{person}{Jeffrey Zhao}, \bibinfo{person}{Izhak Shafran}, \bibinfo{person}{Tom Griffiths}, \bibinfo{person}{Yuan Cao}, {and} \bibinfo{person}{Karthik Narasimhan}.} \bibinfo{year}{2024}\natexlab{}.
\newblock \showarticletitle{Tree of thoughts: Deliberate problem solving with large language models}.
\newblock \bibinfo{journal}{\emph{Advances in Neural Information Processing Systems}}  \bibinfo{volume}{36} (\bibinfo{year}{2024}).
\newblock


\bibitem[Yin et~al\mbox{.}(2022)]%
        {yin2022geomlama}
\bibfield{author}{\bibinfo{person}{Da Yin}, \bibinfo{person}{Hritik Bansal}, \bibinfo{person}{Masoud Monajatipoor}, \bibinfo{person}{Liunian~Harold Li}, {and} \bibinfo{person}{Kai-Wei Chang}.} \bibinfo{year}{2022}\natexlab{}.
\newblock \showarticletitle{Geomlama: Geo-diverse commonsense probing on multilingual pre-trained language models}.
\newblock \bibinfo{journal}{\emph{arXiv preprint arXiv:2205.12247}} (\bibinfo{year}{2022}).
\newblock


\bibitem[Yin et~al\mbox{.}(2021)]%
        {yin2021broaden}
\bibfield{author}{\bibinfo{person}{Da Yin}, \bibinfo{person}{Liunian~Harold Li}, \bibinfo{person}{Ziniu Hu}, \bibinfo{person}{Nanyun Peng}, {and} \bibinfo{person}{Kai-Wei Chang}.} \bibinfo{year}{2021}\natexlab{}.
\newblock \showarticletitle{Broaden the vision: Geo-diverse visual commonsense reasoning}.
\newblock \bibinfo{journal}{\emph{arXiv preprint arXiv:2109.06860}} (\bibinfo{year}{2021}).
\newblock


\bibitem[Yuan and Crooks(2019)]%
        {yuan2019assessing}
\bibfield{author}{\bibinfo{person}{Xiaoyi Yuan} {and} \bibinfo{person}{Andrew Crooks}.} \bibinfo{year}{2019}\natexlab{}.
\newblock \showarticletitle{Assessing the placeness of locations through user-contributed content}. In \bibinfo{booktitle}{\emph{Proceedings of the 3rd ACM SIGSPATIAL International Workshop on AI for Geographic Knowledge Discovery}}. \bibinfo{pages}{15--23}.
\newblock


\bibitem[Zhang and Soergel(2014)]%
        {zhang2014towards}
\bibfield{author}{\bibinfo{person}{Pengyi Zhang} {and} \bibinfo{person}{Dagobert Soergel}.} \bibinfo{year}{2014}\natexlab{}.
\newblock \showarticletitle{Towards a comprehensive model of the cognitive process and mechanisms of individual sensemaking}.
\newblock \bibinfo{journal}{\emph{Journal of the Association for Information Science and Technology}} \bibinfo{volume}{65}, \bibinfo{number}{9} (\bibinfo{year}{2014}), \bibinfo{pages}{1733--1756}.
\newblock


\bibitem[Zhao et~al\mbox{.}(2017)]%
        {zhao2017supporting}
\bibfield{author}{\bibinfo{person}{Jian Zhao}, \bibinfo{person}{Michael Glueck}, \bibinfo{person}{Petra Isenberg}, \bibinfo{person}{Fanny Chevalier}, {and} \bibinfo{person}{Azam Khan}.} \bibinfo{year}{2017}\natexlab{}.
\newblock \showarticletitle{Supporting handoff in asynchronous collaborative sensemaking using knowledge-transfer graphs}.
\newblock \bibinfo{journal}{\emph{IEEE transactions on visualization and computer graphics}} \bibinfo{volume}{24}, \bibinfo{number}{1} (\bibinfo{year}{2017}), \bibinfo{pages}{340--350}.
\newblock


\bibitem[Zhou et~al\mbox{.}(2018)]%
        {zhou2018dataset}
\bibfield{author}{\bibinfo{person}{Kangyan Zhou}, \bibinfo{person}{Shrimai Prabhumoye}, {and} \bibinfo{person}{Alan~W Black}.} \bibinfo{year}{2018}\natexlab{}.
\newblock \showarticletitle{A dataset for document grounded conversations}.
\newblock \bibinfo{journal}{\emph{arXiv preprint arXiv:1809.07358}} (\bibinfo{year}{2018}).
\newblock


\end{thebibliography}

\appendix

\section{Prompt}
\label{appen:prompt}

``Act as a lifelong resident of [City, State] with deep local knowledge. Engage in natural, casual conversation as if chatting with a friend—avoid any mention of being AI. Prioritize brevity (1-2 sentences max), and if listing items (e.g., recommendations), weave them organically into sentences (e.g., 'I love Spot X, Café Y, and Park Z—each has its own vibe!'). Use colloquial language (contractions, slang) and local references where appropriate. Subtly ask follow-up questions (e.g., 'Have you tried…?') to deepen the dialogue. If unsure about a detail, respond vaguely or redirect (e.g., 'I’ve heard mixed things—what’s your take?'). Never use markdown or structured lists.''

\section{hierarchical localness conceptual framework}
\label{appen_frame}

\begin{longtable}{%
  >{\raggedright\arraybackslash}p{\dimexpr0.12\linewidth-2\tabcolsep}%
  >{\raggedright\arraybackslash}p{\dimexpr0.16\linewidth-2\tabcolsep}%
  >{\raggedright\arraybackslash}p{\dimexpr0.22\linewidth-2\tabcolsep}%
  >{\raggedright\arraybackslash}p{\dimexpr0.4\linewidth-2\tabcolsep}%
  >{\raggedright\arraybackslash}p{\dimexpr0.1\linewidth-2\tabcolsep}%
}
\caption{Framework of Localness Construct and their Frequencies} \label{tab:localness} \\

\toprule
\textbf{Domain} & \textbf{Dimension} & \textbf{Component} & \textbf{Sub-component} & \textbf{Count} \\
\midrule
\endfirsthead

\multicolumn{5}{c}%
  {{\tablename\ \thetable{} -- continued from previous page}} \\
\toprule
\textbf{Domain} & \textbf{Dimension} & \textbf{Component} & \textbf{Sub-component} & \textbf{Count} \\
\midrule
\endhead

\midrule
\multicolumn{5}{r}{{Continued on next page}} \\
\endfoot

\bottomrule
\endlastfoot

\multirow[t]{33}{*}{Cognitive} & \multirow[t]{9}{*}{Cultural} & \multirow[t]{3}{*}{Food Culture} & Awareness of food sources         & 6  \\
                              &                           &                               & Knowledge of local cuisine        & 12 \\
                              &                           &                               & Understanding food traditions     & 9  \\
\cmidrule{3-5}
                              &                           & \multirow[t]{3}{*}{Language/Dialect} & Knowledge of local expressions & 20 \\
                              &                           &                                     & Understanding local humor      & 8  \\
                              &                           &                                     & Understanding regional accents & 7  \\
\cmidrule{3-5}
                              &                           & \multirow[t]{3}{*}{Local Customs/Norms} & Knowing local etiquette       & 6  \\
                              &                           &                                     & Understanding community values & 23 \\
                              &                           &                                     & Understanding unwritten rules  & 5  \\
\cmidrule{2-5}
                              & \multirow[t]{12}{*}{Environmental} & \multirow[t]{4}{*}{Ecological Understanding} & Awareness of seasonal patterns       & 6  \\
                              &                                     &                                             & Environmental of environmental cycles & 5  \\
                              &                                     &                                             & Knowledge of local watersheds           & 5  \\
                              &                                     &                                             & Understanding of environmental issues   & 5  \\
\cmidrule{3-5}
                              &                                     & \multirow[t]{4}{*}{Geographic Familiarity}   & Familiarity of spatial organization  & 13 \\
                              &                                     &                                             & Knowledge of physical boundaries     & 8  \\
                              &                                     &                                             & Recognition of landmarks             & 10 \\
                              &                                     &                                             & Understanding of land features       & 8  \\
\cmidrule{3-5}
                              &                                     & \multirow[t]{4}{*}{Natural Environment}       & Awareness of ecological systems  & 4  \\
                              &                                     &                                             & Environmental literacy           & 8  \\
                              &                                     &                                             & Flora/fauna knowledge            & 7  \\
                              &                                     &                                             & Understanding natural patterns   & 6  \\
\cmidrule{2-5}
                              & \multirow[t]{20}{*}{Knowledge}       & \multirow[t]{4}{*}{Change Awareness}          & Awareness of future plans                  & 6  \\
                              &                                     &                                             & Memory of what used to exist               & 17 \\
                              &                                     &                                             & Recognition of developments and changes    & 13 \\
                              &                                     &                                             & Understanding of ongoing transformations   & 11 \\
\cmidrule{3-5}
                              &                                     & \multirow[t]{4}{*}{Hidden Gems}             & Access to insider information  & 6  \\
                              &                                     &                                             & Awareness of unofficial landmarks    & 6  \\
                              &                                     &                                             & Knowledge of local secrets             & 9  \\
                              &                                     &                                             & Understanding special tricks/hacks     & 8  \\
\cmidrule{3-5}
                              &                                     & \multirow[t]{4}{*}{Historical Knowledge}    & Awareness of local history        & 22 \\
                              &                                     &                                             & Familiarity with traditions         & 13 \\
                              &                                     &                                             & Knowledge of past events and impacts & 11 \\
                              &                                     &                                             & Understanding historical context    & 11 \\
\cmidrule{3-5}
                              &                                     & \multirow[t]{4}{*}{Local Recommendations}   & Ability to make informed recommendations   & 28 \\
                              &                                     &                                             & Awareness of local options and alternatives & 49 \\
                              &                                     &                                             & Knowledge of best places for various needs  & 31 \\
                              &                                     &                                             & Understanding local services and amenities    & 21 \\
\cmidrule{3-5}
                              &                                     & \multirow[t]{4}{*}{Navigation/Wayfinding}   & Familiarity with transportation system    & 14 \\
                              &                                     &                                             & Independent navigation without GPS/maps  & 12 \\
                              &                                     &                                             & Knowledge of shortcuts and alternative routes & 9  \\
                              &                                     &                                             & Understanding spatial layout               & 14 \\
\midrule
\multirow[t]{12}{*}{Physical}  & \multirow[t]{3}{*}{Environmental}     & Ecological Understanding                      & Connection to local ecosystem              & 9  \\
\cmidrule{3-5}
                              &                                     & Geographic Familiarity                        & Navigation proficiency                       & 8  \\
\cmidrule{3-5}
                              &                                     & Natural Environment                           & Connection to natural spaces                 & 4  \\
\cmidrule{2-5}
                              & \multirow[t]{9}{*}{Temporal}            & \multirow[t]{3}{*}{Being Born/Native}         & Automatic claim to localness                 & 30 \\
                              &                                     &                                             & Deep historical connection                   & 25 \\
                              &                                     &                                             & Lifelong familiarity                         & 52 \\
\cmidrule{3-5}
                              &                                     & \multirow[t]{3}{*}{Formative Years}         & Core memory formation                        & 50 \\
                              &                                     &                                             & Deep cultural absorption                     & 28 \\
                              &                                     &                                             & Early life experiences                       & 58 \\
\cmidrule{3-5}
                              &                                     & \multirow[t]{3}{*}{Long-term Residence}     & Accumulated experience                       & 76 \\
                              &                                     &                                             & Investment in place                          & 38 \\
                              &                                     &                                             & Witnessing area changes                      & 59 \\
\midrule
\multirow[t]{40}{*}{Relational} & \multirow[t]{15}{*}{Emotional}        & \multirow[t]{5}{*}{Feeling of Home}           & Comfort in environment         & 23 \\
                              &                                     &                                             & Desire to return/stay          & 35 \\
                              &                                     &                                             & Emotional investment           & 54 \\
                              &                                     &                                             & Emotional security             & 46 \\
                              &                                     &                                             & Place attachment               & 61 \\
\cmidrule{3-5}
                              &                                     & \multirow[t]{5}{*}{Identity Connection}     & Identification with local character  & 42 \\
                              &                                     &                                             & Personal investment in place         & 25 \\
                              &                                     &                                             & Place as part of self-definition     & 38 \\
                              &                                     &                                             & Pride in local association           & 55 \\
                              &                                     &                                             & Shared identity with community       & 50 \\
\cmidrule{3-5}
                              &                                     & \multirow[t]{5}{*}{Sense of Belonging}      & Feeling socially connected          & 83 \\
                              &                                     &                                             & Feeling attached                & 86 \\
                              &                                     &                                             & Feeling accepted                 & 28 \\
                              &                                     &                                             & Natural fit with place           & 49 \\
                              &                                     &                                             & Sense of rightful presence       & 37 \\
\cmidrule{2-5}
                              & \multirow[t]{25}{*}{Social/Community}   & \multirow[t]{5}{*}{Active Participation}      & Attending local events           & 42 \\
                              &                                     &                                             & Contributing to local life       & 42 \\
                              &                                     &                                             & Engaging in community initiatives& 32 \\
                              &                                     &                                             & Participating in local organizations & 28 \\
                              &                                     &                                             & Volunteering in community        & 13 \\
\cmidrule{3-5}
                              &                                     & \multirow[t]{5}{*}{Civic Engagement}        & Advocacy for local issues        & 7  \\
                              &                                     &                                             & Attending community meetings     & 8  \\
                              &                                     &                                             & Engaging with local government   & 6  \\
                              &                                     &                                             & Political participation          & 8  \\
                              &                                     &                                             & Voting in local elections        & 3  \\
\cmidrule{3-5}
                              &                                     & \multirow[t]{5}{*}{Community Investment}    & Care about local issues          & 25 \\
                              &                                     &                                             & Contributing to community improvement & 14 \\
                              &                                     &                                             & Involvement in local decisions   & 17 \\
                              &                                     &                                             & Long-term commitment             & 49 \\
                              &                                     &                                             & Supporting local businesses      & 18 \\
\cmidrule{3-5}
                              &                                     & \multirow[t]{5}{*}{Personal Relationships}  & Building social networks         & 46 \\
                              &                                     &                                             & Having local friends             & 52 \\
                              &                                     &                                             & Knowing neighbors                & 27 \\
                              &                                     &                                             & Recognition by community members & 51 \\
                              &                                     &                                             & Regular social interactions      & 63 \\
\end{longtable}

\bigskip
\section{Significant results of localness interpretation differences}
\label{appen_inte}

\renewcommand{\arraystretch}{1.2}

\begin{longtable}{%
  >{\raggedright\arraybackslash}p{\dimexpr0.18\linewidth-2\tabcolsep}%
  >{\raggedright\arraybackslash}p{\dimexpr0.27\linewidth-2\tabcolsep}%
  >{\raggedright\arraybackslash}p{\dimexpr0.08\linewidth-2\tabcolsep}%
  >{\raggedright\arraybackslash}p{\dimexpr0.08\linewidth-2\tabcolsep}%
  >{\raggedright\arraybackslash}p{\dimexpr0.15\linewidth-2\tabcolsep}%
  >{\raggedright\arraybackslash}p{\dimexpr0.15\linewidth-2\tabcolsep}%
  >{\raggedright\arraybackslash}p{\dimexpr0.08\linewidth-2\tabcolsep}%
}
\caption{Significant localness interpretation differences} \label{tab:stats} \\

\toprule
\textbf{Category} & \textbf{Name} & \textbf{Mean (Yes)} & \textbf{Mean (No)} & \textbf{CI} & \textbf{MW-p (Corrected)} & \textbf{Cohensd} \\
\midrule
\endfirsthead

\multicolumn{7}{c}%
  {{\tablename\ \thetable{} -- continued from previous page}} \\
\toprule
\textbf{Category} & \textbf{Name} & \textbf{Mean (Yes)} & \textbf{Mean (No)} & \textbf{CI} & \textbf{MW-p (Corrected)} & \textbf{Cohensd} \\
\midrule
\endhead

\midrule
\multicolumn{7}{r}{{Continued on next page}} \\
\endfoot

\bottomrule
\endlastfoot

Domain      & Physical           & 1.537  & 3.113  & [-2.329, -0.812] & 0.001 & -0.656 \\
\addlinespace
Domain      & Relational         & 6.963  & 4.493  & [0.782, 4.103]   & 0.038 & 0.403  \\
\addlinespace
Dimension   & Temporal           & 1.485  & 3.056  & [-2.322, -0.830] & 0.001 & -0.663 \\
\addlinespace
Component   & Active Participation & 1.037 & 0.254  & [0.445, 1.108]   & 0.005 & 0.561  \\
\addlinespace
Component   & Being Born/Native  & 0.291  & 0.958  & [-0.958, -0.372] & 0.000 & -0.745 \\
\addlinespace
Component   & Formative Years    & 0.455  & 1.056  & [-0.923, -0.272] & 0.002 & -0.571 \\
\addlinespace
Sub-component & Attending local events     & 0.269 & 0.085  & [0.085, 0.282]   & 0.018 & 0.465  \\
\addlinespace
Sub-component & Automatic claim to localness & 0.067 & 0.296 & [-0.341, -0.115] & 0.001 & -0.676 \\
\addlinespace
Sub-component & Awareness of local options and alternatives & 0.306 & 0.113 & [0.085, 0.300] & 0.018 & 0.462 \\
\addlinespace
Sub-component & Contributing to local life & 0.276 & 0.070  & [0.109, 0.299]   & 0.007 & 0.523  \\
\addlinespace
Sub-component & Core memory formation      & 0.164 & 0.394  & [-0.359, -0.102] & 0.005 & -0.552 \\
\addlinespace
Sub-component & Deep cultural absorption   & 0.090 & 0.225  & [-0.244, -0.029] & 0.049 & -0.401 \\
\addlinespace
Sub-component & Deep historical connection & 0.067 & 0.225  & [-0.266, -0.052] & 0.011 & -0.494 \\
\addlinespace
Sub-component & Early life experiences     & 0.201 & 0.437  & [-0.369, -0.098] & 0.005 & -0.536 \\
\addlinespace
Sub-component & Engaging in community initiatives & 0.216 & 0.042 & [0.092, 0.255] & 0.011 & 0.490 \\
\addlinespace
Sub-component & Flora/fauna knowledge      & 0.000 & 0.056  & [-0.113, -0.014] & 0.041 & -0.413 \\
\addlinespace
Sub-component & Lifelong familiarity       & 0.157 & 0.437  & [-0.410, -0.146] & 0.001 & -0.672 \\
\addlinespace
Sub-component & Participating in local organizations & 0.187 & 0.042 & [0.058, 0.226] & 0.035 & 0.427 \\
\end{longtable}

\section{Different Category Distributions in LD's Questions during Conversations}
\label{appen:question}

\begin{figure*}
\centering
\begin{subfigure}[b]{0.45\textwidth}
    \includegraphics[width=\textwidth]{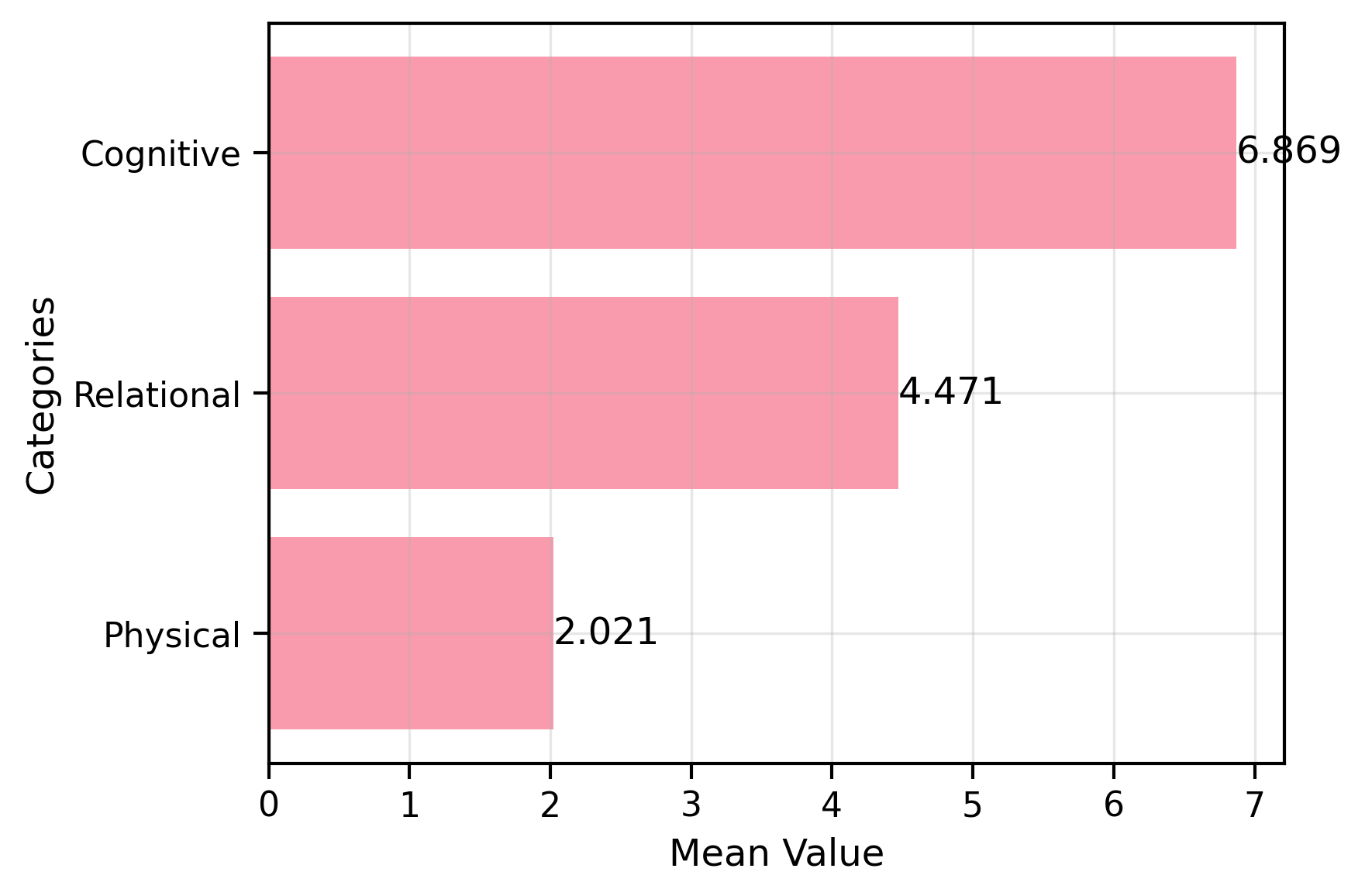}
    \caption{Distribution of Domain Categories}
    \label{fig:dom_selected}
\end{subfigure}
\hfill
\begin{subfigure}[b]{0.45\textwidth}
    \includegraphics[width=\textwidth]{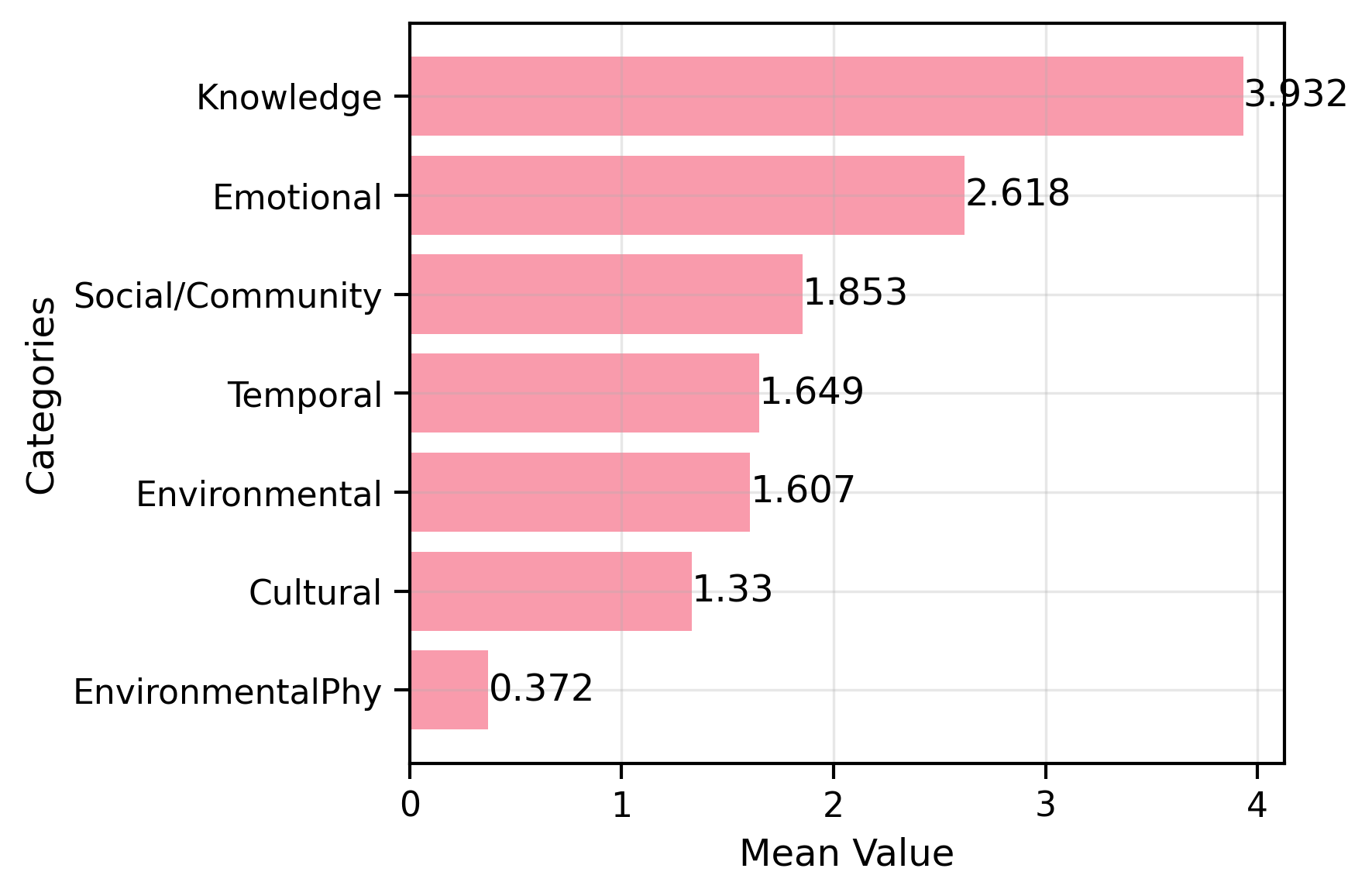}
    \caption{Distribution of Dimension Categories}
    \label{fig:dim_selected}
\end{subfigure}

\begin{subfigure}[b]{0.45\textwidth}
    \includegraphics[width=\textwidth]{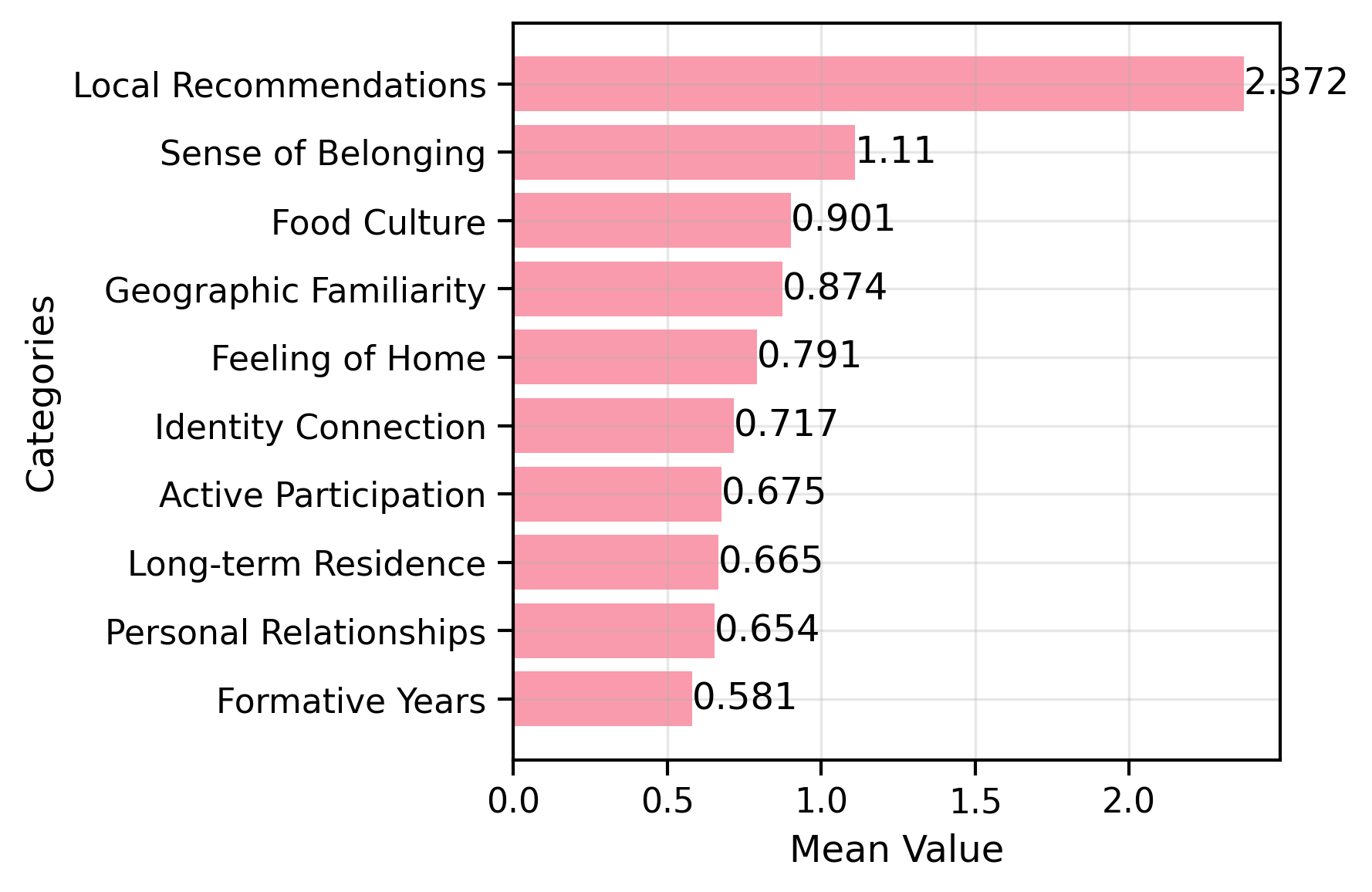}
    \caption{Distribution of Top 10 Components Categories}
    \label{fig:com_selected}
\end{subfigure}
\hfill
\begin{subfigure}[b]{0.45\textwidth}
    \includegraphics[width=\textwidth]{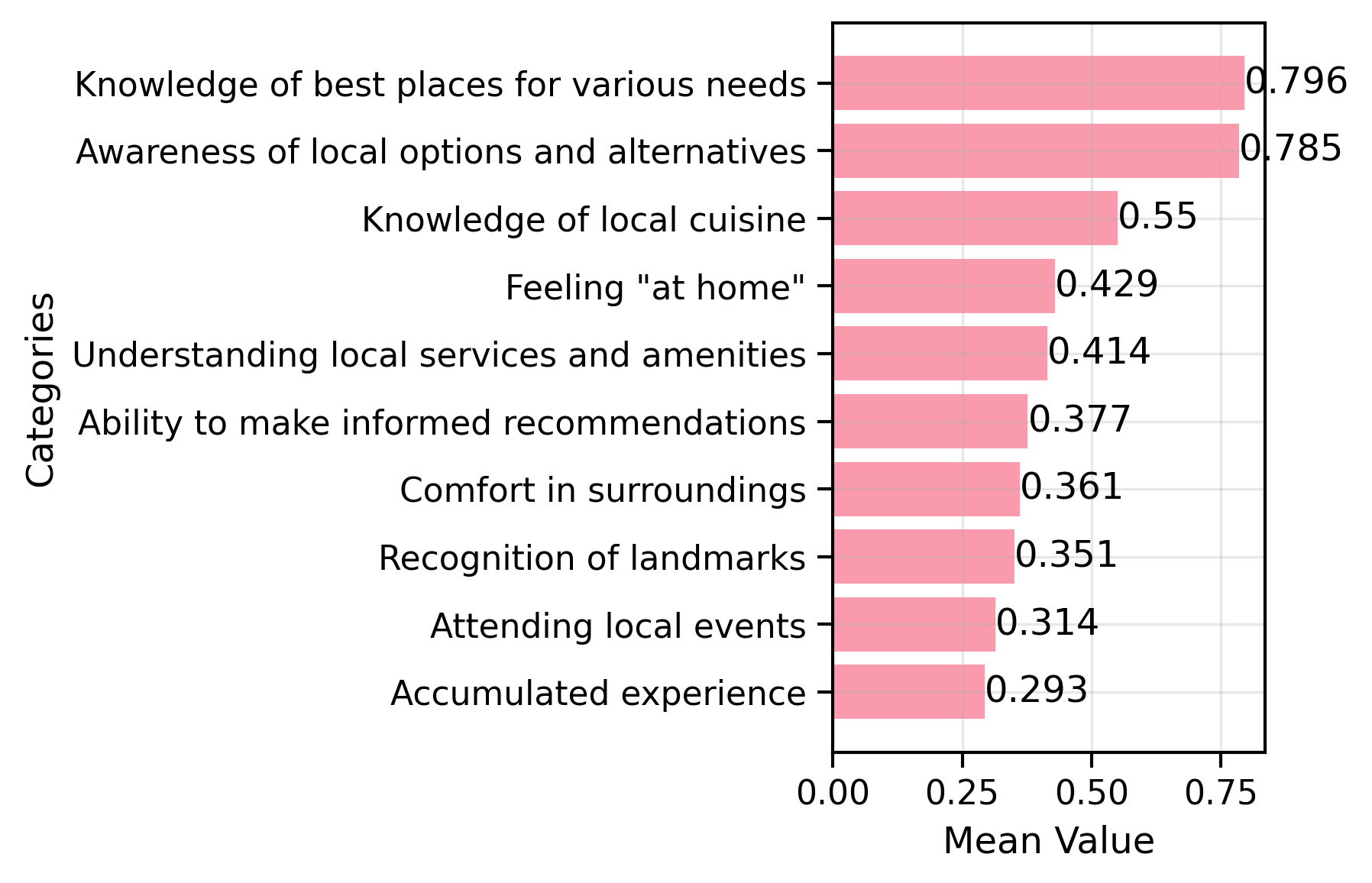}
    \caption{Distribution of Top 10 Sub-components Categories}
    \label{fig:sub_selected}
\end{subfigure}
\caption{Different Category Distributions in LD's Questions during Selected Conversations for Localness Judgment}
\label{fig:all_selected}
\end{figure*}

\begin{figure*}
\centering
\begin{subfigure}[b]{0.45\textwidth}
    \includegraphics[width=\textwidth]{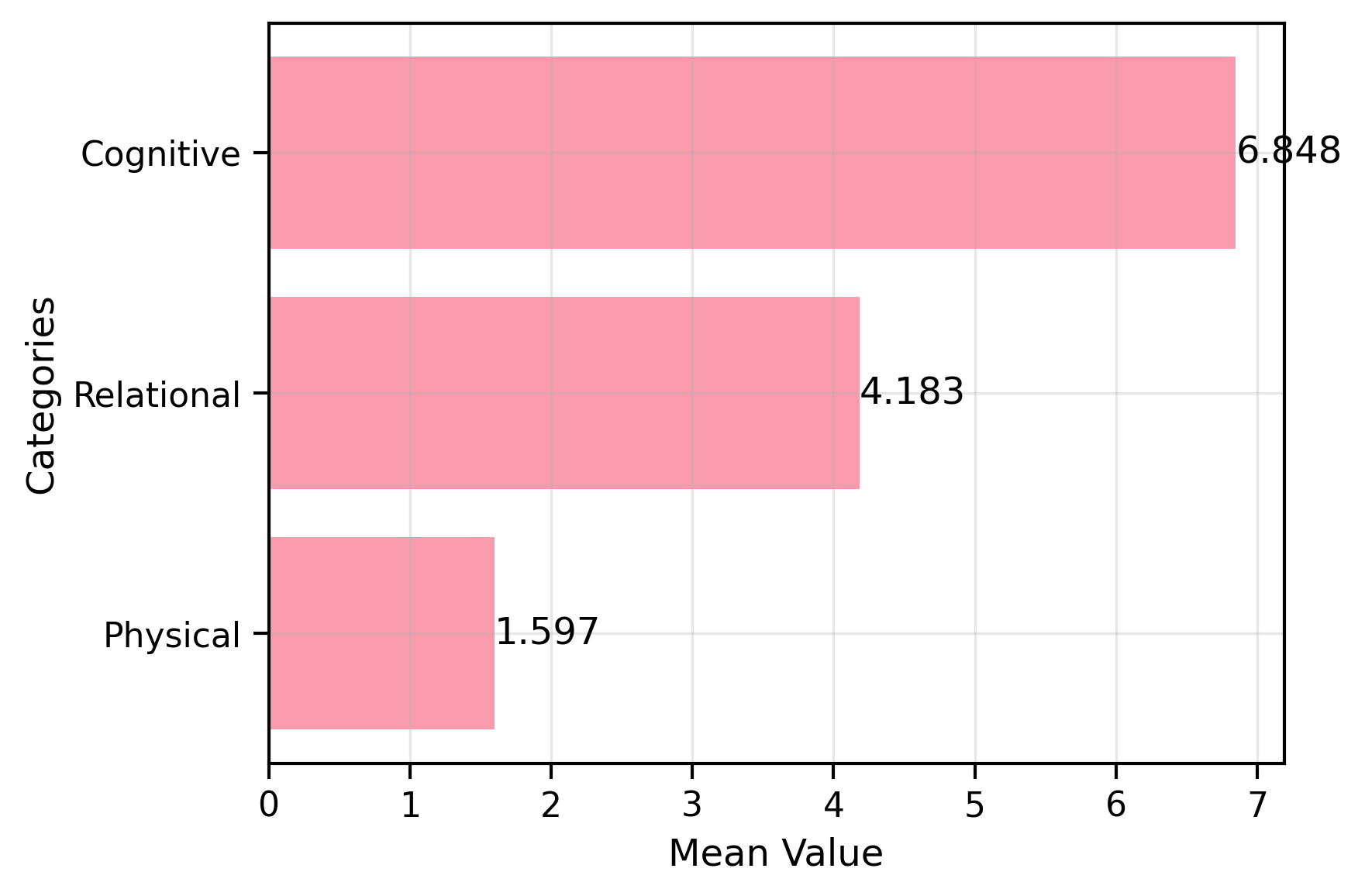}
    \caption{Distribution of Domain Categories}
    \label{fig:dom_bot}
\end{subfigure}
\hfill
\begin{subfigure}[b]{0.45\textwidth}
    \includegraphics[width=\textwidth]{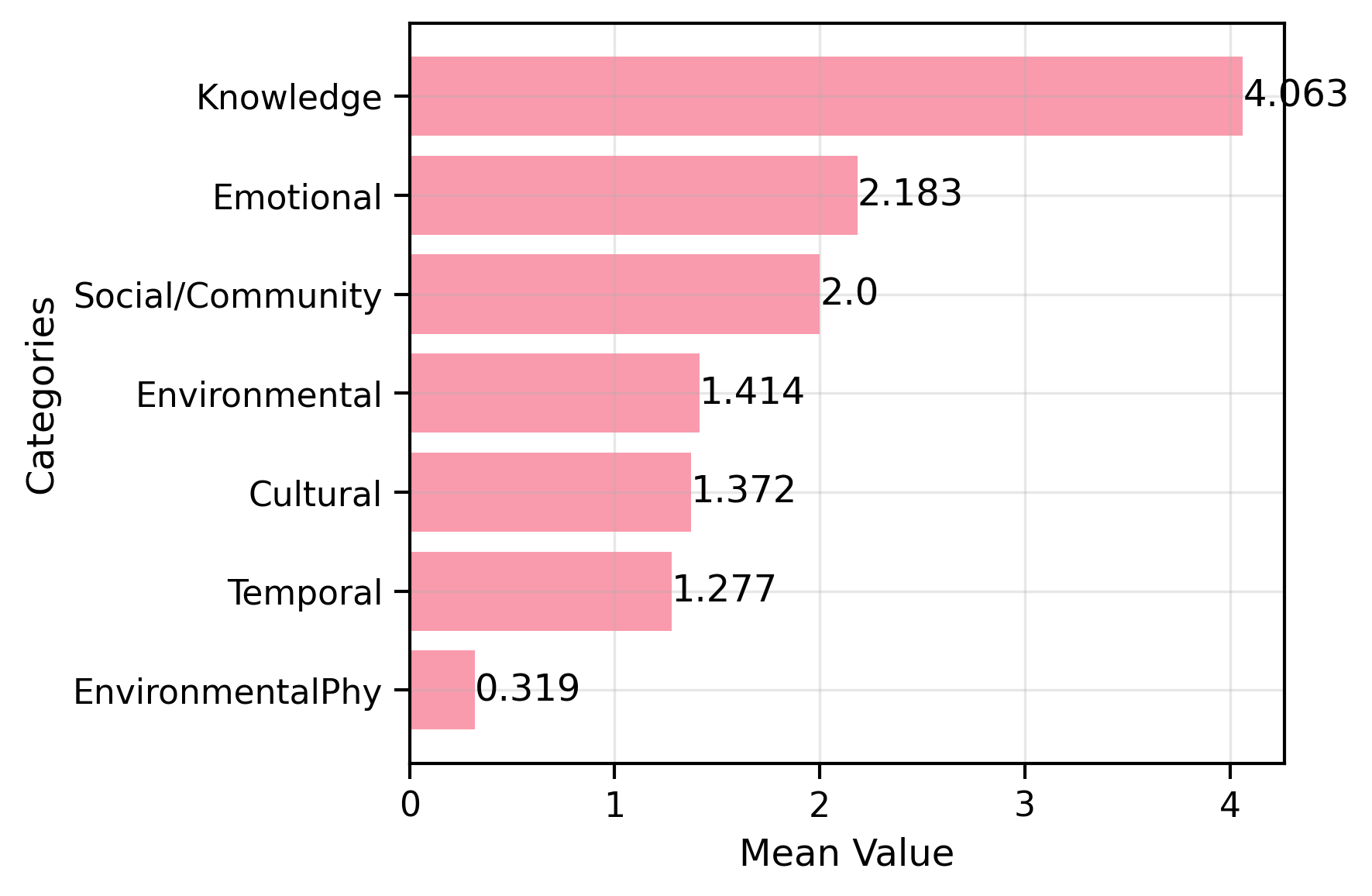}
    \caption{Distribution of Dimension Categories}
    \label{fig:dim_bot}
\end{subfigure}

\begin{subfigure}[b]{0.45\textwidth}
    \includegraphics[width=\textwidth]{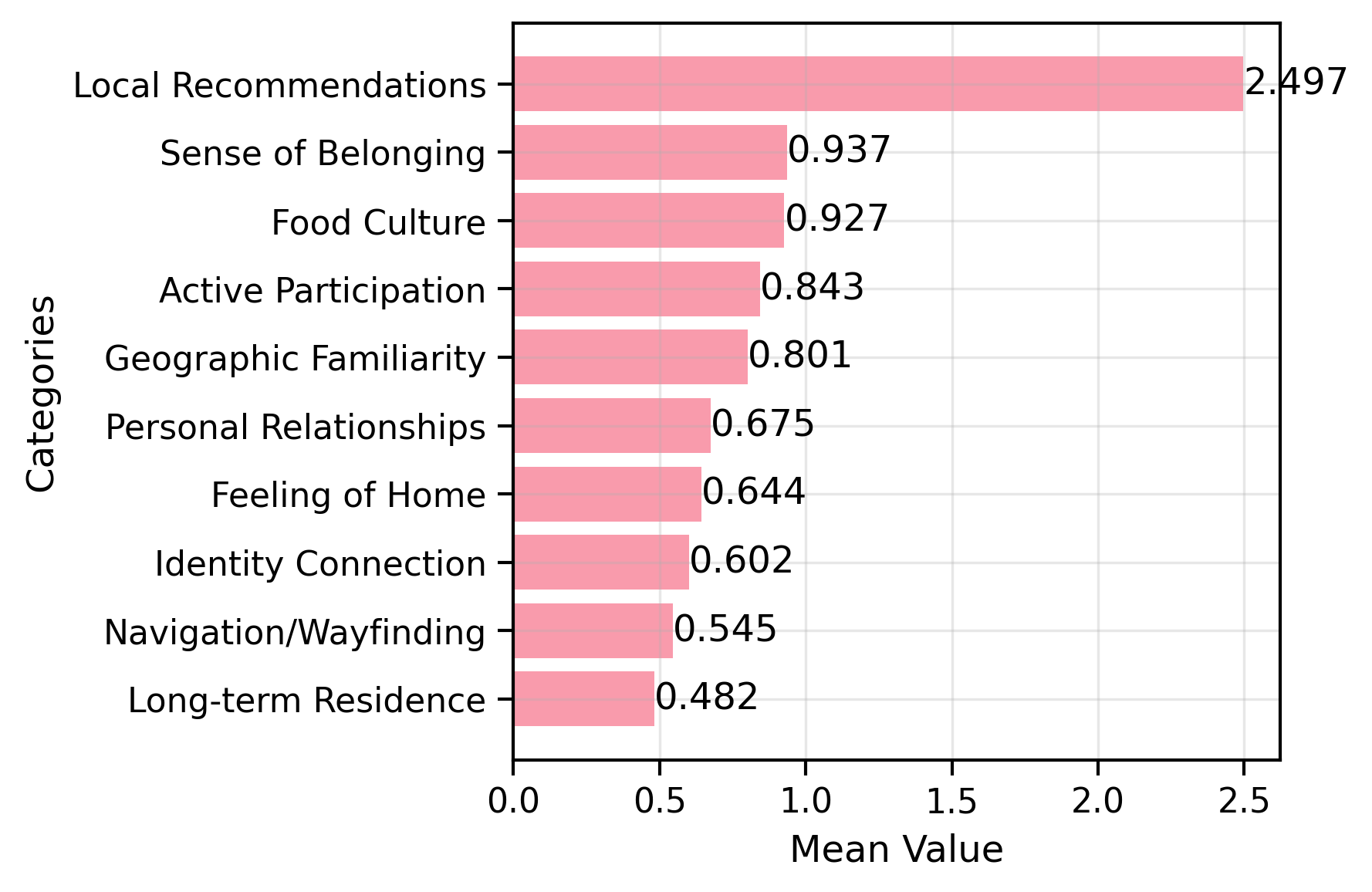}
    \caption{Distribution of Top 10 Components Categories}
    \label{fig:com_bot}
\end{subfigure}
\hfill
\begin{subfigure}[b]{0.45\textwidth}
    \includegraphics[width=\textwidth]{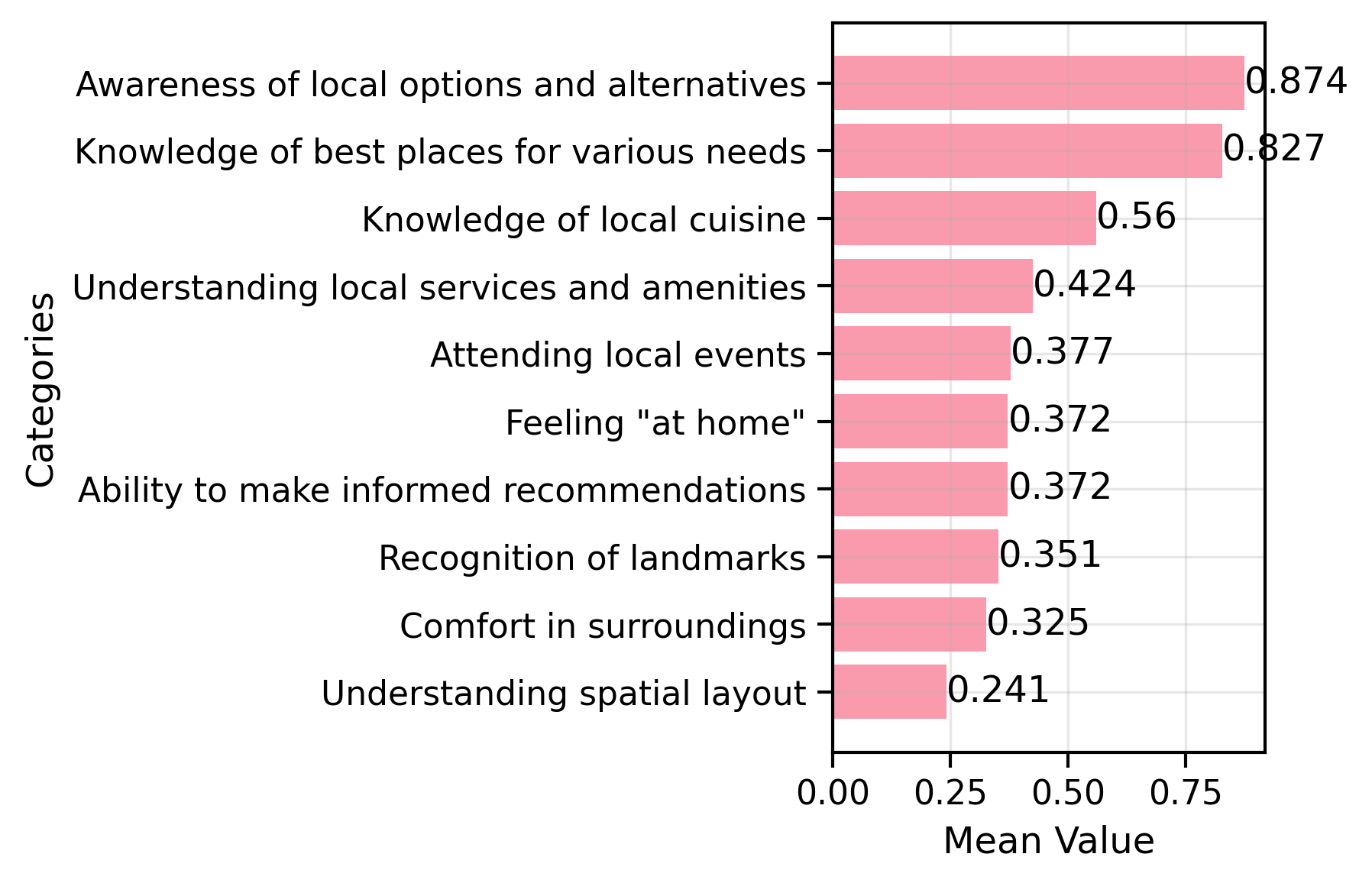}
    \caption{Distribution of Top 10 Sub-components Categories}
    \label{fig:sub_bot}
\end{subfigure}
\caption{Different Category Distributions in LD's Questions during Selected Conversations for LLM Judgment}
\label{fig:all_bot}
\end{figure*}

\section{Results of Bayesian Mixed-effect ZINB Model for Conversation and Selected Conversation Comparison}
\label{appen:zinb}

\begin{table*}
    \centering
    \caption{Results of Count Component Using \texttt{beta\_accuracy\_count} and \texttt{beta\_round\_count} Parameters in Bayesian Mixed-effect ZINB Model for Conversation Comparison. Pink highlights indicate significant results where HDI (Highest Density Interval) bounds do not cross zero}
    \label{tab:count_conv}
    \begin{tabular}{l l r r r r r r}
        \toprule
        \multirow{2}{*}{Levels} & \multirow{2}{*}{Features Name} & \multicolumn{3}{c}{\texttt{beta\_accuracy\_count}} & \multicolumn{3}{c}{\texttt{beta\_round\_count}} \\
        \cmidrule(lr){3-5} \cmidrule(lr){6-8}
        & & Mean & HDI 3\% & HDI 97\% & Mean & HDI 3\% & HDI 97\% \\
        \midrule
        Domain & Cognitive & -0.04 & -0.31 & 0.25 & \cellcolor{pink}-0.22 & \cellcolor{pink}-0.35 & \cellcolor{pink}-0.09 \\
        Domain & Physical & 0.08 & -0.25 & 0.40 & \cellcolor{pink}-0.28 & \cellcolor{pink}-0.53 & \cellcolor{pink}-0.04 \\
        Domain & Relational & 0.16 & -0.14 & 0.48 & \cellcolor{pink}-0.23 & \cellcolor{pink}-0.45 & \cellcolor{pink}-0.03 \\
        Dimension & Cultural & \cellcolor{pink}0.35 & \cellcolor{pink}0.04 & \cellcolor{pink}0.64 & 0.10 & -0.08 & 0.28 \\
        Dimension & Emotional & 0.15 & -0.19 & 0.48 & -0.19 & -0.45 & 0.08 \\
        Dimension & Environmental & 0.15 & -0.16 & 0.47 & -0.21 & -0.45 & 0.04 \\
        Dimension & Knowledge & 0.05 & -0.23 & 0.34 & -0.06 & -0.20 & 0.08 \\
        Dimension & Social/Community & 0.30 & -0.04 & 0.60 & -0.02 & -0.23 & 0.21 \\
        Dimension & Temporal & 0.07 & -0.28 & 0.40 & -0.25 & -0.53 & 0.04 \\
        Dimension & EnvironmentalPhy & 0.10 & -0.28 & 0.46 & -0.02 & -0.37 & 0.36 \\
        Component & Active Participation & 0.23 & -0.13 & 0.57 & 0.04 & -0.27 & 0.34 \\
        Component & Being Born/Native & 0.08 & -0.29 & 0.43 & -0.06 & -0.43 & 0.27 \\
        Component & Change Awareness & 0.08 & -0.31 & 0.43 & 0.00 & -0.34 & 0.36 \\
        Component & Civic Engagement & 0.00 & -0.37 & 0.38 & -0.00 & -0.38 & 0.37 \\
        Component & Community Investment & 0.13 & -0.25 & 0.50 & -0.03 & -0.37 & 0.31 \\
        Component & Ecological Understanding & 0.07 & -0.33 & 0.44 & -0.02 & -0.40 & 0.33 \\
        Component & Feeling of Home & 0.13 & -0.24 & 0.48 & -0.07 & -0.41 & 0.26 \\
        Component & Food Culture & 0.28 & -0.03 & 0.61 & 0.09 & -0.14 & 0.31 \\
        Component & Formative Years & 0.07 & -0.30 & 0.43 & -0.09 & -0.43 & 0.27 \\
        Component & Geographic Familiarity & 0.17 & -0.17 & 0.50 & -0.08 & -0.34 & 0.21 \\
        Component & Hidden Gems & 0.05 & -0.31 & 0.40 & -0.00 & -0.36 & 0.37 \\
        Component & Historical Knowledge & 0.08 & -0.28 & 0.46 & 0.00 & -0.36 & 0.36 \\
        Component & Identity Connection & 0.15 & -0.20 & 0.51 & -0.05 & -0.37 & 0.28 \\
        Component & Language/Dialect & 0.08 & -0.30 & 0.44 & 0.02 & -0.38 & 0.37 \\
        Component & Local Customs/Norms & 0.02 & -0.36 & 0.38 & 0.01 & -0.36 & 0.39 \\
        Component & Local Recommendations & 0.22 & -0.07 & 0.50 & 0.03 & -0.11 & 0.17 \\
        Component & Long-term Residence & 0.10 & -0.29 & 0.45 & -0.06 & -0.39 & 0.28 \\
        Component & Natural Environment & 0.07 & -0.31 & 0.44 & -0.04 & -0.41 & 0.31 \\
        Component & Navigation/Wayfinding & 0.14 & -0.24 & 0.49 & -0.04 & -0.37 & 0.29 \\
        Component & Personal Relationships & 0.12 & -0.26 & 0.47 & -0.02 & -0.38 & 0.33 \\
        Component & Sense of Belonging & 0.18 & -0.15 & 0.52 & -0.07 & -0.34 & 0.21 \\
        Component & Geographic FamiliarityPhy & 0.01 & -0.38 & 0.39 & 0.00 & -0.37 & 0.38 \\
        Component & Natural EnvironmentPhy & 0.07 & -0.30 & 0.42 & -0.01 & -0.37 & 0.36 \\
        Component & Ecological UnderstandingPhy & 0.03 & -0.35 & 0.41 & -0.00 & -0.36 & 0.40 \\
        \bottomrule
    \end{tabular}
\end{table*}

\begin{table*}
    \centering
    \caption{Results of Zero-inflation Component Using \texttt{beta\_accuracy\_zero} and \texttt{beta\_round\_zero} Parameters in Bayesian Mixed-effect ZINB Model for Selected Conversation Comparison. Pink highlights indicate significant results where HDI (Highest Density Interval) bounds do not cross zero}
    \label{tab:zero_selected}
    \begin{tabular}{l l r r r r r r}
        \toprule
        \multirow{2}{*}{Levels} & \multirow{2}{*}{Features Name} & \multicolumn{3}{c}{\texttt{beta\_accuracy\_zero}} & \multicolumn{3}{c}{\texttt{beta\_round\_zero}} \\
        \cmidrule(lr){3-5} \cmidrule(lr){6-8}
        & & Mean & HDI 3\% & HDI 97\% & Mean & HDI 3\% & HDI 97\% \\
        \midrule
        Domain & Cognitive & -0.02 & -0.34 & 0.31 & 0.18 & -0.06 & 0.41 \\
        Domain & Physical & -0.05 & -0.37 & 0.28 &0.11 & -0.11 & 0.35 \\
        Domain & Relational & -0.08 & -0.42 & 0.23 & 0.06 & -0.17 & 0.29 \\
        Dimension & Cultural & -0.21 & -0.53 & 0.12 & -0.02 & -0.26 & 0.20 \\
        Dimension & Emotional & -0.13 & -0.47 & 0.19 & -0.04 & -0.26 & 0.20 \\
        Dimension & Environmental & -0.17 & -0.50 & 0.16 & 0.21 & -0.02 & 0.46 \\
        Dimension & Knowledge & 0.00 & -0.31 & 0.32 & 0.06 & -0.17 & 0.29 \\
        Dimension & Social/Community & -0.10 & -0.41 & 0.24 & 0.10 & -0.13 & 0.33 \\
        Dimension & Temporal & -0.06 & -0.39 & 0.28 & 0.08 & -0.15 & 0.33 \\
        Dimension & EnvironmentalPhy & -0.20 & -0.50 & 0.13 & -0.08 & -0.32 & 0.16 \\
        Component & Active Participation & -0.17 & -0.48 & 0.17 & -0.13 & -0.36 & 0.11 \\
        Component & Being Born/Native & -0.17 & -0.48 & 0.17 & -0.01 & -0.25 & 0.24 \\
        Component & Change Awareness & -0.17 & -0.50 & 0.16 & -0.08 & -0.31 & 0.16 \\
        Component & Civic Engagement & -0.19 & -0.51 & 0.14 & -0.13 & -0.37 & 0.11 \\
        Component & Community Investment & -0.17 & -0.50 & 0.16 & 0.03 & -0.22 & 0.27 \\
        Component & Ecological Understanding & -0.24 & -0.57 & 0.09 & -0.01 & -0.24 & 0.23 \\
        Component & Feeling of Home & -0.17 & -0.51 & 0.14 & -0.08 & -0.32 & 0.15 \\
        Component & Food Culture & -0.15 & -0.46 & 0.18 & -0.02 & -0.26 & 0.21 \\
        Component & Formative Years & -0.09 & -0.42 & 0.25 & 0.02 & -0.22 & 0.26 \\
        Component & Geographic Familiarity & -0.17 & -0.50 & 0.16 & 0.05 & -0.18 & 0.29 \\
        Component & Hidden Gems & -0.23 & -0.56 & 0.09 & -0.17 & -0.39 & 0.08 \\
        Component & Historical Knowledge & -0.18 & -0.50 & 0.16 & -0.12 & -0.36 & 0.12 \\
        Component & Identity Connection & -0.19 & -0.51 & 0.14 & -0.13 & -0.38 & 0.10 \\
        Component & Language/Dialect & -0.29 & -0.62 & 0.04 & -0.07 & -0.31 & 0.17 \\
        Component & Local Customs/Norms & -0.25 & -0.58 & 0.07 & -0.17 & -0.41 & 0.06 \\
        Component & Local Recommendations & -0.07 & -0.42 & 0.24 & -0.09 & -0.31 & 0.14 \\
        Component & Long-term Residence & -0.14 & -0.47 & 0.19 & -0.01 & -0.24 & 0.23 \\
        Component & Natural Environment & -0.21 & -0.55 & 0.11 & -0.03 & -0.26 & 0.21 \\
        Component & Navigation/Wayfinding & -0.19 & -0.50 & 0.16 & -0.07 & -0.30 & 0.18 \\
        Component & Personal Relationships & -0.18 & -0.50 & 0.16 & 0.00 & -0.24 & 0.24 \\
        Component & Sense of Belonging & -0.11 & -0.44 & 0.22 & -0.04 & -0.29 & 0.19 \\
        Component & Geographic Familiarity(Phy) & -0.20 & -0.53 & 0.13 & -0.15 & -0.39 & 0.08 \\
        Component & Natural Environment(Phy) & -0.22 & -0.54 & 0.12 & -0.07 & -0.31 & 0.17 \\
        Component & Ecological Understanding(Phy) & -0.24 & -0.58 & 0.07 & -0.12 & -0.37 & 0.11 \\
        \bottomrule
    \end{tabular}
\end{table*}

\begin{table*}[t]
    \centering
    \caption{Results of Count Component Using \texttt{beta\_accuracy\_count} and \texttt{beta\_round\_count} Parameters in Bayesian Mixed-effect ZINB Model for Selected Conversation Comparison. Pink highlights indicate significant results where HDI (Highest Density Interval) bounds do not cross zero}
    \label{tab:count_selected}
    \begin{tabular}{l l r r r r r r}
        \toprule
        \multirow{2}{*}{Levels} & \multirow{2}{*}{Features Name} & \multicolumn{3}{c}{\texttt{beta\_accuracy\_count}} & \multicolumn{3}{c}{\texttt{beta\_round\_count}} \\
        \cmidrule(lr){3-5} \cmidrule(lr){6-8}
        & & Mean & HDI 3\% & HDI 97\% & Mean & HDI 3\% & HDI 97\% \\
        \midrule
        Domain & Cognitive & 0.01 & -0.28 & 0.32 & -0.12 & -0.27 & 0.01 \\
        Domain & Physical & -0.04 & -0.37 & 0.34 & -0.21 & -0.51 & 0.09 \\
        Domain & Relational & 0.01 & -0.33 & 0.34 & -0.17 & -0.42 & 0.08 \\
        Dimension & Cultural & 0.22 & -0.09 & 0.56 & 0.02 & -0.20 & 0.22 \\
        Dimension & Emotional & 0.01 & -0.34 & 0.38 & -0.07 & -0.39 & 0.24 \\
        Dimension & Environmental & 0.08 & -0.27 & 0.42 & -0.21 & -0.50 & 0.08 \\
        Dimension & Knowledge & 0.02 & -0.29 & 0.30 & 0.05 & -0.09 & 0.20 \\
        Dimension & Social/Community & 0.08 & -0.27 & 0.41 & -0.10 & -0.37 & 0.15 \\
        Dimension & Temporal & -0.04 & -0.38 & 0.34 & -0.15 & -0.46 & 0.19 \\
        Dimension & EnvironmentalPhy & 0.04 & -0.33 & 0.41 & -0.01 & -0.37 & 0.34 \\
        Component & Active Participation & 0.07 & -0.29 & 0.45 & 0.01 & -0.32 & 0.34 \\
        Component & Being Born/Native & 0.01 & -0.35 & 0.38 & -0.02 & -0.39 & 0.38 \\
        Component & Change Awareness & 0.03 & -0.35 & 0.42 & -0.01 & -0.36 & 0.34 \\
        Component & Civic Engagement & 0.00 & -0.37 & 0.37 & -0.00 & -0.38 & 0.37 \\
        Component & Community Investment & 0.03 & -0.33 & 0.40 & -0.05 & -0.42 & 0.30 \\
        Component & Ecological Understanding & 0.03 & -0.32 & 0.43 & -0.03 & -0.37 & 0.36 \\
        Component & Feeling of Home & 0.02 & -0.34 & 0.41 & -0.01 & -0.37 & 0.35 \\
        Component & Food Culture & 0.11 & -0.23 & 0.45 & 0.01 & -0.27 & 0.28 \\
        Component & Formative Years & 0.00 & -0.35 & 0.38 & -0.03 & -0.41 & 0.32 \\
        Component & Geographic Familiarity & 0.06 & -0.29 & 0.41 & -0.08 & -0.39 & 0.25 \\
        Component & Hidden Gems & 0.02 & -0.35 & 0.40 & 0.00 & -0.37 & 0.37 \\
        Component & Historical Knowledge & 0.03 & -0.35 & 0.40 & 0.00 & -0.36 & 0.37 \\
        Component & Identity Connection & 0.03 & -0.33 & 0.41 & -0.00 & -0.38 & 0.35 \\
        Component & Language/Dialect & 0.04 & -0.33 & 0.40 & -0.01 & -0.36 & 0.35 \\
        Component & Local Customs/Norms & 0.01 & -0.36 & 0.40 & 0.00 & -0.37 & 0.37 \\
        Component & Local Recommendations & 0.10 & -0.21 & 0.39 & 0.07 & -0.08 & 0.22 \\
        Component & Long-term Residence & 0.01 & -0.34 & 0.39 & -0.05 & -0.39 & 0.32 \\
        Component & Natural Environment & 0.03 & -0.36 & 0.39 & -0.03 & -0.39 & 0.34 \\
        Component & Navigation/Wayfinding & 0.04 & -0.31 & 0.42 & -0.02 & -0.36 & 0.33 \\
        Component & Personal Relationships & 0.04 & -0.33 & 0.42 & -0.02 & -0.38 & 0.35 \\
        Component & Sense of Belonging & 0.03 & -0.34 & 0.39 & -0.05 & -0.40 & 0.29 \\
        Component & Geographic FamiliarityPhy & 0.00 & -0.37 & 0.37 & 0.00 & -0.37 & 0.39 \\
        Component & Natural EnvironmentPhy & 0.03 & -0.35 & 0.41 & -0.01 & -0.40 & 0.36 \\
        Component & Ecological UnderstandingPhy & 0.01 & -0.36 & 0.38 & -0.00 & -0.41 & 0.36 \\
        \bottomrule
    \end{tabular}
\end{table*}

\end{document}